\documentclass[11pt]{article}

\renewcommand{\arraystretch}{1.3}
\makeatletter
\newdimen\normalarrayskip              
\newdimen\minarrayskip                 
\normalarrayskip\baselineskip
\minarrayskip\jot
\newif\ifold             \oldtrue            \def\new{\oldfalse}
\def\arraymode{\ifold\relax\else\displaystyle\fi} 
\def\eqnumphantom{\phantom{(\theequation)}}     
\def\@arrayskip{\ifold\baselineskip\z@\lineskip\z@
     \else
     \baselineskip\minarrayskip\lineskip2\minarrayskip\fi}
\def\@arrayclassz{\ifcase \@lastchclass \@acolampacol \or
\@ampacol \or \or \or \@addamp \or
   \@acolampacol \or \@firstampfalse \@acol \fi
\edef\@preamble{\@preamble
  \ifcase \@chnum
     \hfil$\relax\arraymode\@sharp$\hfil
     \or $\relax\arraymode\@sharp$\hfil
     \or \hfil$\relax\arraymode\@sharp$\fi}}
\def\@array[#1]#2{\setbox\@arstrutbox=\hbox{\vrule
     height\arraystretch \ht\strutbox
     depth\arraystretch \dp\strutbox
     width\z@}\@mkpream{#2}\edef\@preamble{\halign
\noexpand\@halignto
\bgroup \tabskip\z@ \@arstrut \@preamble \tabskip\z@ \cr}%
\let\@startpbox\@@startpbox \let\@endpbox\@@endpbox
  \if #1t\vtop \else \if#1b\vbox \else \vcenter \fi\fi
  \bgroup \let\par\relax
  \let\@sharp##\let\protect\relax
  \@arrayskip\@preamble}
\def\eqnarray{\stepcounter{equation}%
              \let\@currentlabel=\theequation
              \global\@eqnswtrue
              \global\@eqcnt\z@
              \tabskip\@centering
              \let\\=\@eqncr
 \halign to \displaywidth\bgroup
    \eqnumphantom\@eqnsel\hskip\@centering
    $\displaystyle \tabskip\z@ {##}$%
    \global\@eqcnt\@ne \hskip 2\arraycolsep
         $\displaystyle\arraymode{##}$\hfil
    \global\@eqcnt\tw@ \hskip 2\arraycolsep
         $\displaystyle\tabskip\z@{##}$\hfil
         \tabskip\@centering
    &{##}\tabskip\z@\cr}
\begingroup\ifx\undefined\newsymbol \else\def\input#1 {\endgroup}\fi

\newcounter{app}

\def\app{\setcounter{equation}{0}
\def\theequation{A\Roman{app}.\arabic{equation}}\par
   \addvspace{4ex}
   \@afterindentfalse
  \secdef\@app\@dapp}
\newcommand\@app{\@startsection {app}{1}{0ex}%
                                   {-3.5ex \@plus -1ex \@minus -.2ex}%
                                   {2.3ex \@plus.2ex}%
                                   {\normalfont\Large\bf}}

\def\@dapp#1{%
{\parindent \z@ \raggedright  \bf #1}\par\nobreak}
\def\l@app#1#2{\ifnum \c@tocdepth >\z@
    \addpenalty\@secpenalty
    \addvspace{1.0em \@plus\p@}%
    \setlength\@tempdima{8.5em}%
    \begingroup
      \parindent \z@ \rightskip \@pnumwidth
      \parfillskip -\@pnumwidth
      \leavevmode \bfseries
      \advance\leftskip\@tempdima
      \hskip -\leftskip
      #1\nobreak\hfil \nobreak\hb@xt@\@pnumwidth{\hss #2}\par
    \endgroup\fi}
\newcounter{sapp}[app]

\def\sapp{\def\theequation{A\arabic{app}.\arabic{equation}}\par
   \@afterindentfalse
  \secdef\@sapp\@dsapp}
\newcommand\@sapp{\@startsection{sapp}{2}{\z@}%
                                     {-3.25ex\@plus -1ex \@minus -.2ex}%
                                     {1.5ex \@plus .2ex}%
                                     {\normalfont\large\bfseries}}

\def\@dsapp#1{%
{\parindent \z@ \raggedright  \bf #1}\par\nobreak}
\newcommand{\l@sapp}{\@dottedtocline{2}{1.5em}{3em}}

\catcode`\@=11
\def\marginnote#1{}

\newcount\hour
\newcount\minute
\newtoks\amorpm
\hour=\time\divide\hour by60
\minute=\time{\multiply\hour by60 \global\advance\minute by-\hour}
\edef\standardtime{{\ifnum\hour<12 \global\amorpm={am}%
        \else\global\amorpm={pm}\advance\hour by-12 \fi
        \ifnum\hour=0 \hour=12 \fi
        \number\hour:\ifnum\minute<10 0\fi\number\minute\the\amorpm}}
\edef\militarytime{\number\hour:\ifnum\minute<10 0\fi\number\minute}

\def\draftlabel#1{{\@bsphack\if@filesw {\let\thepage\relax
      \xdef\@gtempa{\write\@auxout{\string
          \newlabel{#1}{{\@currentlabel}{\thepage}}}}}\@gtempa \if@nobreak
    \ifvmode\nobreak\fi\fi\fi\@esphack} \gdef\@eqnlabel{#1}}
    \def\@eqnlabel{}
\def\@vacuum{}
\def\draftmarginnote#1{\marginpar{\raggedright\scriptsize\tt#1}}

\def\draft{
  \oddsidemargin -.5truein
  \def\@oddfoot{\footnotesize \sl preliminary draft \hfil
    \rm\thepage\hfil\sl\today\quad\militarytime}
  \let\@evenfoot\@oddfoot \overfullrule 3pt
    \let\label=\draftlabel
    \let\marginnote=\draftmarginnote
  \def\@eqnnum{(\theequation)\rlap{\kern\marginparsep\tt\@eqnlabel}%
    \global\let\@eqnlabel\@vacuum}

  }

\voffset= - 1.2in
\hoffset= - 1.0in
\def\be{\begin{eqnarray}}
\def\ee{\end{eqnarray}}
\def\nn{\nonumber}
\def\mJ{{\bf{J}}}
\def\p{\partial}

\def\beq{\begin{equation}}
\def\eeq{\end{equation}}
\def\ba{\beq\new\begin{array}{c}}
\def\ea{\end{array}\eeq}
\def\be{\ba}
\def\ee{\ea}

\def\theequation{\arabic{section}.\arabic{equation}}

\newfont{\alef}{msbm10 at 11pt}
\newfont {\goth}{eufm10 at 11pt}
\def\mathbb#1{\hbox{{\alef #1}}}
\def\gotht{\hbox{{\goth t}}}

\let\@@savethanks\thanks
\def\thanks#1{\gdef\thefootnote{\alph{footnote}}\@@savethanks{#1}}

\title{{\bf Instantons and Merons in Matrix Models
}
\vspace{.5cm}}
\author{{\bf A.Alexandrov}\thanks{E-mail: \ al@itep.ru}
\date{ } \\ {\small
{\it ITEP, Moscow, Russia}}\\ \\
{\bf A. Mironov}\footnote{E-mail:
\ mironov@itep.ru; mironov@lpi.ru}
\date{ } \\
{\small {\it Lebedev Physics Institute}
and {\it ITEP, Moscow, Russia}}\\ \\
{\bf A.Morozov}\thanks{E-mail: \ morozov@itep.ru}
\date{ } \\ {\small {\it ITEP, Moscow, Russia}}
}

\begin{document}

\maketitle

\vspace{-10.5cm}

\begin{center}
\hfill FIAN/TD-2/06\\
\hfill ITEP/TH-107/05\\
\end{center}

\vspace{9.0cm}

\begin{abstract}
\noindent Various branches of matrix model partition function can be
represented as intertwined products of universal elementary
constituents: Gaussian partition functions $Z_G$ and Kontsevich
$\tau$-functions $Z_K$. In physical terms, this decomposition is the
matrix-model version of multi-instanton and multi-meron
configurations in Yang-Mills theories. Technically, decomposition
formulas are related to representation theory of algebras of
Krichever-Novikov type on families of spectral curves with
additional Seiberg-Witten structure. Representations of these
algebras are encoded in terms of ``the global partition functions".
They interpolate between $Z_G$ and $Z_K$ associated with different
singularities on spectral Riemann surfaces. This construction is
nothing but M-theory-like unification of various matrix models with
explicit and representative realization of dualities.
\end{abstract}
\newpage

\tableofcontents

\newpage
\def\thefootnote{\arabic{footnote}}

\section{Introduction. The origin and the essence of decomposition formulas}

This paper is a part of a new attack \cite{amm1,M} on the bastions
of {\it Hermitean matrix model} \cite{HM}-\cite{UFN3}, began in the
recent papers \cite{DV}-\cite{EyCh}, and targeted at exhaustive
(reference-book-like) description of properties of {\it partition
function} $Z(t)$ -- the basic special function of string theory. In
the present paper, we study relation between $Z(t)$, the partition
function of complex matrix model $Z_C$ and the Kontsevich
$\tau$-function $Z_K$. Moreover, we unify all of them with the use
of (an {\it a priori} different) interpolating {\it global}
partition function associated with Virasoro constraints of the
Krichever-Novikov type. This M-theory like unification of various
matrix models (with the global partition function playing the role
of M-theory partition function) provides an explicit realization of
dualities \cite{amm4} (with three types of matrix models playing the
role of the five superstring theories).

\subsection{Two decomposition formulas}

In the present paper we address the subject of {\it decomposition
formulas}. Any {\it branch} of partition function, which possesses
{\it genus expansion} and is parameterized by a hyperelliptic {\it
spectral curve} $\Sigma_n$ with $n$ pairs of ramification points of
the second order, $a_1^\pm,\ldots,a_{n}^\pm$, can be decomposed into
a product of $n$ {\it Gaussian} partition functions $Z_G$
\cite{Th,amm1} and, further \cite{mmmm}-\cite{Kostov2}, into that of
$2n$ Kontsevich $\tau$-functions \cite{Ko}-\cite{Kom}, associated
with particular ramification points $a_i$:
$$
Z_{\Sigma_n} = \hat{\cal O}_{n}\left(\prod_{I=1}^n Z_G\right) =
\hat U_{\Sigma_n} \left(\prod_{i=1}^{2n} Z_K^{a_i}\right)
$$
Here $\hat{\cal O}_n$ and $\hat U_{\Sigma_n}$ are certain
 intertwining operators, acting on the $t$-arguments of individual
constituents and depending on the branch. For DV branches those operators
simplify and become of Moyal-type.

This decomposition is exact matrix-model counterpart of instanton gas
decomposition in Yang-Mills theories: contribution of every saddle-point
(i.e. a particular branch of partition function) can be decomposed into
elementary constituents, associated with particular instantons \cite{BPST}
($Z_G$) and, further, merons \cite{CGD,mmt} ($Z_K$). Parameters $a_i$
play the role of moduli of multi-instanton and multi-meron configurations.
The first step, decomposition of the DV solution into Gaussian (single-cut) partition
functions $Z_G$ (``instantons") is straightforward, in what follows we
concentrate mostly on the second step: decomposition of $Z_G$ into two
$Z_K$ (of ``instanton" into a pair of ``merons"), which is more
involved and less investigated (both at matrix-model and Yang-Mills levels).

\subsection{Partition function as a D-module}

Partition function $Z(t)$ of countable set of arguments $t_0,t_1,t_2,\ldots$
is defined \cite{UFN3,amm1} as a $D$-module, i.e. as generic solution of the
system of linear differential equations:\footnote{In such framework the
usual matrix integral \cite{HM} is nothing but a particular integral
representation of partition function. As usual with integral representations,
different branches are associated with different choices of integration
hypersurface in the space of matrices of arbitrary size.}
the {\it discrete Virasoro
constraints} \cite{GMMMO,discvir}
$$
\hat L_n Z = 0,\ \ \ n\geq -1,
$$
$$
\hat L_n = \sum_{k>0} kt_k\frac{\p}{\p t_{k+n}} +
g^2\sum_{a+b=n}\frac{\p^2}{\p t_a\p t_b}
$$
Parameter $g$ could be absorbed into rescaling of $t$-variables, but it is
used to select an important class of solutions (branches of $Z$), which
``possess genus expansion" \cite{tH}, i.e. such that
$$g^2\log Z(t) = \sum_{p=0}^\infty g^{2p}{\cal F}^{(p)}(t)$$
is a formal series in non-negative powers of $g^2$ with $g$-independent
genus-$p$ {\it prepotentials} ${\cal F}^{(p)}(t)$.
Such branches are further distinguished and labeled by hyperelliptic Riemann
surfaces (spectral curves)
$$\Sigma_n:\ \ \ y^2 = \prod_{i=1}^{2n} (z-a_i)$$
so that ${\cal F}^{(p)}(t^\Sigma|\Sigma_n)$ are formal series in non-negative
powers of $t^\Sigma$'s with $a_i$'s allowed in denominators.
The variables $t^\Sigma$ are related to original $t$ by $a$-dependent
{\it shifts}:
$$ t_k = -T_k(a) + t_k^\Sigma, \ \ \ T_k=0\ \ {\rm for}\ \ k>n+1$$
See \cite{amm1} and \cite{amm23} for detailed description of this
branch/phase structure of $Z(t)$. Below for the sake of brevity, we
write all the partition functions as functions of $t$, not of
$t^\Sigma$, since for all our calculations up to the Appendices it
makes no difference (we need only commutation relations of these
variables and the derivatives w.r.t. them). However, in the
Appendices, where we obtain partition functions as formal series, we
restore the notations $t^\Sigma$.

\subsection{Krichever-Novikov type algebra and the $\star$-product}

Virasoro operators $\hat L_n$ are components of the Virasoro stress-tensor:
the loop operator
$$\hat L(z) = \sum_{n=-\infty}^{+\infty} \frac{(dz)^2}{z^{n+2}}\hat L_n$$
Virasoro constraint is the ``negative" part of this two-form,
$$\hat L_-(z) = \sum_{n=-1}^{+\infty} \frac{(dz)^2}{z^{n+2}}\hat L_n$$
The language of loop operators has two advantages:
it makes especially simple the Sugawara
embedding of Virasoro operators into the universal enveloping of Kac-Moody
algebra: $\hat L(z) = :\hat J^2(z):$, where
$$
\hat J(z) = \sum_{k\geq 0} \left\{\frac{1}{2}kt_kz^{k-1}dz +
\frac{dz}{z^{k+1}}\frac{\p}{\p t_k}\right\}
$$
and it allows straightforward deformation to arbitrary Riemann surfaces
(above formulas are essentially written on a sphere or, better, in the
vicinity of a point $z=\infty$). Not-surprisingly, when we deal with
a particular branch of partition function, the relevant deformation
is to the spectral curve $\Sigma_n$ \cite{mmmm}.

On a given Riemann surface (complex curve) $\Sigma$ the current
$\hat \mJ(z)$ is an operator-valued analytical ($\p \hat \mJ/\p \bar z = 0$)
1-form, allowed to have singularities of a given type at certain points.
Such $\hat \mJ(z)$ form a Krichever-Novikov-type algebra \cite{KN}-\cite{Grass}.
In the simplest case of the Riemann sphere $\Sigma = CP_1$ with two punctures,
where $\hat \mJ(z)$ is allowed to have poles of arbitrary order, i.e. is an
operated-valued Laurent series, it generates just an ordinary
$\hat U(1)_{k=1}$ Kac-Moody algebra with the commutation relations
$$
\left[ \hat \mJ^{CP_1}(z), \hat \mJ^{CP_1}(z')\right] =
\frac{1}{2}\sum_{k=-\infty}^\infty \frac{kz^{k-1}}{{z'}^{k+1}}dzdz' =
-\frac{1}{2}\sum_{k=-\infty}^\infty \frac{k{z'}^{k-1}}{z^{k+1}}dzdz'
\equiv \frac{1}{2}\delta'(z-z')dzdz'
$$
The sums at the r.h.s. define $z$-derivative of ``holomorphic
$\delta$-function", satisfying $\oint_\infty \delta(z-z')f(z')
= f(z)$ for arbitrary Laurent series $f(z)$.

Since matrix model is associated with {\it Virasoro} constraints,
formed by the {\it square} of $\hat J(z)$, the corresponding
$\hat\mJ(z)$ can have ramifications of degree two and are naturally
defined on hyperelliptic Riemann surfaces. Of actual interest for us
will be $\hat {\bf J}(z)$ on hyperelliptic $\ \Sigma_n:\ \ y^2 =
\prod_{i=1}^{2n}(z-a_i)\ $ with singularities allowed at $z= a_i$
and at $z=\infty_\pm$, which changes sign whenever $z$ goes around
any of the points $a_i$, i.e. is expandable in formal series with
{\it odd} powers of $\sqrt{z-a_i}$ near $z=a_i$ and in arbitrary
Laurent series (with {\it integer} powers of $z^{-1}$) near
$z=\infty_\pm$ (expansions at two infinities coincide). If all
ramification points $a_i$ merge pairwise (all cuts are contracted to
points), then the algebra reduces to that with poles only (no
ramifications). Vice versa, sometime it can make sense to blow up
the poles at infinities into $n$ additional cuts between $2n$ new
ramification points $\Lambda_1,\ldots,\Lambda_{2n}$ \cite{Ito4}. We
give further details about the loop algebras of this type below. Of
principal importance will be emergency of
{\it conjugation operators}, intertwining currents 
in different charts. They reveal the hidden symplectic (quantum)
structure ($\star$-product) of matrix model, which is important feature
of generic string-field theory \cite{SFT}, of which the matrix model is
a {\it representative} toy-example.

\subsection{Seiberg-Witten structure and the $\ast$-product}

The Virasoro-constraint operator is not quite $\hat L(z) =\ :\hat
J^2(z):$ -- it is rather $\hat L_-(z)$. The ``minus-projection" is
actually another structure added to Krichever-Novikov-type algebra.
For $CP_1$ the minus-projection is just $f_-(z) = \oint_0
\frac{f(z')dz'}{z-z'} = \oint_{\infty_\pm; z}\frac{f(z')dz'}{z'-z}$.
However, since $\hat L(z)$ is a 2-differential, it can not be
integrated along a contour without additional comments. In fact,
$$
\hat{\bf L}_-(z) = :\hat \mJ * \hat \mJ:(z) = \oint_C {\cal K}(z,z'):\hat \mJ^2(z'):
$$
where contour $C$ encircles $\infty_\pm$ and $z$ or, alternatively,
all the ramification points $a_i$, and ${\cal K}(z,z')$ is a certain
$(1,-1)$-bidifferential. Note that we define $\hat{\bf L}_-(z)$ as a
1-form, not a 2-differential(!), making use of peculiar $*$-product
in the space of analytic 1-forms (see \cite{MMMWDVV} for
introductory discussion of such products). ${\cal K}(z,z')$ can be
obtained from the readily available $(m,n)$-bidifferentials with
$m,n\geq 0$ -- the free-field Green functions on Riemann surfaces
\cite{FFRS} -- by division over some 1-differential, which is the
above-mentioned additional structure on $\Sigma_n$.

This structure is nothing but the celebrated Seiberg-Witten
structure \cite{SW}-\cite{GM}, and its relevant version is the
Dijkgraaf-\-Vafa differential \cite{DV} $\Omega_{DV} = y(z)dz$, so
that\footnote{See \cite{ChMMV2,Ey} For more details about this ratio
of $d{\cal G}(z,z')$, the one differential w.r.t. the first argument
and the primitive of the Bergmann kernel w.r.t. the second argument,
and the Dijkgraaf-Vafa differential $\Omega_{DV}$. In simplest
particular case of the Gaussian model, $\Sigma_G:\ y^2_G(z)=z^2-4S$
$$
{\cal K}(z,z')= \frac{dz}{dz'}\frac{1}{z-z'}\left(\frac{1}{y_G(z)} -
\frac{1}{y_G(z')}\right)
$$}
$$
{\cal K}(z,z') = \frac{\langle \p\phi(z)\ \phi(z')\rangle_{\Sigma_n}}
{\Omega_{DV}(z')}
$$

An exact relation between the $\star$-structure behind the
conjugation operators and the $*$-structure, their connection to
Kontsevich quantization, Batalin-Vilkovissky and
string-\-field-\-theory formalisms remain beyond the scope of the
present paper. These subjects are relevant for discussion of cubic
Eynard-type actions \cite{Ey,EyCh} and the background-independence
phenomenon.

\subsection{Local Virasoro constraints}

Operator $\hat{\bf L}_-(z)$ reduces to $\hat L_-(z)$, used in the
definition of $Z(t)$, in the vicinity of $z=\infty$, and the basic
identity
$$
\hat L_-^G(z)\sim\hat{\bf L}_-(z) = \oint_{\infty\cup z} {\cal
K}(z,z'):\hat \mJ^2(z'): = - \sum_{i=1}^{2n} \oint_{a_i} {\cal
K}(z,z'):\hat \mJ^2(z'): \equiv - \sum_{i=1}^{2n} \hat{\bf
L}^{(a_i)}_-(z)\sim -
$$
implies that $Z(t)$ -- the zero mode of $\hat L_-(z)$ -- is a product
of zero-modes of the {\it commuting} 
operators $ \hat{\bf L}^{(a_i)}_-(z)$. Operators  $\hat{\bf
L}^{(a_i)}_-(z)$ are defined in the vicinities of particular
ramification points, where $\hat \mJ(z)$ behaves as an {\it odd}
current, therefore, up to a common conjugation, $\hat{\bf
L}^{(a_i)}_-(z)$ are ``continuous Virasoro constraints", which
annihilate the Kontsevich $\tau$-function  $Z_K(\tau^{(a_i)}_k)$
\cite{MMM}. The time-variables $\tau^{(a_i)}_k$ are introduced
below.

\subsection{Generic construction. Summary}

Algebraic formalism described in this paper is easily expandable to
zero-modes ($D$-modules) of $W$-algebras, annihilated by projected powers
of $\hat J(z)$-operators, not obligatory squares.
For $\hat W^{(q)} \sim \hat J^q$ 
spectral surfaces
are not obligatory hyperelliptic. The relevant Seiberg-Witten structures
and associated Whitham hierarchies, quasiclassical $\tau$-functions and
WDVV equations deserve investigation.

\bigskip

We end our conceptual explanation of decomposition formulas
with the following generic plan for their investigation:

\bigskip

{\bf I. Currents}
\begin{itemize}
\item Riemann surface \ \ $\Sigma\ \ \rightarrow\ \ $ loop current algebra
$\hat{\bf J}^\Sigma(z)$
\item
Global realization of the current algebra on $\Sigma$
\item
Its projections to standard realizations in vicinities of singularities
\item
Gaussian case: two poles at $\alpha$ and $\infty$ with $\alpha$
blown up into a cut between $a_-$ and $a_+$, $a_\pm = \alpha \pm
\sqrt{S}$
\end{itemize}

\bigskip

{\bf II. Sugawara realization of constraints}\begin{itemize}
\item
Generic Virasoro ($q=2$) case: $n+1$ poles at $\alpha_I$ and $\infty$ with each
$\alpha_I$ blown up into a cut between $a_{2I-1}$ and $a_{2I}$, $I =
1,\ldots, n$
 \item $\hat {\bf W}^{(q)}(z) = \oint_C {\cal K}_q(z,z')\hat{\bf
J}^q(z')$\ \
%
and current is allowed to have ramification singularities of orders
which are divisors of $q$
\item
The contour $C = C_\infty = \sum_i^{2n} C_i$
\end{itemize}

\bigskip

{\bf III. Partition functions}\begin{itemize}
\item
Partition function satisfies $\ \ \hat {\bf W}^{(q)}(z){\bf Z}_q  =
\left(\oint_{C_\infty} {\cal K}_q\hat{\bf J}^q\right){\bf Z}_q = 0$
\item
Particular equations
$\left\{\hat U^{-1}\left(\oint_{C_i} {\cal K}_q\hat{\bf J}^q\right)\hat
U\right\}Z_{gK}^{(i)} = 0$\ \
with a {\it common} conjugation operator
define the generalized Kontsevich $\tau$-functions $Z_{gK}$
\item
It follows that  ${\bf Z}_q = \hat U
\left(\prod_i^{2n}Z_{gK}^{(i)}\right)$
\end{itemize}

\bigskip

{\bf IV. Shifts}\begin{itemize}
\item Shifts of $t$-variables
$t \rightarrow t^\Sigma$, normal ordering, background independence,
representation of $Z_q$ and $Z_{gK}$ as formal series in $t^\Sigma$
variables.
\end{itemize}

\bigskip

The remaining part of this paper describes realization of this plan
for $q=2$ in the simplest Gaussian (1-cut, $n=1$) case, thus
completing the program, originally outlined in \cite{mmmm}. On our
way, we reproduce numerous particular results obtained during the
last decade. The central decomposition formula in this case was
originally obtained in \cite{Ch} "by brute force". Certain
controversy, persisted since the second paper of \cite{Kostov2}, is
resolved below in s.7 (especially formula (\ref{dfk})). Moment
variables and related simplifications of genus expansion of
Kontsevich $\tau$-function were deeply investigated in \cite{MomK},
for moment variables in Hermitean matrix model see \cite{ACKM}.
Group-theory approach and connections to topological theory were
considered in \cite{Giv,Hori}.

Our approach should be useful for study of decomposition formulas for different
topological string theories. In particular, the basic example of the decomposition
formula considered in this paper, corresponding to
the simplest branch of the Hermitian matrix model, that is, to the Gaussian branch is
in a sense dual to the simplest example of the Givental-style decomposition
formula in topological strings. That formula is for the $\mathbb{CP}_1$ model,
its partition function having a matrix model representation similar to the
Gaussian matrix model, \cite{Eguchi}. This partition function is known to be
decomposed into a Moyal-like product of two Kontsevich $\tau$-function \cite{TSTCP1}.

\newpage

\section{Summary of the Gaussian example}

\subsection{Ingredients of the construction}

The starting point in this example is the Gaussian curve \be
\Sigma_G:  \ \ \ y_G^2 = (z-a_+)(z-a_-) \ee This is the Riemann
sphere in hyperelliptic parametrization with two quadratic
ramification points $a_\pm$ and two infinities $\infty_\pm$.
Throughout the paper, we often (but not always) put $a_- = -a_+ =
-2\sqrt{S}$. Since this is the sphere, there is a globally defined
coordinate $w$, \be\label{w} z=\sqrt{S}\left(w+{1\over w}\right),\ \
\ dz=\sqrt{S}\left(1-{1\over w^2}\right)dw,\ \ \
y_G(z)=\sqrt{S}\left(w-{1\over w}\right) \ee

The second ingredient is the {\it current} $\hat{\bf J}$,
allowed to have singularities at these four punctures,
\be\label{gcurrent}
\hat {\bf J}=\hat {\bf J}^{e}(z,y_G)+\hat {\bf J}_G^{o}(z,y_G)\\
\hat {\bf J}^{e}(z,y_G|R,Q) \equiv \sum_{k=0}^\infty \left\{
\frac{1}{2}\left(k+\frac{1}{2}\right)(R_k + zQ_k)y_G^{2k}dz +g^2
\frac{dz}{y_G^{2k+2}}\left(\frac{\p}{\p \tilde R_k} +
z\frac{\p}{\p\tilde Q_k}\right)\right\}\\
\hat {\bf J}^{o}(z,y_G|T,S) \equiv \sum_{k=0}^\infty \left\{
\frac{1}{2}\left(k+\frac{1}{2}\right)(T_k + zS_k)y_G^{2k-1}dz +g^2
\frac{dz}{y_G^{2k+3}}\left(\frac{\p}{\p \tilde T_k} +
z\frac{\p}{\p\tilde S_k}\right)\right\} \ee where we separate even
and odd parts of the current w.r.t. the $\mathbb{Z}_2$-symmetry $y_G
\rightarrow -y_G$ of $\Sigma_G$. Here the convenient parameter
(coupling constant) $g$ is introduced by appropriate rescaling of
time-variables, and commutation relations for the current require
that \be \frac{\partial}{\partial\tilde T_k} =
2\left(4S{\partial\over\partial T_k}+{k+1\over k+{3\over
2}}{\partial\over\partial T_{k+1}}\right),\ \ \ \
{\partial\over\partial\tilde S_k}=2{\partial \over\partial S_k} \ee
Indeed, the operator product expansions of the global currents
(\ref{gcurrent}) are \be
\hat {\bf J}^e(z)\hat {\bf J}^e(z')=g^2{dzdz'\over 2(z-z')^2}+\hbox{regular part}\\
\hat {\bf J}^o(z)\hat {\bf J}^o(z')=g^2{dzdz'\over 4(z-z')^2}
{2zz'-(a_++a_-)(z+z')+2a_+a_-\over y(z)y(z')}+
\hbox{regular part}
\ee
These formulas are consistent with the standard $U(1)$ Kac-Moody algebra.
In order to check it, one suffices to make a change of coordinates to the
globally defined variable $w$, (\ref{w}) and to see that
\be
\hat {\bf J}(z)\hat {\bf J}(z')=
\hat {\bf J}^e(z)\hat {\bf J}^e(z')+\hat {\bf J}^o(z)\hat {\bf J}^o(z')+
\hbox{regular part}=\\
=g^2{dzdz'\over 4(z-z')^2}
\left({2zz'-(a_++a_-)(z+z')+2a_+a_-\over y(z)y(z')}+2\right)+\hbox{regular part}=
g^2{dwdw'\over 2(w-w')^2}+\hbox{regular part}
\ee

The variables in (\ref{gcurrent}) are chosen in a way convenient for
the symmetric model, i.e. that with $a_+=-a_-$. In the asymmetric
case, one better puts \be \hat {\bf J}^{o}(z,y_G|T^\pm)=
\sum_{k=0}^\infty
\left\{\frac{1}{2}\left(k+\frac{1}{2}\right)\Big(T_k^+(z-a_+)+
T_k^-(z-a_-)\Big)y_G^{2k-1}dz +
\right.\\\left.+g^2\frac{dz}{y_G^{2k+3}}
\left((z-a_+)\frac{\partial}{\partial \tilde T_k^+} +
(z-a_-)\frac{\partial}{\partial\tilde
T_k^-}\right)\right\},\label{ascurrent} \ee etc.

The remaining ingredient is the $D$-module: the set of Virasoro
constraints, e.g., \be \hat{\bf L}^o_-(z) = \ :\hat {\bf J}^o * \hat
{\bf J}^o:(z) = \oint_C {\cal K}_G(z,z') :\left(\hat {\bf
J}^o(z)\right)^2: \ee and its solution (zero mode) -- the
Gaussian global partition function ${\bf Z}(T,S)$, -- satisfying
\be\label{ZGl} \hat {\bf L}^o_-(z){\bf Z}(T,S) = 0 \ee

One may restrict the even and odd global currents in (\ref{gcurrent})
further, making use of another $\mathbb{Z}_2$-symmetry of $\Sigma_G$,
$z\to -z+a_++a_-$ (or $z\to -z$ in the symmetric model with $a_+=-a_-$).
This transformation exchanges, say, $T_+$ and $T_-$ in (\ref{ascurrent})
and allows one to construct global partition functions like ${\bf
Z}(T_++T_-)$ etc (see table 1).

\subsection{Familiar special functions}

${\hat {\bf L}}_-(z)$ and ${\bf Z}(T^\pm)$ are new personages in the
theory. As explained in the Introduction, they interpolate between a
variety of well known quantities and provide non-linear relations
among them. In this paper we consider, in addition to ${\bf Z}$,
three such familiar functions:

\begin{itemize}

\item The Gaussian branch $Z_G(t)$ of the Hermitean partition function
\cite{amm1,amm23}, annihilated by the "discrete Virasoro
constraints" \cite{discvir}, \be\label{G}
\hat L_-^G(z) Z_G(t) = 0, \\
\hat L_-^G(z) = 
\left(:\hat J_G^2(z):\right)_- = g^2\sum_{n=-1}^{+\infty}
\frac{(dz)^2}{z^{n+2}}\left( \sum_{k>0} kt_k\frac{\p}{\p t_{k+n}} +
g^2\sum_{a+b=n}\frac{\p^2}{\p t_a\p t_b}\right)
\\
\hat J_G(z|t) = d\hat\Omega_G(z) = {1\over 2}dv(z) +
g^2\hat\nabla(z) = \sum_{k=0}^\infty \left\{{1\over 2}kt_kz^{k-1}dz
+ g^2\frac{dz}{z^{k+1}}\frac{\partial}{\partial t_k}\right\} \ee
This partition function $Z_G$ has the Hermitean matrix integral
representation \be\label{Gint} Z_G= \frac{1}{\hbox{Vol}_{U(N)}}\int
D\phi\ \exp \left(\frac{1}{g} \sum_{k\in Z_+} t_{k}\ {\rm Tr}\
\phi^{k}\right)
 \ee where measure is the
invariant measure on Hermitean matrices of the size $N\times N$, the
volume of the unitary group $U(N)$ \be\label{volume}
\hbox{Vol}_{U(N)} = \int_{U(N)}DU \sim \prod_{k=1}^{N-1} k! \ee and
the integral (\ref{Gint}) is understood as a perturbative power
series in $t^\Sigma_k$'s, $t_k=t^\Sigma_k-{1\over 2}\delta_{k,2}$.

Note that ${\p\over\p t_0}$-term in the current $\hat J_G$ commutes with the
whole current (there is no $t_0$ in the current) and, therefore,
one may deal with ${\p\over\p t_0}$ as with $c$-number. We, indeed,
throughout the paper put ${\p Z_G\over\p t_0}={S\over g^2}Z_G$. In
particular, in (\ref{Gint}) $S$ is associated with the size of matrix $N$.
A generic case is discussed in s.6.2.

Other branches of $Z_G$, including the Dijkgraaf-Vafa ones and
generic branches with genus expansion \cite{amm1} can be also
treated by the method of this paper. However, this will be done
elsewhere.

\item Kontsevich $\tau$-function \cite{Ko,GKM} $Z_K(\tau)$,
annihilated by "continuous Virasoro constraints" \cite{MMM},
\be\label{K}
\hat L^K_-(\xi) Z_K(\tau) = 0, \\
\hat L^K_-(\xi) = 
\left(:\hat J_K^2(\xi):\right)_-=\\= g^2\sum_{n=-1}^{+\infty}
\frac{(d\xi)^2}{\xi^{2n+2}}\left( \sum_{k>0} \left(k+{1\over
2}\right)\tau_k\frac{\p}{\p \tau_{k+n}} +
g^2\sum_{a+b=n-1}\frac{\p^2}{\p \tau_a\p \tau_b}+
\frac{\delta_{n,0}}{16}+\frac{\delta_{n,-1}\tau_0^2}{16g^2}
\right)\\
{\hat J}_K(\xi|\tau) = d\hat\Omega_K(\xi) =
\sum_{k=0}^\infty \left\{{1\over 2}\Big(k+\frac{1}{2}\Big)\tau_k
\xi^{2k}d\xi + g^2\frac{d\xi}{\xi^{2k+2}}\frac{\partial}{\partial
\tau_k}\right\} \ee

This partition function $Z_K$ can be presented as the Hermitean
matrix integral depending on the external matrix $A$, \be
\label{Kint}Z_K= {\int DX\ \exp\left(-{4g^2\over
3}\hbox{Tr}X^3-{2g\over\sqrt{3}}\hbox{Tr}AX^2\right)\over \int DX\
\exp\left(-{2g\over\sqrt{3}}\hbox{Tr}AX^2\right)}\ee where the
integral is understood as a perturbative power series in
$\tau^\Sigma_k\equiv g\displaystyle{{3^{k+{1\over 2}}\over k+{1\over
2}}\hbox{Tr} A^{-2k-1}}$, $\tau_k=\tau^\Sigma_k-{2\over
3}\delta_{k,1}$. Note that this integral does not depend on the size
of matrices $X$ and $A$ being considered as a function of
$\tau^\Sigma_k$ \cite{GKM}. By the shift of the integration
variable, it can be also reduced to the form \be Z_K=
\exp\left(-{1\over 3g}\hbox{Tr}\Lambda^{3\over 2}\right){\int DX\
\exp\left(-{4g^2\over 3}\hbox{Tr}X^3+\hbox{Tr}\Lambda X\right)\over
\int DX\ \exp\left(-{2g\over\sqrt{3}}\hbox{Tr}AX^2\right)} \ee where
$3\Lambda=A^2$, i.e. $\tau^\Sigma_k\equiv \displaystyle{{g\over
k+{1\over 2}}\hbox{Tr} \Lambda^{-k-{1\over 2}}}$\ .\footnote{In
order to introduce an arbitrary shifted first time,
$\tau_k=\tau^\Sigma_k-{2M\over 3}\delta_{k,1}$, where $M$ is a
parameter, one should consider instead of (\ref{Kint}) the integral
\be Z_K= {\int DX\ \exp\left(-{4g^2\over
3M^2}\hbox{Tr}X^3-{2g\over\sqrt{3M}}\hbox{Tr}AX^2\right)\over \int
DX\ \exp\left(-{2g\over\sqrt{3M}}\hbox{Tr}AX^2\right)}=
\exp\left(-{M\over 3g}\hbox{Tr}\Lambda^{3\over 2}\right){\int DX\
\exp\left(-{4g^2\over 3M^2}\hbox{Tr}X^3+\hbox{Tr}\Lambda
X\right)\over \int DX\
\exp\left(-{2g\over\sqrt{3M}}\hbox{Tr}AX^2\right)}\ee which is a
function of the same $\tau^\Sigma_k=g\displaystyle{{3^{k+{1\over
2}}\over k+{1\over 2}}\hbox{Tr} A^{-2k-1}}=\displaystyle{{g\over
k+{1\over 2}}\hbox{Tr} \Lambda^{-k-{1\over 2}}}$.}

\item Partition function of the complex matrix model \cite{cmamo},
annihilated by the truncated ``discrete Virasoro constraints"
\cite{mmmm} \be\label{C}
\hat L^C_-(z) Z_C(\gotht) = 0, \\
\hat L^C_-(z) = 
\left(:\hat J_C^2(z):\right)_- = g^2\sum_{n=0}^{+\infty}
\frac{(dz)^2}{z^{2n+2}}\left( \sum_{k>0} k\gotht_k\frac{\p}{\p
\gotht_{k+n}} + g^2\sum_{a+b=n}\frac{\p^2}{\p \gotht_a\p
\gotht_b}\right)
\\
\hat J_C(z|\gotht) = \sum_{k=0}^\infty \left\{{1\over
2}k\gotht_kz^{2k-1}dz +
g^2\frac{dz}{z^{2k+1}}\frac{\partial}{\partial \gotht_k}\right\} \ee

This partition function $Z_C$ has the complex matrix integral
representation \be Z_C=\frac{1}{\hbox{Vol}^2_{U(N)}}\int D\phi\
D\phi^{\dag}\ \exp \left(\frac{1}{g} \sum_{k\in Z_+} \gotht_{k}\
{\rm Tr}\ \left(\phi\phi^{\dag}\right)^{k}\right) \ee where the
integral which goes over $N\times N$ complex matrices is understood
as a perturbative power series in $\gotht^\Sigma_k$'s,
$\gotht_k=\gotht^\Sigma_k-\delta_{k,1}$. Similarly to the Hermitean
case (\ref{G}), we put ${\p Z_C\over\p \gotht_0}={S\over g^2}Z_C$.

\item To deal with decomposition formula for the complex matrix model,
we will also need a partition function $\tilde Z_K({\cal T})$ that
solves the Virasoro constraints (\ref{K}) only with $n\ge 0$ and
with the zeroth time shifted. At the moment, we have no formula for
its matrix model representation available, therefore, it can be only
recurrently calculated from the Virasoro constraints.
\end{itemize}

In applications, first terms of the $(g^2,t)$-expansions of the
partition functions are often needed, see Appendix I.

\subsection{Web of dualities between matrix models}

\bigskip


\begin{tabular}{|c|c|c|c|c|}
\hline
&&&&\\
Type & Reduction& Global& Vicinity of & Vicinity of \\
of ${\rm\bf Z}$& of ${\rm\bf Z}$& current &$\infty_{\pm}$&$a_{\pm}$\\
&&&&\\
\hline
&&&&\\
${\rm\bf Z}(T,S)$ & $\mathbb{Z}_2$-odd & ${\hat {\bf J}}^o(z)$& $Z_G (t)$ & $Z_K(\tau_{\pm})$\\
&($y\to -y$)&&&\\
&&&&\\
\hline
&&&&\\
${\rm\bf Z}(T)$ & ${\mathbb{Z}}_2\times{\mathbb{Z}}_2$-(odd,odd) &
${{\hat {\bf J}}^o(z)-{\hat {\bf J}}^o(-z)\over 2}$& $Z_C(\pm t^e)$
&
$Z_K(\tau_{\pm})$\\
&($z\to -z$, $y\to -y$) &&&\\
& &&&\\
\hline
&&&&\\
${\rm\bf
Z}(S)$&${\mathbb{Z}}_2\times{\mathbb{Z}}_2$-(even,odd)&${{\hat {\bf
J}}^o(z)+{\hat {\bf J}}^o(-z)\over 2}$&$Z_K(\pm t^o)$ &
$Z_K(\tau_{\pm})$\\
&($z\to -z$, $y\to -y$)&&&\\
&&&&\\
\hline
&&&&\\
${\rm\bf Z}(R,Q)$&${\mathbb{Z}}_2$-even &${\hat {\bf J}}^e(z)$& $Z_G (t)$ & $Z_C(\tau_{\pm})$ \\
&($y\to -y$)&&&\\
&&&&\\
\hline
&&&&\\
${\rm\bf Z}(R)$&${\mathbb{Z}}_2\times{\mathbb{Z}}_2$-(odd,even)
&${{\hat {\bf J}}^e(z)-{\hat {\bf J}}^e(-z)\over 2}$&$Z_K(t^o)$&
$Z_C(\tau_{\pm})$\\
&($z\to -z$, $y\to -y$)&&&\\
&&&&\\
\hline
&&&&\\
${\rm\bf Z}(Q)$&${\mathbb{Z}}_2\times{\mathbb{Z}}_2$-(even,even) &${{\hat {\bf
J}}^e(z)+{\hat {\bf J}}^e(-z)\over 2}$&
                        $Z_C(t^e)$      &$Z_C(\tau_{\pm})$\\
&($z\to -z$, $y\to -y$)&&&\\
&&&&\\
\hline
\end{tabular}

\bigskip

\bigskip

\noindent {\bf Table 1. Relations between global currents and
partition functions, their symmetry and partition functions emerging
at singularities}

\phantom{a}

\bigskip

In the vicinities of punctures $\infty_\pm$ and $a_\pm$ the global
current $\hat {\bf J}$ (or its reductions to $\hat {\bf J}^o$ etc)
is isomorphic to $\hat J_G^{\infty_{\pm}}$ (whose reductions gives
rise to $\hat J$ etc) and $\hat J^{a_{\pm}}$ (whose reductions gives
rise to $\hat J_K$ etc) respectively. Isomorphism is realized by
certain conjugation operators, depending on the choice of local
coordinates and associated relation between the time-variables.
Accordingly, the above special functions, which are the zero-modes
of the thus related Virasoro constraints, are expressed by action of
these operators on ${\bf Z}$. Moreover, one could start with reduced
global currents, say, with ${\bf J}^o$ and consider {\it its}
projections to the local patches, which also generates proper
projections of the partition functions.

In the table above, we consider these projections of various global
partition functions to the local partition functions in vicinities
of the four singular points. Those local partition functions are the
Hermitean matrix model, $Z_G$, the complex matrix model, $Z_C$ and
the Kontsevich matrix model, $Z_K$. In the table, we denote through
$t^o$ the odd set $\{t_{2k+1}\}$ and through $t^e$ the even set
$\{t_{2k+1}\}$ of variables $\{t_k\}$. The table is fully formulated
in terms of the symmetric currents (\ref{gcurrent}), though
(\ref{ascurrent}) is needed at intermediate steps in some
derivations.

In order to illustrate our general construction, we also consider
below an example of the sphere with two more singular points placed
in $z=0$ on the both sheets. In this case, we deal with the global
partition function of the type ${\bf Z}(T)$, it has as its local
projection in the vicinity of the new singular point the partition
function $\tilde Z({\cal T})$.

\subsection{Moment variables}

In fact, we are going to add to the list of special functions three
more that are associated with the different set of variables, {\it
moment variables}, defined so that, in these variables, the finite
genus contribution to the partition function ${\cal F}^{(p)}$ is a
polynomial. (Instead, discrete and continuous Virasoro constraints
lose their simple form in these coordinates.) Besides, these
variables are graded and every finite genus contribution involves
only a few {\it first} moment variables, their number increasing
with genus (the degree of gradation grows with genus). Therefore,
the moments, first, depend on the concrete genus expansion and,
second, admit arbitrary triangle change of variables, which does not
change their defining property).

\subsubsection{Changing local parameter}

The simplest way to introduce moment variables is to consider
changing local parameters in vicinities of the four singular points.
This way one introduce two new sets of variables:

\begin{itemize}

\item Moments for the Kontsevich $\tau$-function
\be\label{ttau} \tilde\tau_m = G^{-\frac{2m+1}{3}} \sum_{k=0}^\infty
\frac{\Gamma(m+k+3/2)\ u^k}{\Gamma(m+3/2)\ k!} \tau_{m+k},
\ee corresponding to the change of the local parameter \be \tilde
\xi = G^{1/3}\sqrt{\xi^2-u} \ee where $u$ and $G$ are some specific
functions of $\tau_k$ fixed by the conditions \be
\tilde\tau_0^\Sigma=\tilde\tau_1^\Sigma=0\ee

Note that the shift in this case is the
same in both $\tau$ and $\tilde\tau$ variables, \be \tau_m \equiv
\tau^\Sigma_m - \frac{2}{3}\delta_{m,1}, \ \ \ \tilde\tau_m \equiv
\tilde\tau^\Sigma_m - \frac{2}{3}\delta_{m,1} \ee

\item Moments for the Gaussian branch of $Z_G$
\be\label{tt} m\tilde t_m=\oint_{\infty}\frac{d v(z)}{\tilde z^{m}}=
a \oint_\infty\frac{\sum_k
kt_k(az+b)^{n-1}dz}{z^m}=a^m\sum_k{k!b^{k-m}\over (m-1)!(k-m)!}t_k
\ee that correspond to the change of the local parameter \be \tilde
z=A\tilde z+B \ee where $A,B$ are functions of $t_k$ fixed by the
conditions \be \tilde t_1^\Sigma=\tilde t_2^\Sigma=0\ee

Again the shift is
the same in both $t$ and $\tilde t$ variables, \be
t_k=t_k^\Sigma-\frac{\delta_{k,2}}{2},\ \ \  \tilde t_k=\tilde
t_k^\Sigma-\frac{\delta_{k,2}}{2} \ee
\end{itemize}

While the Kontsevich moments correspond to the standard genus
expansion of $Z_K$, the moment variables for $Z_G$ are associated
with a non-standard expansion, see s.2.5.1.

\subsubsection{Moment variables for the standard genus expansion of
$Z_G$}

One can also express $Z_G$ it terms of moments that provide the {\it
standard} genus expansion for it \cite{ACKM}, \be\label{ACKMmom}
t_k^\pm = \oint_{A_-,A_+} \frac{v(z)dz}{(z-A_\mp(t))^k y_G(z)},\ \ \
\ v(z)=\sum t_kz^k,\ \ \ \  y_G(z)\equiv (z-A_+(t))(z-A_-(t)) \ee
The simplest way then to calculate the corresponding partition
function, \be\label{ZACKM}Z_{ACKM}(t^\pm)\equiv Z_G(t)\ee is to use
the relation (see s.2.7 and s.7) between $Z_G(t)$ and
$Z_K(\tau^\pm)$ parameterized by an arbitrary branching points
$a_\pm$, to rescale $Z_K(\tau^\pm)$ and time variables
$\tau^\pm\to\tilde\tau^\pm$, and, at the final stage of the
calculation, to make\footnote{We use capital $A$ in order to stress
that it becomes a function of times.} $a_\pm=A_\pm(\tau)$ depending
on $\tau^\pm$ in such a way that
$\tau^\pm_0=\tau^\pm_1=0$.\footnote{The idea of this calculation has
first appeared in the improved version of \cite{Kostov2}.} This can
be formulated as a kind of normal ordering: one can consider, by
definition, \be :\ e^{\hat {\tilde
U}_{GK}}Z_K(\tilde\tau_+)Z_K(\tilde\tau_-)\ :\ee as being calculated
at constant $a_\pm=A_\pm(\tau)$.

\subsection{Partition functions: genus and polynomial expansions}

In order to illuminate the role of moment variables, let us discuss
structure of the partition functions as power series in $g$ and
$t^\Sigma$.

\subsubsection{$Z_G$}

Let us assume existence of the genus expansion, \be \log Z_G(t) =
\sum_{p=0}^\infty g^{2p-2}{F}_N^{(p)}(t) \ee The index $N$ stresses
that the free energy also depends on $N$. Now we note that the
Virasoro constraints (\ref{G}) are invariant under simultaneous
rescalings \be t_k \rightarrow \lambda^{k+s}t_k,\ \ \ M \rightarrow
\lambda^{2+s} M, \ \ \
g \rightarrow \lambda^s g
\ee with arbitrary $s$. Here we introduced a parameter $M$ that
controls the shift of times, $t_k=t^\Sigma_k-{M\over
2}\delta_{k,2}$. Since we look for the partition function as a power
series in $g$ and $t^\Sigma$, ${F}_N^{(p)}\left(\lambda^{2+s} M,\
\lambda^{k+s}t_k\right) = \lambda^{2s(1-p)}{F}_N^{(p)}(M,t_k)$, and
\be {F}_N^{(p)}(M,t) = \sum_r \sum_{k_1,\ldots,k_r}
C^{(p)}_{k_1\ldots k_r} \frac{t^\Sigma_{k_1}\ldots t^\Sigma_{k_r}}
{M^{\frac{1}{2}(k_1+\ldots+k_r)}}\ \delta\Big(\sum_{i=1}^r(k_i-2) -
4(p-1)\Big) \label{selrule} \ee The topological correlators
$C^{(p)}_{k_1\ldots k_r}$ can be found from recurrent relations
emerging from (\ref{G}) upon inserting there the anzatz
(\ref{selrule}).

Selection rule, expressed by the $\delta$-factor in (\ref{selrule}),
implies that for $t_1^\Sigma = t_2^\Sigma = 0$, the spherical and
toric free energies ${F}_N^{(0)}-Nt_0/g = {F}_N^{(1)} = 0$ and all
higher ${F}_N^{(p)}$ with $p\geq 2$ are polynomials of finite
($p$-dependent) degree in remaining time-variables $t_k^\Sigma$,
$k\geq 3$. This is exactly what moment variables provide: using
variables (\ref{tt}) and fixing specific time-dependent $a(t)$ and
$b(t)$ in their definition, one comes to $Z_G(\tilde t)$ with
$\tilde t_1^\Sigma=\tilde t_2^\Sigma=0$, i.e. to the polynomial
representations of $F^{(p)}_N(\tilde t^\Sigma)$'s.

These moment variables have only one drawback -- they are associated
with the genus expansion that is done at constant $N$, while the
standard matrix model genus expansion is done at constant $S=gN$.
This another expansion immediately destroys the simple scheme above
and forces one to use more tricky moment variables in order to get a
polynomial representation for the standardly genus expanded free
energy, see s.2.4.2 and s.7.

\subsubsection{$Z_K$}

Similarly to the case of $Z_G$, we expand $Z_K$ into power series in
$g$ and $\tau^\Sigma$, \be\label{loopexK} \log\ Z_K(\tau) =
\sum_{p=0}^\infty g^{2p-2}{\cal F}^{(p)}(\tau),\ \ \
\tau=\tau^\Sigma -{2M\over 3}\delta_{k,1} \ee Again note that the
Virasoro constraints (\ref{K}) are invariant under simultaneous
rescalings \be \tau_k \rightarrow \lambda^{2k+1+s}\tau_k,\ \ \ M
\rightarrow \lambda^{3+s} M, \ \ \
g \rightarrow \lambda^s g
\ee with arbitrary $s$.
This means that $
{\cal F}^{(p)}\left(\lambda^{3+s} M,\ \lambda^{2k+1+s}\tau_k\right)
= \lambda^{2s(1-p)}{\cal F}^{(p)}(M,\tau_k)$, and \be {\cal
F}^{(p)}(M,\tau) = \sum_r \sum_{k_1,\ldots,k_r}
{\cal C}^{(p)}_{k_1\ldots k_r} \frac{\tau^\Sigma_{k_1}\ldots
\tau^\Sigma_{k_r}} {M^{\frac{r}{3} + \frac{2}{3}(k_1+\ldots+k_r)}}\
\delta\Big(\sum_{i=1}^r(k_i-1) - 3(p-1)\Big) \label{selruleK} \ee
and topological correlators ${\cal C}^{(p)}_{k_1\ldots k_r}$ can be
found from recurrent relations emerging from (\ref{K}) upon
inserting there the anzatz (\ref{selruleK}).

This time the selection rule, expressed by the $\delta$-factor in
(\ref{selruleK}), implies that for $\tau_0^\Sigma = \tau_1^\Sigma =
0$, the spherical and toric free energies ${\cal F}^{(0)} = {\cal
F}^{(1)} = 0$ and all higher ${\cal F}^{(p)}$ with $p\geq 2$ are
polynomials of finite ($p$-dependent) degree in remaining
time-variables $\tau_k^\Sigma$, $k\geq 2$.

Again, this is what moment variables provide: using variables
(\ref{ttau}) and fixing specific time-dependent $G(t)$ and $u(t)$ in
their definition, one comes to $Z_K(\tilde \tau^\Sigma)$ with
$\tilde \tau_1^\Sigma=\tilde \tau_2^\Sigma=0$, i.e. to the
polynomial representations of ${\cal F}^{(p)}(\tilde
\tau^\Sigma)$'s.

\subsubsection{Non-standard genus expansions}

One can also use another non-standard genus expansion for $Z_K$
which allows one to deal with polynomials at each order in the
standard variables without making the special change of variables to
$\tilde\tau_0^\Sigma=\tilde\tau_1^\Sigma=0$. To this end, one
suffices to rescale the time variables $\tau_k^\Sigma =
g{\tau'}_k^\Sigma$. Then, \be\label{nsgenusK} \log
Z_K(\tau^\Sigma)=\sum_{p=0}^\infty g^{2p-2}{\cal
F}'_{(p)}({\tau'}^\Sigma) \ee Then, indeed, the only non-polynomial
part of the genus $g$ free energy, that is, that containing power
series in ${\tau'}_0^\Sigma$ and ${\tau'}_1^\Sigma$,
(\ref{selruleK}) becomes a polynomial.

The similar change of variables in $Z_G$, $t_k= g{t'}_k^\Sigma$
gives rise to the genus expansion \be\label{nsgenusG}\log
Z_G(t^\Sigma)=\sum_{p=0}^\infty g^{2p-2}F'_{(p)}({t'}^\Sigma) \ee
Let us now represent \cite{amm1} \be
{{F}}^{(p)}(t^\Sigma)=\sum_{k=1}^{\infty}\frac{1}{k!}\oint\dots\oint\rho^{(p|k)}(z_1,\dots,z_k|S)v(z_1)\dots
v(z_k) \ee with some coefficient functions
$\rho^{(p|k)}(z_1,\dots,z_k|S)$ (which are $k$-point resolvents, or
loop operators). From this expansion, it is clear that $F'_{(1)}$ is
linear in (infinitely many) times, $F'_{(2)}$ is quadratic,
$F'_{(3)}$ gets contributions from $F_{(3)}$ and $F_{(1)}$ and
contains both linear and cubic terms etc: \be\label{nsgenusG1}
{{F}}_{(1)}({{t'}})=\oint\rho^{(0|1)}(z){{v'}}(z)\nn\\
{{F}}_{(2)}({{t'}})=\frac{1}{2!}\oint\rho^{(0|2)}(z_1,z_2){{v'}}(z_1){{v'}}(z_2)\nn\\
{{F}}_{(3)}({{t'}})=\frac{1}{3!}\oint\rho^{(0|3)}(z_1,z_2,z_3){{v'}}(z_1)
{{v'}}(z_2){{v'}}(z_3)+\oint\rho^{(1|1)}(z){{v'}}(z)\nn\\
{{F}}_{(4)}({{t'}})=\frac{1}{4!}\oint\rho^{(0|4)}(z_1,z_2,z_3,z_4){{v'}}(z_1)
{{v'}}(z_2){{v'}}(z_3){{v'}}(z_4)+\frac{1}{2!}\oint\rho^{(1|2)}(z_1,z_2){{v'}}(z_1){{v'}}(z_2)
\ee where $v'(z)$ denotes $\sum t'_kz^k$.

These polynomials $F_{(k)}({t'}^\Sigma)$ containing infinitely many
times are polynomials of only finite number of other variables,
$\{{\tau'}_k^\Sigma\}$, due to relations between $Z_G$ and $Z_K$
(see Table 2). We discuss this in Appendix I.

\subsection{Conjugation operators}

Let us explain now how one construct the conjugation operator in a
manifest way. To this end, first of all one constructs the
two-differentials \be f_{\hat J}(z,z'|g^2)=\hat J(z|g^2)\hat
J(z'|g^2) - :\hat J(z|g^2)\hat J(z'|g^2):=d_z h_{\hat J}(z,z'|g^2)
\ee that can be made of both local and global currents. In
particular, \be\label{fGKC}
f_G(z,z'|g^2)=g^2\frac{dzdz'}{2(z-z')^2}=g^2d_z\frac{dz}{2(z'-z)}\\
f_K(z,z'|g^2)=g^2\frac{(z^2+z'^2)dzdz'}{4(z^2-z'^2)^2}=g^2d_z\frac{zdz'}{4(z'^2-z^2)}\\
f_C(z,z'|g^2)=g^2\frac{zz'dzdz'}{2(z^2-z'^2)^2}=g^2d_z\frac{z'dz'}{4(z'^2-z^2)}
\ee Depending on its symmetry, the global current in the vicinity of
a singular point $\xi$ is equivalent to one or another canonical
local current listed in s.2.2, \be \hat
\textbf{J}(z|g^2)\sim\alpha_\xi\hat J^\xi(x|\beta_\xi g^2) \ee for
some constants $\alpha_\xi$ and $\beta_\xi$. This means that in the
local coordinate in the vicinity of the point $\xi$, it is equal to
\be\label{jeq} \alpha\left(\hat J_\xi(x|\beta g^2)+\Delta_\xi \hat
J^\xi(x|\beta_\xi g^2)\right)=\alpha_\xi e^{\hat U}\hat
J^\xi(x|\beta_\xi g^2)e^{-\hat U} \ee In its turn, it implies that
\be \label{AJ}
f_{\hat\textbf{J}}^{\xi\xi}(z(x),z'(x')|g^2)-\alpha_\xi^2f_{\hat
J^\xi}(x,x'|\beta_\xi
g^2)=A_{\hat\textbf{J}}^{\xi\xi}(x,x'|g^2)=\sum_{k,m=0}^\infty
\alpha^{\xi\xi}_{km}x^kx'^mdxdx' \ee

Consider, e.g., the local currents at two infinities. They coincide
with each other, and the conjugation operator is \be \label{Uinf}
U_{\infty}=\frac{1}{2\alpha_\infty^2g^4c_\infty^2\beta_\infty^2}\oint_\infty\oint_\infty
A_{\hat\textbf{J}}^{\infty\infty}(x,x'|g^2)\hat
\Omega^\infty(x|\beta_\infty g^2)\hat\Omega^\infty(x'|\beta_\infty
g^2)+\oint_\infty k^\infty(z|\beta_\infty g^2)\hat
\Omega^\infty(z|\beta_\infty g^2) \ee where, depending on the local
current associated with infinity (i.e. on the projection of the
global current), $c_\infty$ is ${1\over 4}$ for the complex matrix
model and the Kontsevich currents and $\frac{1}{2}$ for the Gaussian
current\footnote{While $\Delta \hat J^\infty(s|\beta_\infty g^2)$ is
manifestly \be \Delta \hat J^\infty(s|\beta_\infty g^2)=-\left[\hat
J^\infty(s|\beta g^2),\hat U_\infty\right]
=\frac{1}{\alpha_\infty^2g^4c_\infty^2\beta^2}
\oint_\infty\oint_\infty
A_{\hat\textbf{J}}^{aa}(x,x'|g^2)\hat\Omega^\infty(x|\beta_\infty
g^2)\left[\hat J^\infty(s|\beta_\infty
g^2),\hat\Omega^\infty(x'|\beta_\infty g^2)\right]+\\
+\oint_\infty k^\infty(x|\beta_\infty g^2)\left[\hat J^\infty
(s|\beta_\infty g^2),\hat \Omega^\infty(x|\beta_\infty g^2)\right]
=\frac{1}{\alpha_\infty^2g^4c_\infty^2\beta_\infty^2}
\oint_\infty\oint_\infty A_{\hat\textbf{J}}^{aa}(x,x'|\beta_\infty
g^2)\hat\Omega^\infty(x|\beta_\infty g^2)h_\infty(x',s|\beta_\infty
g^2)+\\+ \oint_\infty k^\infty(x|\beta_\infty
g^2)h_\infty(x,s|\beta_\infty g^2) =g^2c_\infty\beta_\infty
\left(\frac{1}{\alpha_\infty^2g^4c_\infty^2\beta_\infty^2}
\oint_\infty
A_{\hat\textbf{J}}^{aa}(x,s|g^2)\hat\Omega^\infty(x|\beta_\infty
g^2)+ k^\infty(s|\beta_\infty g^2)\right) \ee}. Emerging
$k^\infty(s|\beta_\infty g^2)$ here is due to different shifts of
global and local currents and it can be got from the difference of
these currents\footnote{I.e. from \be \left(\hat
J^\infty(x|\beta_\infty g^2)+\Delta \hat J^\infty(x|\beta_\infty
g^2)\right)\left(\hat J^\infty(x'|\beta_\infty g^2)+\Delta \hat
J^\infty(x'|\beta_\infty g^2)\right)-\\
-:\left(\hat J^\infty(x|\beta_\infty g^2)+\Delta \hat J^\infty(x|\beta_\infty g^2)\right)\left(\hat J^\infty(x'|\beta_\infty g^2)+\Delta \hat J^\infty(x'|\beta_\infty g^2)\right):=\nn\\
=f_{\hat J^\infty}(x,x'|\beta_\infty g^2)+
\hat J^\infty(x|\beta_\infty g^2)\Delta J^\infty(x'|\beta_\infty g^2)-:\hat J^\infty(x|\beta_\infty g^2)\Delta J^\infty(x'|\beta_\infty g^2):+\nn\\
+\Delta J^\infty(x|\beta_\infty g^2)\hat J^\infty(x'|\beta_\infty
g^2)-:\Delta J^\infty(x|\beta_\infty g^2)\hat
J^\infty(x'|\beta_\infty g^2): \ee \be \hat J^\infty(x|\beta_\infty
g^2)\Delta J^\infty(x'|\beta_\infty g^2)
-:\hat J^\infty(x|\beta_\infty g^2)\Delta J^\infty(x'|\beta_\infty g^2):=\nn\\
=\frac{1}{\alpha_\infty^2g^2c_\infty \beta_\infty } \oint_\infty
A_{\hat\textbf{J}}^{aa}(x',s|g^2)\left(\hat J^\infty(x|\beta_\infty
g^2)\hat\Omega^\infty(s|\beta_\infty g^2)-:\hat
J^\infty(x|\beta_\infty g^2)\hat \Omega^\infty(s|\beta_\infty
g^2):\right)=
\frac{1}{\alpha_\infty^2}A_{\hat\textbf{J}}^{aa}(x,x'|g^2) \ee \be
\Delta \hat J^\infty(x|\beta_\infty g^2)\hat
J^\infty(x'|\beta_\infty g^2)-:\Delta J^\infty(x|\beta_\infty
g^2)\hat J^\infty(x'|\beta_\infty g^2):=0 \ee}. Calculations for the
branching points looks similar, the main difference being that there
can be inequivalent points, as in examples s.5-8, and with help of
the operator $\hat U$ one should compensate impacts of $f^{ab}_{\hat
\textbf{J}}$ with $a\neq b$. An accurate calculation shows that this
changes the common sign, and one gets the result \be \label{U1}\hat
U=\sum_{a,b}\frac{1}{2\alpha_a\alpha_bg^4c_ac_b
\beta_a\beta_b}\oint_a\oint_b A_{\hat\textbf{J}}^{ab}(x,x'|g^2)\hat
\Omega^a(x|\beta_a g^2)\hat\Omega^b(x'|\beta_b g^2)+\sum_a\oint_a
k^a(z|\beta_a g^2)\hat \Omega^a(z|\beta_a g^2) \ee where the sum
runs over (inequivalent) branching points. In particular, in
examples below these are two branching points. One finally gets as a
result \be Z_\infty(t^\infty|\beta_\infty g^2)=e^{U_\infty}e^{\hat
U}\prod_{a}Z_a(t^a|\beta_a g^2) \ee

\subsection{Summary: Relations between special functions}

Given a couple of projections of the same ${\bf Z}$ (say, $Z(T,S)$
or $Z(R,Q)$) from Table 1, one can exclude the global partition
function and obtain an additional relation between the other two
special functions. In this way, any two special functions $Z_G$,
$Z_K$ and $Z_C$ (and $Z_{ACKM}$) can be related.

Relations between these special functions and the global partition
functions are summarized in the following tables\footnote{We do not
need more than two global partition functions, say, ${\bf Z}(T,S)$
and ${\bf Z}(T)$ in order to give rise to all interrelations between
local partition functions.}, where we write down only necessary part
of interrelations, while all the remaining relations can be read off
from these. Note that these tables are not symmetric, sometimes one
trivially inverse the relation (say, between $Z_G(t)$ and
$Z_G(\tilde t)$), while sometimes it is impossible (like the
relation between $Z_G$ and $Z_K$). In the former case we put in the
corresponding cell of the table "S", in the latter case, we put the
cross "X".

\bigskip

\subsubsection*{Table 2. Relations between local partition functions
emerging from ${\bf Z}(T^\pm)$}

\bigskip

\footnotesize{
\begin{tabular}{|c|c|c|c|c|c|c|}
\hline
&&&&&&\\
&${\bf Z}(T^\pm)$& $Z_G(t)$ & $Z_K(\tau)$
&$Z_G(\tilde t)$ & $ Z_K(\tilde\tau)$ & $Z_{ACKM}(t^\pm)$\\
&eq.(\ref{ZGl})&eq.(\ref{G})&eq.(\ref{K})&eq.(\ref{tt})&eq.(\ref{ttau})&
eq.(\ref{ZACKM})\\
\hline && $=e^{-U_G}Z_G(t)$&
$=e^{\hat U_K}Z_K(\tau_+)Z_K(\tau_-)$&&&\\
${\bf Z}(T^\pm)$& \rule{0.5cm}{0.5cm}& $t_k:$ (\ref{Tt}) & $\tau_k:$
(\ref{Ttau}),(\ref{61}),(\ref{62})
 && &\\
&&$U_G:$ (\ref{UG})&$\hat U_K:$ (\ref{UK5}),(\ref{UK})&&&\\
\hline & && $=e^{\hat U_{GK}}Z_K(\tau_+)Z_K(\tau_-)$ &
$=e^{U_{GG}} Z_G(\tilde t)$&&\\
$Z_G(t)$&S&\rule{0.5cm}{0.5cm}&$\tau^\pm_k:$ (\ref{62}),(\ref{60})&$t_k:$ (\ref{t-tt})&&\\
&&&$\hat U_{GK}=U_G+\hat U_K$&$U_{GG}:$ (\ref{UGG})&&\\
\hline
&&&&&$=e^{\hat U_{KK}}Z_K(\tilde\tau)$&\\
$Z_K(\tau)$&X& X & \rule{0.5cm}{0.5cm} & X&$\tilde\tau_k$:
(\ref{eta-teta})& X\\
&&&&&$\hat U_{KK}$: (\ref{UKK})&\\
\hline
&&&&&&\\
$Z_G(\tilde t)$& S& S & &\rule{0.5cm}{0.5cm} &&\\
&&&&&&\\
\hline
&&&&&&\\
$Z_K(\tilde\tau)$&X & X & S & X & \rule{0.5cm}{0.5cm}& \\
&&&&&&\\
\hline
&&&$=:e^{\hat {U}_{GK}}Z_K(\tau_+)Z_K(\tau_-):$ (\ref{dfk})&&&\\
$Z_{ACKM}(t^\pm)$&S & S &$t^\pm_k:$ (\ref{usl})&S&  &\rule{0.5cm}{0.5cm} \\
&&&$\hat U_{GK}$: (\ref{UGKo}),(\ref{UKKo1}),(\ref{UKKo2}),(\ref{dfk})&&&\\
\hline
\end{tabular}}
\normalsize

\bigskip

\bigskip

Table 2 deals with projections of the global partition function
${\bf Z}(T^\pm)$ and relations between the four special functions
which these projections generate. There are also similar relations
for other projections of the global partition function. For example,
the projection ${\bf Z}(T)$ gives rise to relations between the
complex matrix model and the Kontsevich partition function. We
consider it for the case of the sphere with two additional singular
points.

Table 3 deals with projections of the global partition function
${\bf Z}(T)$ and special functions $Z_K$ and $Z_C$.


\subsubsection*{Table 3. Relations between local partition functions
emerging from ${\bf Z}(T)$}

\bigskip

\begin{tabular}{|c|c|c|c|}
\hline
&&&\\
&${\bf Z}(T)$& $Z_C(\gotht)$ &
$Z_K(\tau)\tilde Z_K({\cal T})$\\
&&eq.(\ref{C})&\\
\hline &&$=e^{-V_C}Z_C(\gotht)$&
$=e^{V_C+\hat V_{ram}}Z_K(\tau)\tilde Z_K({\cal T})$\\
${\bf Z}(T)$& \rule{0.5cm}{0.5cm} &$\gotht_k:$ (\ref{gotht})
&$\tau_k:$ (\ref{tauC}); ${\cal
T}_k:$ (\ref{TauC})\\
&&${V_C}:$ (\ref{VC}) &${\hat V_{ram}}:$ (\ref{Vram1}),(\ref{Vram2})\\
\hline
\end{tabular}

\bigskip

\bigskip

These projections of ${\bf Z}(T)$ give rise to yet another
decomposition formula, $Z_C(\gotht)=e^{\hat V_{CK}}Z_K(\tau)\tilde
Z_K({\cal T})$ with $\hat V_{CK}=V_C+\hat V_K$.

Remaining sections will be devoted to detailed description of
particular cells in this table. They will be ordered so that simpler
calculations precede the more sophisticated ones. The simplest are
the relations involving vicinity of a single puncture and well known
functions, i.e. between $Z_G(t)$ and $Z_G(\tilde t)$ and between
$Z_K(\tau)$ and $Z_K(\tilde\tau)$. The next level of complexity
involves the more sophisticated ${\bf Z}(T,S)$, but relations are
still in vicinities of a single point: i.e. between ${\bf Z}(T,S)$
and other functions. After that, ${\bf Z}(T,S)$ can be eliminated to
provide relations involving vicinities of different punctures. The
same procedure is then repeated with the global function ${\bf
Z}(T)$ leading to relations involving $Z_C(\gotht)$, $Z_K(\tau)$ and
$\tilde Z_K({\cal T})$. Further generalization to arbitrary
(multi-cut, $n>1$) hyperelliptic curves and to Dijkgraaf-Vafa
branches of Hermitean-model partition function, as well as to
generic branches with genus expansion, \cite{amm1} described in
terms of the check-operators \cite{amm23} will be considered
elsewhere.

\newpage

\section{$Z_K(\tau) \rightarrow  Z_K(\tilde\tau)$}
\setcounter{equation}{0}

\subsection{Summary}

We start with the current \be \hat J_K(\xi|\tau) = \sum_{m=0}^\infty
\left(\frac{2m+1}{4} \tau_m
\xi^{2m} d\xi +
g^2\frac{d\xi}{\xi^{2m+2}}\frac{\partial}{\partial\tau_m} \right)
\label{Kocur} \ee A change of coordinate \be \tilde\xi =
G^{1/3}\sqrt{\xi^2-u} \ \ \ {\rm or}\ \ \ \xi =
\sqrt{G^{-2/3}\tilde\xi^2+u}, \label{xi-txi} \ee supplemented by
transform of time-variables \be
\tilde\tau_m = G^{-\frac{2m+1}{3}} \sum_{k=0}^\infty
\frac{\Gamma(m+k+3/2)\ u^k}{\Gamma(m+3/2)\ k!} \tau_{m+k},
\label{eta-teta} \ee converts this current into itself, \be \hat
J_K(\tilde\xi|\tilde\tau) = \sum_{m=0}^\infty \left(\frac{2m+1}{4}
\tilde\tau_m
\tilde\xi^{2m} d\tilde\xi +
g^2\frac{d\tilde\xi}{\tilde\xi^{2m+2}}\frac{\partial}{\partial\tilde\tau_m}
\right) \ee plus an additive correction\footnote{$\hat\nabla(\xi)$
is defined in (\ref{G}).} \be \hat J_K(\xi|\tau) = \hat
J_K(\tilde\xi|\tilde\tau) + \sum_{m<0} \frac{2m+1}{4}\tilde\tau_m
\tilde\xi^{2m} d\tilde\xi\ + \ee
$$
+ \ g^2\Big(\hat\nabla(\xi)u(\tau)\Big) \Big(\hat {\tilde{\cal
L}}_{-1}(\tilde\tau) - \frac{1}{16g^2}\Big)
-\frac{2g^2}{3}\Big(\hat\nabla(\xi)\log G(\tau)\Big)
\Big(\hat{\tilde{\cal L}}_{0}(\tilde\tau) - \frac{1}{16}\Big)
$$
which can be partly eliminated by twisting, \be
\Big(G^{1/12}e^{U_{KK}}\Big)\hat J_K(\xi|\tau)
\Big(G^{-1/12}e^{-U_{KK}}\Big) = \hat J_K(\tilde\xi|\tilde\tau) +
g^2(\hat\nabla u)\hat{\tilde{\cal L}}_{-1} + g^2(\hat\nabla\log G)
\hat{\tilde{\cal L}}_{0}, \label{j-tj} \ee where {\it function}
\be\label{A} U_{KK} = \frac{1}{2g^2}\sum_{k,l=0}^\infty A_{kl}(u,G)
\tau_k\tau_l
\\\sum_{k,l=0}^\infty A_{kl}(u,G) \tau_l
\frac{d\xi}{\xi^{2k+2}} = \frac{1}{2}\sum_{m=0}^\infty
(m+1/2)\frac{d\tilde\xi}{\tilde\xi^{2m+2}} \tilde\tau_{-m-1} \
\stackrel{(\ref{eta-teta})}{=}\\={1\over 2}G^{2m+1\over
3}\sum_{k,m=0}^\infty \left(m+{1\over
2}\right){\Gamma\left(k-m+{1\over 2}\right)\over\Gamma\left({1\over
2}-m\right)}{u^k\over
k!}{d\tilde\xi\over\tilde\xi^{2m+2}}\tau_{k-m-1} \ee \be \label{UKK}
A_{kl} = \frac{\Gamma(k+3/2)\Gamma(l+3/2)\ u^{k+l+1}}
{2\Gamma(1/2)^2k!l!(k+l+1)} = A_{lk} \ee Since $Z_K(\tilde\tau)$ is
annihilated by Virasoro constraints, including
$\hat{\tilde{L}}^K_{-1}$ and $\hat{\tilde{L}}^K_{0}$, it follows
(see comments below) that \be Z_K(\tau) =
G^{-1/24}e^{U_{KK}(\tau)}Z_K(\tilde\tau) \label{taueta-tauteta} \ee
for arbitrary choice of $u$ and $G$. Since (\ref{taueta-tauteta}) is
an identity in $u$ and $G$, these parameters are allowed to depend
on $\tau$ in an arbitrary way. This freedom can be used to impose
two arbitrary constraints on $\tilde\tau$. In particular, one can
define moment variables as follows (see s.2.5.2): \be
{\rm put} \ \ \tilde\tau_0=\tilde\tau_1^\Sigma=0, \nn \\
{\rm then} \ \  U_{KK} = \frac{1}{g^2}{\cal F}^{(0)}(\tau),\ \ \
-\frac{1}{24}\log G = {\cal F}^{(1)}(\tau),
\nn \\
{\rm while} \ \
{\cal F}^{(0)}(\tilde\tau) = {\cal F}^{(1)}(\tilde\tau) = 0,\ \ 
{\rm and} \ \ {\cal F}^{(p)}(\tilde\tau)\ {\rm with}\
p\geq 2\  {\rm are\ polynomials} \nn \\
\hbox{of a finite ($p$-dependent) order in finite ($p$-dependent)
number of a few first}\ \tilde\tau-{\rm variables.} \ee

\subsubsection{Comments on the derivation}

{\it Triangular} relation (\ref{eta-teta}) \be \tilde\tau_m
= G^{-\frac{2m+1}{3}} \sum_{k=0}^\infty \frac{\Gamma(m+k+3/2)\
u^k}{\Gamma(m+3/2)\ k!} \tau_{m+k}
\label{eta-teta2} \ee follows from equality of terms $\xi^{2m} d\xi$
with $m\geq 0$ in $\hat J_K(\xi)$ and $\hat J_K(\tilde\xi)$, which
are singular at infinity $\xi \sim \tilde\xi \sim \infty$.

In terms, which vanish at infinity, it is convenient first to
consider the case of $\tau$-independent $u$ and $G$. Then \be
\hat\nabla(\xi) \equiv \sum_{m\geq 0}^\infty \frac{d\xi}{\xi^{2m+2}}
\frac{\partial}{\partial\tau_m} = \sum_{m\geq 0}^\infty
\frac{d\tilde\xi}{\tilde\xi^{2m+2}}
\frac{\partial}{\partial\tilde\tau_m} \equiv
\hat{\tilde\nabla}(\tilde\xi) \label{nab-tnab-const} \ee and the
discrepancy between the two currents, \be \hat J_K(\xi|\tau) - \hat
J_K(\tilde\xi|\tilde\tau) = \frac{1}{4}\sum_{m<0} (2m+1)\tilde\tau_m
\tilde\xi^{2m} d\tilde\xi = -\frac{1}{4}\sum_{m= 0}^\infty
(2m+1)\tilde\tau_{-m-1} \frac{d\tilde\xi}{\tilde\xi^{2m+2}}, \ee is
eliminated by twisting with $U_{KK}$.

Relation (\ref{nab-tnab-const}) is a corollary of (\ref{eta-teta2})
for constant $u$ and $G$ \be \frac{\partial}{\partial\tau_n} =
\sum_{m = 0}^\infty \frac{\partial\tilde\tau_m}{\partial\tau_n}
\frac{\partial}{\partial\tilde\tau_m} =  \sum_{m = 0}^n
G^{-\frac{2m+1}{3}}\frac{\Gamma(n+3/2)\ u^{n-m}}{\Gamma(m+3/2)\
(n-m)!} \frac{\partial}{\partial\tilde\tau_m} \label{peta-pteta} \ee
Inverting this triangular relation, we obtain \be
\frac{\partial}{\partial\tilde\tau_m} = G^{\frac{2m+1}{3}}
\sum_{0\leq k \leq m} \frac{\Gamma(m+3/2)}{\Gamma(m-k+3/2)}
\frac{(-)^ku^k}{k!} \frac{\partial}{\partial \tau_{m-k}}
\label{pteta-peta} \ee

If $u$ and $G$ are $\tau$-dependent, additional terms are added to
the r.h.s. of (\ref{peta-pteta}): \be \frac{\partial u}{\partial
\tau_n} \sum_{m=0}^\infty \frac{\partial \tilde \tau_m}{\partial u}
\frac{\partial}{\partial\tilde\tau_m} + \frac{\partial \log
G}{\partial \tau_n} \sum_{m=0}^\infty \frac{\partial \tilde
\tau_m}{\partial\log G} \frac{\partial}{\partial\tilde\tau_m} = \nn
\\ = \frac{\partial u}{\partial \tau_n} \Big(\sum_{m=0}^\infty
(m+3/2) \tilde\tau_{m+1}
\frac{\partial}{\partial\tilde\tau_m} \Big) + \frac{\partial \log
G}{\partial \tau_n} \Big(-\frac{2}{3}\sum_{m=0}^\infty (m+1/2)
\tilde\tau_m
\frac{\partial}{\partial\tilde\tau_m}\Big) = \nn \\ = \frac{\partial
u}{\partial \tau_n} \Big(\tilde{\cal L}_{-1} -
\frac{1}{16}\tilde\tau_0^2\Big) -\frac{2}{3}\frac{\partial \log
G}{\partial \tau_n} \Big(\tilde{\cal L}_{0} - \frac{1}{16}\Big),
\label{pteta-teta-corr} \ee so that instead of
(\ref{nab-tnab-const}) we have: \be \hat\nabla(\xi) =
\hat{\tilde\nabla}(\tilde\xi) +
\Big(\hat\nabla(\xi)u\Big)\Big(\hat{\tilde{\cal L}}_{-1}(\tilde\tau)
- \frac{\tilde\tau_0^2}{16g^2}\Big)
-\frac{2}{3}\Big(\hat\nabla(\xi)\log G\Big) \Big(\hat{\tilde{\cal
L}_{0}}(\tilde\tau) - \frac{1}{16}\Big) \label{nabla-tnabla} \ee
When acting on $Z_K(\tilde\tau)$, which is annihilated by
$\hat{\tilde{\cal L}}_n$ with $n\geq -1$, including $n=-1$ and
$n=0$, additional terms produce \be
-\frac{\tilde\tau_0^2}{16g^2}\Big(\hat\nabla(\xi)u\Big) +
\frac{1}{24}\Big(\hat\nabla(\xi)\log G\Big) \ee which coincides with
additional terms \be
\Big(\hat\nabla(\xi)u\Big)\frac{\partial U_{KK}}{\partial u} +
\Big(\hat\nabla(\xi)\log G\Big) \left(
\frac{\partial U_{KK}}{\partial \log G} + \frac{1}{24}\right) \ee
coming from twisting the l.h.s. of (\ref{nabla-tnabla}) with
$G^{1/24}e^{-U_{KK}}$. Here we used that \be\label{*} {\partial
U_{KK}\over\partial \log G}=0 \ee \be\label{**} {\partial
U_{KK}\over\partial u}={\tilde\tau_0^2\over 16g^2} \ee


Taking a square of (\ref{j-tj}) we have: For $\hat J_K =
d\hat\varphi$ \be \hat{\tilde J_K}^2(\tilde\xi) = \tilde
T(\tilde\xi) +
d\left(\Big(\hat{\tilde\nabla}_{\tilde\xi}\tilde\xi\Big) \hat{\tilde
J_K}(\tilde\xi)
\right) \ee \be \left[ \hat{\tilde{\cal L}}_{-1}, \hat{\tilde
J_K}(\tilde\xi)\right] = d \left(\hat{\tilde
J_K}(\tilde\xi)\Big(\frac{1}{\tilde\xi} +
\frac{1}{\tilde\xi}\left[\hat{\tilde{\cal L}}_{-1},u\right] +
\tilde\xi\left[\hat{\tilde{\cal L}}_{-1},\log G\right]\Big)\right)
\ee \be \left[ \hat{\tilde{\cal L}}_{0}, \hat{\tilde
J_K}(\tilde\xi)\right] = d \left(\hat{\tilde
J_K}(\tilde\xi)\Big(\tilde\xi +
\frac{1}{\tilde\xi}\left[\hat{\tilde{\cal L}}_{0},u\right] +
\tilde\xi\left[\hat{\tilde{\cal L}}_{0},\log G\right]\Big)\right)
\ee so that \be \Big(\hat{J_K}(\tilde\xi|\tilde\tau) +
g^2(\hat\nabla u)\hat{\tilde{\cal L}}_{-1} + g^2(\hat\nabla\log G)
\hat{\tilde{\cal
L}}_{0}\Big)\hat{J_K}(\tilde\xi|\tilde\tau) Z_K(\tilde\tau) = \nn \\
= \left(\hat{\tilde T}(\tilde \xi) + d\left\{\hat{\tilde
J_K}(\tilde\xi) \Big[\hat\nabla- \hat{\tilde\nabla} + g^2(\hat\nabla
u)\Big(\hat{\tilde{\cal L}}_{-1} - \frac{1}{16g^2}\Big)
-\right.\right.\\-\left.\left. \frac{2g^2}{3} (\hat\nabla \log
G)\Big(\hat{\tilde{\cal L}}_{0} - \frac{1}{16}\Big),\
\frac{1}{\tilde \xi} u + \tilde\xi \log G\Big]\right\}
\right)Z_K(\tilde\tau) = \nn \hat{\tilde T}(\tilde
\xi)Z_K(\tilde\tau) \ee since operator in the commutator vanishes
due to (\ref{nabla-tnabla}). Note that $c$-numbers $1/16g^2$ and
$1/16$ do not affect the value of the commutator, but are needed to
make the operator vanishing.

One more (almost identical) way to check consistency is to check
that the r.h.s. of (\ref{taueta-tauteta}) is indeed independent of
$u$ and $G$: \be \frac{1}{2g^2}\left(\sum_{k,l=0}^\infty
\frac{\partial A_{kl}}{\partial u} \tau_k\tau_l\right)
Z_K(\tilde\tau) + \sum_{m=0}^\infty \frac{\partial
\tilde\tau_m}{\partial u}\frac{\partial Z_K(\tilde\tau)} {\partial
\tau_m} = \\=
\frac{1}{2g^2}\left(\sum_{k,l=0}^\infty
\frac{\partial A_{kl}}{\partial u} \tau_k\tau_l\right)
Z_K(\tilde\tau) + \left(\tilde{\cal L}_{-1} -
\frac{1}{16g^2}\tilde\tau_0^2\right)Z_K(\tilde\tau)=0 \ee because of
(\ref{**}).
and \be -\frac{1}{24}Z_K(\tilde\tau) +
\frac{1}{2}\left(\sum_{k,l=0}^\infty \frac{\partial
A_{kl}}{\partial\log G} \tau_k\tau_l\right)
Z_K(\tilde\tau) + \sum_{m=0}^\infty \frac{\partial
\tilde\tau_m}{\partial\log G}\frac{\partial Z_K(\tilde\tau)}
{\partial \tau_m} = \nn \\ = -\frac{1}{24}Z_K(\tilde\tau) +
\frac{1}{2}\left(\sum_{k,l=0}^\infty \frac{\partial
A_{kl}}{\partial\log G} \tau_k\tau_l\right)
Z_K(\tilde\tau) -\frac{2}{3} \Big(\tilde{\cal L}_{0} -
\frac{1}{16}\Big)Z_K(\tilde\tau) =0\ee because of (\ref{*}).

\subsubsection{Comments on generic change of variables}

Make now arbitrary change of variables in the vicinity of $\xi =
\infty$, consistent with the $\hat J_K(-\xi) = -\hat J_K(\xi)$ of
the current: \be \label{tilde}\xi \rightarrow \tilde\xi(\xi) =
G^{1/3}\xi \left(1 + \frac{b_1}{G^{2/3}\xi^2} + \ldots +
\frac{b_k}{G^{2k/3}\xi^{2k}} + \ldots\right) \ee The coefficients
$b_k$ can be either constants or functions of time-variables $\tau$.
We omit $G$ below.

Introduce non-linear combinations $C_{m,k}$ and $B_{m,k}$ of the
coefficients $b_k$: \be \tilde\xi^{2m+1} = \sum_{k=0}^\infty B_{m,k}
\xi^{2m+1-2k} \ee and \be \xi^{2m+1} = \sum_{k=0}^\infty C_{m,k}
\tilde\xi^{2m+1-2k} \ee Then, from
\be \sum_{m=0}^\infty \tau_m d\xi^{2m+1} = \sum_{m=0}^\infty
\tilde\tau_m d\tilde\xi^{2m+1} \ee we read \be \tilde\tau_m =
\sum_{k\geq 0} C_{m+k,k} \tau_{m+k} \ee and \be
\frac{\partial}{\partial \tau_m} = \sum_{0\leq k \leq m}
C_{m,m-k}\frac{\partial}{\partial\tilde \tau_k} +
O\Big(\frac{\partial C}{\partial\tau}\Big) \ee For
$\tau$-independent coefficients $B$ (and thus $C$) one obtains \be
\sum_{m=0}^\infty \frac{d\xi}{\xi^{2m+2}}
\frac{\partial}{\partial\tau_m} = \sum_{m=0}^\infty
\frac{d\tilde\xi}{\tilde\xi^{2m+2}}
\frac{\partial}{\partial\tilde\tau_m} \ee since \be\label{***}
\sum_{m\geq k}^\infty \frac{C_{m,m-k}\xi^{-2m-1}}{2m+1} =
\frac{\tilde\xi^{-2k-1}}{2k+1} \ee or \be (m+1/2)C_{m+k,k} =
(m+k+1/2)B_{-m-1,k} \label{CBrelation} \ee 
For the square-root change of the local parameter (\ref{xi-txi}) with $G=1$ \be
C_{m,k} = \frac{\Gamma(m+3/2)}{\Gamma(m-k+3/2)}\frac{u^k}{k!}, \ \ \
B_{m,k}(u) = C_{m,k}(-u) \ee and one can easily check
(\ref{CBrelation}) is fulfilled. More generally, put
$$
\tilde\xi = \xi\Big(1 + \frac{a}{\xi^2} + \frac{b}{\xi^4} +
\ldots\Big), \ \ \ \hbox{i.e.}\ \ \ \ \xi = \tilde\xi\Big(1 -
\frac{a}{\tilde\xi^2} - \frac{a^2+b}{\tilde\xi^4} - \ldots\Big)
$$
Then \be \label{generic} C_{m,0} = B_{m,0} = 1;\ \ \ C_{m,1} =
-(2m+1)a,\ \ \ B_{-m-1,1} = -(2m+1)a; \\ C_{m,2} = (2m+1)((m-1)a^2 -
b),\ \ \ B_{-m-1,2} = (2m+1)((m+1)a^2-b);\ \ \ \hbox{etc}\ee

One can also easily check these coefficients satisfy
(\ref{CBrelation}).

Using definition (\ref{A}), one can construct the matrix $A_{kl}$
for this generic change of variables. The result is \be\label{A1}
A_{kl}={1\over 2}\sum_{n=0}^l \left( l+{1\over
2}\right)C_{k,n+k+1}B_{-n-1,l-n}\ee Using (\ref{CBrelation}), it can
be rewritten as \be\label{A2} A_{kl}={1\over
2}\sum_{n=0}^l\left(n+{1\over 2}\right) C_{l,l-n}C_{k,n+k+1}\ee

This matrix $A_{kl}$ is to be symmetric for the conjugation operator
to exist. Let us check this is really the case\footnote{It is easy
to check with the explicit expressions (\ref{generic}) for several
first entries that $A_{kl}$ is symmetric: $A_{0,1}=A_{1,0}$,
$A_{2,0}=A_{0,2}$ etc}. To this end, one can convolute $A_{kl}$ in
the form (\ref{A2}) with ${x^{-2k-1}\over (2k+1)}{y^{-2l-1}\over
(2l+1)}$ to produce the generating function $A(x,y)$. Then, one can
use formula (\ref{***}) both for positive and negative $k$ to obtain
\be \label{Axy} A(x,y)\equiv \sum_{k,l=0}^\infty A_{kl}
{x^{-2k-1}\over (2k+1)}{y^{-2l-1}\over (2l+1)}={1\over 16}\left(\log
\left[{ x+y\over x-y}\right]^2-\log \left[{\tilde x(x)+\tilde
y(y)\over\tilde x(x)-\tilde y(y)}\right]^2\right)\ee where $\tilde
x(x)$ and $\tilde y(y)$ are accordingly functions of $x$ and $y$
given by (\ref{tilde}). It follows from this general formula that
$A(x,y)$ is a symmetric function of $x$ and $y$ and, therefore, the
matrix $A_{kl}$ is symmetric.

One can also ``independently" derive formula (\ref{Axy}) using the
tools developed in s.2.6. Indeed, using formulas (\ref{U1}) and
(\ref{AJ}), one can easily understand that the second derivative of
$A(x,y)$ w.r.t. $x$ and $y$ is equal to the difference of the two
$f_K$'s, up to the factor of ${1\over g^2}$, \be {1\over
g^2}{\partial^2 A(x,y)\over\partial x\partial
y}=f_K(x,y|g^2)-f_K(\tilde x (x),\tilde y (y)|g^2) \ee These latter
can be read off from the second formula of (\ref{fGKC}), and finally
the result is \be {\partial^2 A(x,y)\over\partial x\partial
y}={1\over 4}\left[{(x^2+y^2)dxdy\over (x^2-y^2)^2} - {(\tilde
x^2+\tilde y^2)d\tilde xd\tilde y\over (\tilde x^2-\tilde
y^2)^2}\right]\ee Upon integrating twice, this formula immediately
leads to (\ref{Axy}).

Thus, we have proved that, for the generic change of the local
parameter, there exists a conjugation operator that intertwines
between the currents $\hat J_K(\xi|\tau)$ and $\hat
J_K(\tilde\xi|\tilde\tau)$. However, this is not the case for the
corresponding Virasoro algebras: the projector generally is not
invariant under this change of the local parameter. If one requires
the Virasoro algebras would map to each other, this restricts the
generic changes of variables to the square-root one (\ref{xi-txi}).


\newpage

\section{${Z}_G(t) \rightarrow  Z_G(\tilde t)$\label{Gmom}}
\setcounter{equation}{0}

\subsection{Summary}

Now we start with the current \be \hat J_G
(z|t)=\sum_{k=0}\frac{1}{2}k {t}_k
z^{k-1}dz+\frac{Sdz}{z}+g^2\sum_{k=1}\frac{dz}{z^{k+1}}\frac{\p}{\p
t_k} \ee A change of coordinate \be z= a\tilde z+b \ee supplemented
by transform of time-variables \be \label{t-tt}m\tilde
t_m=\oint_{\infty}\frac{d v(z)}{\tilde z^{m}}= a
\oint_\infty\frac{\sum_k
kt_k(az+b)^{n-1}dz}{z^m}=a^m\sum_k{k!b^{k-m}\over (m-1)!(k-m)!}t_k
\ee converts this current into itself, \be \hat J_G (\tilde z|\tilde
t)=\sum_{k=0}\frac{1}{2}k \tilde{t}_k \tilde z^{k-1}d\tilde
z+\frac{Sd\tilde z}{\tilde z}+g^2\sum_{k=1}\frac{d\tilde z}{\tilde
z^{k+1}}\frac{\p}{\p \tilde t_k} \ee plus an additive correction \be
\hat J_G (z|t)=\hat J_G (\tilde z|\tilde t)+S{d\tilde z\over\tilde
z}\sum_{k=0}^\infty\left(-{b\over a\tilde
z}\right)^k+g^2\left[\hat\nabla (z)\log a\right]\left[\hat{\tilde
L}_0(\tilde t)-S^2\right]+g^2\left[\hat\nabla (z)
b\right]\hat{\tilde L}_{-1}(\tilde t) \ee which can be partly
eliminated by twisting \be \left(a^{S^2}e^{U_{GG}}\right)\hat J_G
(z|t)\left(a^{-S^2}e^{-U_{GG}}\right)=\hat J_G (\tilde z|\tilde
t)+g^2\left(\hat\nabla\log a\right)\tilde {\hat
L}_0+g^2\left(\hat\nabla b\right)\tilde {\hat L}_{-1}\ee where {\it
function} \be\label{UGG} U_{GG} = \frac{S}{g^2}\sum_{k=1}^\infty b^k
t_k =\frac{S}{g^2}\sum_{k=1}^\infty \left(-{b\over a}\right)^k
\tilde t_k\ee
 Since $Z_G(\tilde t)$ is
annihilated by the Virasoro constraints, including $\hat{\tilde
L}^G_{-1}$ and $\hat{\tilde L}^G_{0}$, it follows (see comments
below) that \be Z_G(t) = a^{S^2}e^{U_{GG}(t)}Z_G(\tilde t)
\label{Zt-Ztt} \ee for arbitrary choice of $a$ and $b$. Since
(\ref{Zt-Ztt}) is an identity in $a$ and $b$, these parameters are
allowed to depend on $t$ in an arbitrary way. This freedom can be
used to impose two arbitrary constraints on $\tilde t$.

\subsection{Comments}

Let us check that $Z_G$ in fact does not depend on $a$ and $b$: \be
Z_G^{-1}\p_b Z_G=S\sum_{k=1}k b^{k-1}
t_k+Z_G(\tilde t)^{-1}\sum_{k=1}\frac{\p \tilde t_k}{\p b}\frac{\p}{\p
\tilde t_k}Z_G(\tilde t)=\frac{1}{aZ_G(\tilde t)}\left(\sum_{k=1}(k+1)\tilde
t_{k+1}\frac{\p}{\p \tilde t_k}+S\tilde t_1\right)Z_G(\tilde t_t)=0 \ee and \be
\p_a Z_G=\frac{S^2}{a}+\frac{1}{a}\sum_{k=1}k \tilde
t_k\frac{\p}{\p \tilde t_k}Z_G(\tilde t)=0 \ee




\newpage

\section{Projections of ${\bf Z}(T,S)$}
\setcounter{equation}{0}

For the sake of simplicity, we consider here the symmetric model.
The generic model is considered in s.7.

\subsection{${\bf Z}(T,S)$ to $Z_G(t)$. Summary}

Projecting means one should establish equality, up to conjugation,
of the global current and the local current $\hat J_G$ at the
vicinity of infinity, i.e. \be \frac{1}{2}kt_k \cong \oint_\infty
z^{-k}\hat{\bf J^o}, \ \ \ \frac{\p}{\p t_k} \cong \oint_\infty
z^k\hat{\bf J^o} \label{tvsT} \ee where the sign $\cong$ means
equality modulo possible conjugation, which does not affect
commutation relations. In fact, the first equality is exact and
gives rise to the change of time variables, \be \frac{1}{2}kt_k =
\oint_\infty z^{-k}\hat{\bf J^o}\ee and, inversely (see notations in
(\ref{G})), \be\label{Tt}
\left(k+\frac{1}{2}\right)T_k=2\oint_\infty\frac{z\hat
J_G(z)}{y^{2k+1}}, \ \ \ \
\left(k+\frac{1}{2}\right)S_k=2\oint_\infty\frac{\hat J_G(z)}{y^{2k+1}}\\
\widetilde{T}_k=\oint_\infty\frac{v(z)dz}{y^{2k+3}},\ \ \ \
\widetilde{S}_k=\oint_\infty\frac{zv(z)dz}{y^{2k+3}} \ee In
accordance with these formulas, $T_0$ contains ${\p\over\p t_0}$.
However, as we mentioned above, one may deal with ${\p\over\p t_0}$
as with a $c$-number (equal to ${S\over g^2}$ when acting on $Z_G$)
and, moreover, to make the Legandre transform to replace it with
$t'_0$.

In order to reproduce the second equality in (\ref{tvsT}), one needs
a non-trivial conjugation operator, \be \frac{\p}{\p t_k} = e^{
U_{G}}\left(\oint_\infty z^k\hat{\bf J^o}\right) e^{-U_{G}} \ee We
construct it following s.2.6, the result being the sum of two
pieces, the first one is
\be U_{G,1}=\frac{1}{2}\oint_\infty\oint_\infty A_{{\bf
J}}^{\infty\infty}(z,z')\hat\Omega_G(z)\hat\Omega_G(z')dzdz'
=\frac{1}{2}\oint_\infty\oint_\infty\rho^{(0|2)}_G(z_1,z_2)
v^\Sigma(z_1)v^\Sigma(z_2)dz_1dz_2 \ee where \be A_{{\bf
J}}^{\infty_+ \infty_+}(z,z')=-A_{{\bf
J}}^{\infty_+\infty_-}(z,z')=A_{{\bf J}}^{\infty_-\infty_-}(z,z')=
\rho^{(0|2)}_G(z,z')dzdz' \ee and
\be\rho^{(0|2)}_G(z_1,z_2)=\frac{dz_1 dz_2}{2(z_1-z_2)^2}\left(
\frac{z_1z_2-4S}{y(z_1)y(z_2)}-1\right) \ee is the two-point
resolvent, \cite{amm1}, $v^\Sigma(z)\equiv \sum t^\Sigma_kz^k$, for
other notations see (\ref{G}).

The second piece, as explained in s.2.6, comes from noting that the
shift for the global current is to be proportional to the
1-form\footnote{Considering $y^n(z)dz$ instead would lead to higher
critical points of the same Gaussian branch.} $\Omega_{DV}=y(z)dz$
in order to reproduce the corresponding shift in $J_G$. It is \be
\delta \hat J_G=-\frac{y(z)dz}{2} \ee at the vicinity of infinity,
can be also reproduced by the (still $c$-number) conjugation
operator\footnote{This is {\it a priori} clear from the fact that,
unlike changing the polynomial part of the potential \cite{amm1},
adding any negative powers does not change the structure of
solutions to the loop equations.}: \be
U_{G,2}=\oint_{\infty}\left(-\frac{y(z)dz}{2}+zdz\right)
\hat\Omega^\infty (z)=\oint \rho^{(0|1)}_G v^\Sigma(z)dz \ee where
$\rho^{(0|1)}_G$ is the one-point resolvent, \cite{amm1}. Thus,
finally one has \be \label{UG} {\bf Z}(T,S)=e^{- U_G}Z_G(t),\\
U_{G}=\frac{1}{g}\oint\rho^{(0|1)}(z){v^\Sigma}(z)+
\frac{1}{2!}\oint\rho^{(0|2)}(z_1,z_2){v}^\Sigma(z_1){v}^\Sigma(z_2)
\ee


\subsection{${\bf Z}(T,S)$ to $Z_K(\tau)$. Summary}

In the vicinities of ramification points we not fix the local
parameters, just put them to be a series \be z\pm
a=\sum_{k=1}^\infty\alpha^\pm_k x_\pm^{2k} \ee with arbitrary
coefficients $\alpha^\pm_k$ and non-zero $\alpha^\pm_1$. Here we choose
$\alpha^+_k=\alpha^-_k$, postponing the generic case until s.7. In the next
section, we consider an explicit example of choice of the local
parameter.

Projecting in the vicinity of ramification points provides us with
the first example when, of two equalities\be\label{Ttau}
(2k+1)\tau^\pm_k \cong \oint_\infty x_\pm^{-(2k+1)}\hat{\bf J^o}, \
\ \ \frac{\p}{\p \tau^\pm_k} \cong \oint_\infty x_\pm^{2k+1}\hat{\bf
J^o} \label{etavsT} \ee the second equality does not require
conjugation, \be \frac{1}{2}\frac{\p}{\p \tau^\pm_k} =\oint_{a_\pm}
x_\pm^{2k+1}\hat{\bf J^o} \ee while the first one does. This leads
to conjugation operator which is an exponential of quadratic form
(with constant coefficients) in time derivatives, not in times as in
the examples above: \be \left(k+\frac{1}{2}\right)\tau^\pm_k  =
e^{\hat U_{K}}\left(\oint_\infty x_\pm^{-k}\hat{\bf J^o}\right)
e^{-\hat U_{K}} \label{etaoint} \ee with \be \hat U_K=\hat
U_{K,1}+U_{K,2}\\
\hat U_{K,1}=\frac{1}{2}\sum_{a,b=a_\pm}\oint_a\oint_b A_{{\bf
J}}^{ab}(z,z')\hat\Omega_K^a(z)\hat\Omega_K^b(z')=\nn\\
=\frac{1}{4}\oint_{a_+}\oint_{a_+}\left(\frac{dz_1
dz_2}{2(z_1-z_2)^2}
\frac{z_1z_2-4S}{y(z_1)y(z_2)}-\frac{dx_1dx_2(x_1^2+x_2^2)}
{(x_1^2-x_2^2)^2}\right)\hat\Omega_K^{a_+}(x_1)\hat\Omega_K^{a_+}(x_2)+\nn\\
+\frac{1}{4}\oint_{a_+}\oint_{a_-}\left(\frac{dz_1
dz_2}{2(z_1-z_2)^2}
\frac{z_1z_2-4S}{y(z_1)y(z_2)}\right)\hat\Omega_K^{a_+}(x_1)
\hat\Omega_K^{a_-}(x_2)=\\=
\frac{1}{4}\sum_{a=a_\pm}\oint_{a_+}\oint_{a}\left(\frac{dz_1
dz_2}{2(z_1-z_2)^2}
\frac{z_1z_2-4S}{y(z_1)y(z_2)}\right)\hat\Omega_K^{a_+}(x_1)
\hat\Omega_K^{a}(x_2)\ee 

Again the shift \be
\delta \hat J_K=-\frac{y(z)dz}{2}
\ee at the vicinities of the ramification points can be reproduced
by the conjugation operator \be \hat U_{K,2}=\sum_{a=a_\pm}\oint_{a}
\frac{1}{2}\left(y(z)dz-x(z)^2dx(z)
\oint_{a}\frac{y(z')dz'}{x(z')^3}\right)\hat \Omega^a_K(x(z)) \ee
Thus, finally one has \be\label{UK5} {\bf Z}(T,S)=e^{\hat
U_K}Z_K(\tau),\\ \hat U_K={1\over 2}\sum_{\xi=a_\pm}\oint_{\xi}
\left[y(z)dz-x^2dx \oint_{a}\frac{y(z')dz'}{x(z')^3}
+\frac{1}{2}\oint_{a_+}\frac{dz' dz}{2(z'-z)^2}
\frac{z'z-4S}{y(z')y(z)}\ \hat\Omega_K^{a_+}(x')\right]
\hat\Omega_K^{\xi}(x)\ee

\subsection{Comments}

Let us check that the $J_G$ and $J_K$ are actually local currents
corresponding to the global current ${\bf J}^o$, (\ref{gcurrent}) at
vicinities of infinity and ramification points. To this end, we
compare the corresponding operator product expansions, which are,
for the local currents, \be \hat J_G(z)\hat J_G(z')= {dzdz'\over
2(z-z')^2}+ \hbox{regular part}\\
\hat J_K(z)\hat J_K(z')= {(x^2+x'^2)dxdx'\over (x^2-x'^2)^2}+
\hbox{regular part}\ee One can immediately check that at the
vicinity of infinity, $z\to\infty$ \be \frac{dz
dz'}{2(z-z')^2}\frac{z z'-4S}{y(z)y(z')}=\frac{dzdz'}
{2(z-z')^2}+\sum_{i,j=2}^\infty \frac{A_{ij}dzdz'}{z^iz'^j}
\label{normcom} \ee which proves the identity of operator product
expansions for $\hat {\bf J}^o$ and $\hat J_G$ at the vicinity of
infinity, while the similar identity for $\hat {\bf J}^o$ and $\hat
J_K$ at the vicinity of ramification points requires some job.
Indeed, one needs to prove that \be \frac{dzdz'}{2(z-z')^2}\frac{z
z'-4S}{y(z)y(z')}=\frac{(x^2+x'^2)dxdx'}{(x^2-x'^2)^2}+\sum_{i,j=0}^\infty
\alpha_{ij}x^ix'^jdxdx' \label{oddcom} \ee for any local coordinate
$x$, \be z=\pm 2\sqrt{S}\left(1+f(x_{\pm}^2)\right) \ee with $f(x)$
an arbitrary power series such that $f(0)=0$. Then, one has to prove
the two identities \be {dzdz'\over 2(z-z')^2}{zz'-4S\over
y(z)y(z')}{(x^2-x'^2)^2\over dxdx'}\stackrel{x^2\to
x'^2}{\longrightarrow} 2x^2 \ee and \be {\partial\over\partial
x}\left({dzdz'\over 2(z-z')^2}{zz'-4S\over
y(z)y(z')}{(x^2-x'^2)^2\over dxdx'}\right)\stackrel{x^2\to
x'^2}{\longrightarrow} 2x \ee The first one is quite obvious, the
second one requires some (straightforward) calculation.

\newpage

\section{$Z_G(t) \rightarrow Z_K(\tau_+)Z_K(\tau_-)$}
\setcounter{equation}{0}

\subsection{Summary}

Let us consider an explicit example of the local coordinate system
at the vicinities of ramification points and manifestly relate $Z_G$
and $Z_K$ in this parametrization. Namely, we choose the local
coordinate in accordance with \cite{Ch}\footnote{In this choice of
local coordinate, the change of time variables can be naturally
formulated in terms of an external matrix $\Lambda\equiv e^\lambda$,
parameterizing the $(T,S)$-time variables \be\label{61}
\left(k+\frac{1}{2}\right)T_k=\hbox{Tr}\frac{\Lambda^{2k}(\Lambda^2+S)}{(\Lambda^2-S)^{2k+1}}-\delta_{k0}\\
\left(k+\frac{1}{2}\right)S_k=\hbox{Tr}\frac{\Lambda^{2k+1}}{(\Lambda^2-S)^{2k+1}}
\ee Then, (\ref{etaoint}) can be represented as \be\label{62}
\tau^\pm_k=\frac{\sqrt{S}}{(2k+1)!}\frac{\p^{2k}}{\p\lambda^{2k}}\hbox{Tr}\frac{1}{e^\lambda\pm\sqrt{S}}
\ee In the same terms, (\ref{Tt}) can be represented as
\be\label{60} t_k={1\over k}\hbox{Tr}{1\over
\left(\Lambda+\sqrt{S}\Lambda^{-1}
\right)^k}\ee} \be
y(z)^2=S(e^{x_\pm}-e^{-x_{\pm}})^2\ \ \ \ \ \ \ \ \
z=\pm \sqrt{S}(e^{x_\pm}+e^{-x_\pm}) \ee the two signs corresponding
to the two ramification points. On the physical plane \be dz= y(z)
dx \ee The (1,1)-differentials in this case are \be
f^{a_+a_+}(z,z')=f^{a_-a_-}(z,z')= \frac{dzdz'}{2(z-z')^2}\frac{z
z'-4S}{y(z)y(z')} -\frac{(x_1^2+x_2^2)dx_1dx_2}
{(x_1^2-x_2^2)^2}=\\=\sum_{i,j=0}^{\infty}(C_{++}^{ik})x_1^{2i}x_2^{2j}(2j+1)dx_1dx_2 \\
f^{a_+a_-}(z,z')=f^{a_-a_+}(z,z')=
\frac{dzdz'}{2(z-z')^2}\frac{z z'-4S}{y(z)y(z')}
=\sum_{i,j=0}^{\infty}(C_{+-}^{ik})x_1^{2i}x_2^{2j}(2j+1)dx_1dx_2
\ee where \be
\sum_{i=0}^{\infty}(C_{++}^{ik})x^{2i}=-\frac{1}{2}\sum_{j=0}^{\infty}\frac{B_{2j+2k+2}x^{2j}}
{(2j+1)!(2k)!(j+k+1)} \ee and \be
\sum_{i=0}^{\infty}(C_{+-}^{ik})x^{2i}=-\frac{1}{2}\sum_{j=0}^{\infty}\frac{B_{2j+2k+2}x^{2j}(2^{2j+2k+2}-1)}
{(2j+1)!(2k)!(j+k+1)} \ee Here $B_{2k}$ are the Bernoulli numbers.

Then local currents are \be \hat{J}^{a_-}=2\hat J_K(x_-|\tau_-)
+\sum_{ij=0}^\infty x^{2i}\left((C_{++}^{ij})\frac{\p}{\p\tau^-_j}-
(C_{+-}^{ij})\frac{\p}{\p\tau^+_j}\right)dx \ee and \be
\hat{J}^{a_+}=-2\hat J_K(x_+|\tau_+) +\sum_{ij=0}^\infty
x^{2i}\left(-(C_{+-}^{ij})\frac{\p}{\p\tau^-_j}+
(C_{++}^{ij})\frac{\p}{\p\tau^+_j}\right)dx \ee
This means that the conjugation operator in the decomposition
formula \be\label{df} Z_G(t)=e^{\hat
U_{GK}}Z_K(\tau^+)Z_K(\tau^-)\ee is \be \hat U_{GK}=U_G+\hat U_K=\\=
U_G+\frac{1}{2}\sum_{i,j=\{+,-\}}\oint_{a_i}\oint_{a_j}f^{a_ia_j}(x_i,x_j)
\hat\Omega_K^{a_i}(x_i)\hat\Omega_K^{a_j}(x_j)-\frac{1}{2}\sum_{i=\{+,-\}}\oint_{a_i}
\left(2Sx^2-\frac{y^2}{2}\right)\hat\Omega_K^{a_i}dx \ee and, since
\be y(x)^2=S\sum_{k=1}^{\infty}\frac{2^{2k+1}x^{2k}}{(2k)!} \ee the
explicit expression for $\hat U_K$ coincides with that given in
\cite{Ch}, \be\label{UK} \hat
U_K=\frac{1}{4}\sum_{i,j=0}\frac{B_{2i+2j+2}}{(i+j+1)}\frac{1}{(2i+1)!(2j+1)!}\left(
\frac{\p}{\p\tau^+_i}\frac{\p}{\p\tau^+_j}+\frac{\p}{\p\tau^-_i}\frac{\p}{\p\tau^-_j}-2
(2^{2i+2j+2}-1)\frac{\p}{\p\tau^+_i}\frac{\p}{\p\tau^-_j}\right)+\\+{2S\over
g}\sum_{i=2}^\infty\frac{2^{2i-1}}{(2i+1)!}
\left(\frac{\p}{\p\tau^-_i}-\frac{\p}{\p\tau^+_i}\right) \ee

\subsection{Comments}

Let us consider here what happens if one does not put ${\p
Z_G\over\p t_0}={S\over g^2} Z_G$ instead dealing with it as with
the differential operator and a partition function $Z_{eG}$ that
generalizes the matrix model partition function $Z_G$. The global
current $\hat {\bf J}^{o}(z,y_G)$ in the vicinity of the infinity is
equivalent to the local current $\hat J_G(z)
-\frac{1}{z}\frac{\p}{\p t_0}$. To give this current the standard
form, one should make the two conjugations: first, \be
e^{4\frac{\p}{\p t_0}\frac{\p}{\p T_0}}\hat {\bf
J}^{o}(z,y_G)e^{-4\frac{\p}{\p t_0}\frac{\p}{\p T_0}}=\hat {\bf
J}^{o}(z,y_G)+\frac{1}{y(z)}\frac{\p}{\p t_0} \ee and then \be
e^{\sum_{k=0}^\infty c_k T_k \frac{\p}{\p t_o}}\left(\hat {\bf
J}^{o}(z,y_G)+\frac{1}{y(z)}\frac{\p}{\p t_0}\right)
e^{-\sum_{k=0}^\infty c_k T_k \frac{\p}{\p t_o}}=\hat {\bf
J}^{o}(z,y_G)+\frac{1}{z}\frac{\p}{\p t_0} \ee which defines the
coefficients $c_k$.

Since $\hat {\bf J}^{o}(z,y_G)=\hat \nabla(z)+ \dots$, this
conjugation operator can be expressed explicitly \be
\sum_{k=0}^\infty c_k T_k \frac{\p}{\p t_0}=(T_{-1}-t_0)\frac{\p}{\p
t_0} \ee with \be\label{T-1} T_{-1}=\oint_\infty \frac{v(z)dz}{y(z)}
\ee This means that the actual decomposition formula for the
Gaussian curve is \be\label{dft0} e^{-4\frac{\p}{\p t_0}\frac{\p}{\p
T_0}}e^{(t_0-T_{-1})\frac{\p}{\p t_0}}e^{-\hat
U_{G}}Z_{eG}=Z(t_0)e^{\hat U_{K}} Z_K(\tau^+)Z_K(\tau^-) \ee where
$Z(t_0)$ is a function (formal series) of $t_0$ with genus
expansion. This formula describes the decomposition of an arbitrary
branch (with genus expansion) of the partition function for the
curve $y_G(z)$. For the Gaussian solution, $e^{-\hat U_{\infty}}Z_G$
does not depend on $t_0$ and one returns back to \be e^{-
U_{G}}Z_{G}=e^{\hat U_{K}} Z_K(\tau^+)Z_K(\tau^-)\ee A nontrivial
consequence of this formula is factorizing out the dependence on
$t_0$. We check this statement explicitly for the perturbative
expansion in $t$ in Appendix II and determine first terms of the
function $Z(t_0)$ manifestly.



\newpage

\section{$Z_{ACKM}$ and recovery of Kostov's formula}
\setcounter{equation}{0}

\subsection{$Z_G(t) \rightarrow Z_K(\tau_+)Z_K(\tau_-)$ in
the non-symmetric case. Summary}

In this section we consider the non-symmetric curve \be
y^2(z)=(z-a_+)(z-a_-) \ee with constant $a_\pm$. Now we, following
the line of s.5-6, repeat all the calculation for non-symmetric case
and for generically chosen coordinates in the vicinities of the two
ramification points (i.e. for the two corresponding local parameters
chosen independently). The global current now is the following \be
\hat {\bf J}^o(z)=
\frac{1}{2}\sum_{k=0}^{\infty}\left(k+\frac{1}{2}\right)
\left((z-a_+)\left(T_+^k+\frac{2S\delta_{k,0}}{a_--a_+}\right)+(z-a_-)\left(T_-^k-\frac{2S\delta_{k0}}{a_--a_+}\right)\right)
y^{2k-1}dz+g^2\nn\\
\sum_{k=0}\left((z-a_+)\frac{\p}{\p \tilde T_+^k}+
(z-a_-)\frac{\p}{\p\tilde T_-^k}\right)y^{-2k-3}dz \ee where \be
\frac{\p}{\p \tilde T_+^k}=\frac{\p}{\p T_+^k}+\frac{\p}{\p
T_-^k}+\frac{4(k+1)}{(2k+3)(a_+-a_-)^2}\left(\frac{\p}{\p T_+^{k-1}}-\frac{\p}{\p T_-^{k-1}}\right)\\
\frac{\p}{\p \tilde T_-^k}=\frac{\p}{\p T_-^k}+\frac{\p}{\p
T_+^k}+\frac{4(k+1)}{(2k+3)(a_+-a_-)^2}\left(\frac{\p}{\p
T_-^{k-1}}-\frac{\p}{\p T_+^{k-1}}\right) \ee and we deal with $S$
as a parameter independent on $a_\pm$. Nearby infinity the global
current is equivalent to the corresponding local current \be \hat
J_G(z)=\frac{1}{2}\sum_{k=0} k t_kz^{k-1}dz
+\frac{Sdz}{z}+\sum_{k=1} {dz\over z^{k+1}}{\p\over\p t_k} \ee By
definition, one has \be
T^k_+=\frac{1}{(k+\frac{1}{2})(a_--a_+)}\oint_\infty\frac{(z-a_+)d v(z)}{y^{2k+1}}\\
T^k_-=\frac{1}{(k+\frac{1}{2})(a_+-a_-)}\oint_\infty\frac{(z-a_-)d
v(z)}{y^{2k+1}} \ee The conjugation operator, as in the symmetric
case, is \be\label{UGKo}
U_{G}=\frac{1}{2g^2}\oint_\infty\oint_\infty\rho^{(0|2)}(z_1,z_2)
v^\Sigma(z_1)v^\Sigma(z_2)+\frac{S}{g^2}
\oint\left(\frac{1}{y}-\frac{1}{z}\right)v^\Sigma(z)dz +f(S,a_{\pm})
\ee where \be
\rho^{(0|2)}(z_1,z_2)=\frac{1}{2(z_1-z_2)^2}\left(\frac{z_1z_2-(a_++a_-)(z_1+z_2)/2+a_+a_-}{y(z_1)y(z_2)}-1\right)
\ee and $f(S,a_{\pm})$ is given in (\ref{f}).
Like above, \be \tau_\pm^k=\frac{1}{(k+1/2)}\oint_{a_\pm}\frac{
d{v}(z)}{x_\pm^{2k+1}} \ee and the operator $\hat U_K$ is now
\be\label{UKKo1} \hat
U_K=8g^2\sum_{i,j=\pm}\sum_{k,m=0}^{\infty}A_{ij}^{km}\frac{\p^2}{\p\tau_k^i\p
\tau_m^j} \ee where \be\label{UKKo2}
A_{++}^{km}=\oint_{a_+}\oint_{a_+}\left(\frac{dz_1dz_2}{2(z_1-z_2)^2}\left(\frac{z_1z_2-(a_++a_-)(z_1+z_2)/2+a_+a_-}{y(z_1)y(z_2)}\right)\right)\frac{1}{(2k+1)(2m+1)x_1^{2k+1}x_2^{2m+1}}\nn\\
A_{--}^{km}=\oint_{a_-}\oint_{a_-}\left(\frac{dz_1dz_2}{2(z_1-z_2)^2}\left(\frac{z_1z_2-(a_++a_-)(z_1+z_2)/2+a_+a_-}{y(z_1)y(z_2)}\right)\right)\frac{1}{(2k+1)(2m+1)x_1^{2k+1}x_2^{2m+1}}\nn\\
A_{+-}^{km}=\oint_{a_+}\oint_{a_-}\frac{dz_1dz_2}{2(z_1-z_2)^2}\left(\frac{z_1z_2-(a_++a_-)(z_1+z_2)/2+a_+a_-}{y(z_1)y(z_2)}\right)\frac{1}{(2k+1)(2m+1)x_1^{2k+1}x_2^{2m+1}}\nn\\
A_{-+}^{km}=\oint_{a_-}\oint_{a_+}\frac{dz_1dz_2}{2(z_1-z_2)^2}\left(\frac{z_1z_2-(a_++a_-)(z_1+z_2)/2+a_+a_-}{y(z_1)y(z_2)}\right)\frac{1}{(2k+1)(2m+1)x_1^{2k+1}x_2^{2m+1}}
\ee and \be\label{ffK} Z_G(t,g^2)=e^{\hat U_{G}}e^{\hat
U_K}Z_K(\tau^+,4g^2)Z_K(\tau^-,4g^2) \ee As usual, this formula is
defined up to an arbitrary factor $f(S,a_\pm)$. Below we prove that,
in the present case, this factor is equal to
\be\label{f}f(S,a_{\pm})=-\left({S^2\over g^2}+{1\over 8}\right)\log
(a_+-a_-)\ee

\subsection{Kostov's formula and ACKM polynomial decomposition}

Now using the decomposition formula (\ref{ffK}), we construct a
polynomial (moment) representation of the Gaussian partition
function $Z_G$. Our strategy is as follows. First of all, we make
the change of variables as in (\ref{eta-teta}) with some functions
$G_{\pm}$ and zero $u$ and then, after calculating all the
derivatives w.r.t. time variables in (\ref{ffK}), fix $G_{\pm}$ and
$a_{\pm}$ to be specific functions of times so that the r.h.s. of
(\ref{ffK}) becomes a polynomial of moment variables (\ref{ACKMmom})
at each genus. This is an allowed procedure, since the r.h.s. of
(\ref{ffK}) does not really depend on $a_{\pm}$! Therefore, one may
put them to be any functions of times.

Thus, we start from the transformation of times (\ref{eta-teta})
with $u=0$ so that \be
Z_K(\tau_0,\tau_1,\tau_2,\dots,g^2)=G^{-\frac{1}{24}}Z_K(\tau_0
G^{-1/3},\tau_1 G^{-1},\tau_2 G^{-5/3},\dots,g^2) \ee and \be
Z_G=e^{U_{G}-\frac{1}{24}\log(G_+G_-)}e^{\hat U_{K}} Z_K(\tau^+_0
G^{-1/3}_+,\tau^+_1 G^{-1/3}_+,\tau^+_2G^{-5/3}_+,\dots,4g^2)
\times\\\times Z_K(\tau^-_0 G^{-1/3}_-, \tau^-_1 G_-^{-1},\tau^-_2
G^{-5/3}_-,\dots,4g^2) \ee

Now one should achieve $\tilde\tau_{0,1}^{\Sigma\pm}=0$ so that
$\log Z_G$ become a finite degree polynomial at each genus in ACKM
moment variables defined as \be
t^-_k=\frac{1}{2}\oint_{A_+,A_-}\frac{dv}{y(z)(z-A_+)^k}=-\frac{1}{2\sqrt{d}}\sum_{m=0}^{k-1}
\frac{\Gamma(1/2)(k-m+1/2)}{\Gamma{(m+1)}\Gamma{(1/2-m)}}
{\tau^+_{k-m}\over d^m}\\
t^+_k=\frac{1}{2}\oint_{A_+,A_-}\frac{dv}{y(z)(z-A_-)^k}=(-1)^{k+1}
\frac{1}{2\sqrt{d}}\sum_{m=0}^{k-1}
\frac{\Gamma(1/2)(k-m+1/2)}{\Gamma{(m+1)}\Gamma{(1/2-m)}}{\tau^-_{k-m}\over
d^m} \label{usl} \ee where $d\equiv A_+-A_-$. To this end, after the
action of the differential conjugation operator, one should consider

\paragraph{1)} special $a_\pm=A_\pm$, namely, those satisfying the condition
\be\label{A+-} T_\pm^0\pm\frac{2S}{A_--A_+}=0\ee
 This fixes
$\tilde\tau_{0}^{\Sigma\pm}=0$. Note that such $A_{\pm}$ coincide
with appropriate ramification points in \cite{ACKM}.

\paragraph{2)} special $G_\pm$ such that $(\tau^{\Sigma\pm}_1-\frac{2}{3})
G_{\pm}^{-1}+\frac{2}{3}=0$. This fixes
$\tilde\tau_{1}^{\Sigma\pm}=0$.

\bigskip

Finally one gets \be\label{dfk}
Z_G(t^{\pm},g^2)=e^{U_{G}-\frac{1}{24}\log(\tau_1^+\tau_1^-)}e^{\hat
U_{K}(a_\pm)}Z_K(\tau^+_k G_+^{-2k+1\over 3},4g^2)Z_K(\tau^-_k
G_-^{-2k+1\over 3},4g^2)
\Big|_{a_\pm=A_\pm,\tau^{\Sigma\pm}_0=0,G_\pm=1-{3\over
2}\tau^{\Sigma\pm}_1} \ee

This formula was first suggested by I.Kostov in the second paper of
\cite{Kostov2}.\footnote{It seems to us that in formula (7.24) of
this paper instead of ${1\over rr'}$ one should write ${2\over
(2r)!!}{2\over (2r')!!}$ .} Here we provided its proof and
applications. In particular, we demonstrate how to use this formula
to obtain effectively explicit formulas for the free energy in the
example of the genus 2 free energy in Appendix I.


\subsection{Comments}

It remains to prove that the r.h.s. of (\ref{ffK}) does not depend
on $a_+$ and $a_-$ so that they can be put equal to any functions of
times $A_{\pm}$ (after taking all the time derivatives). For
simplicity, we consider only dependence on $a_-$: \be\label{161}
\frac{\p}{\p a_-}e^{U_{G}}e^{\hat U_{K}}Z_K(\tau^-)Z_K(\tau^+)
=e^{U_{G}}\frac{\p U_{G}}{\p a_-}e^{\hat
U_{K}}Z_K(\tau^-)Z_K(\tau^+)+\\+ e^{U_{G}}e^{\hat
U_{K}}\left(\frac{\p \hat U_{K}}{\p a_-}+
\sum_{k=0}^\infty\frac{\p\tau^+_k}{\p a_-}\frac{\p}{\p\tau^+_k}+\sum_{k=0}^\infty\frac{\p\tau^-_k}{\p a_-}\frac{\p}{\p\tau^-_k}\right)Z_K(\tau^-)Z_K(\tau^+)=\nn\\
=e^{U_{G}}e^{\hat U_{K}}\left(\p_-U_{G}+[\p_-U_{G},\hat
U_{K}]+\frac{1}{2}[[\p_-U_{G}, \hat U_{K}],\hat
U_{K}]+\right.\\+\left.\p_- \hat U_{K}+
\sum_{k=0}^\infty\frac{\p\tau^+_k}{\p
a_-}\frac{\p}{\p\tau^+_k}+\sum_{k=0}^\infty\frac{\p\tau^-_k}{\p
a_-}\frac{\p}{\p\tau^-_k}\right)Z_K(\tau^-)Z_K(\tau^+) \ee The
expression in the brackets in the last two lines is a second order
differential operator. We prove that it is equal to
$-L_{-1}^K(\tau^-)$: \be
\p_-\rho^{(0|2)}(z_1,z_2)=-\frac{d}{8y(z_1)y(z_2)(z_1-a_-)(z_2-a_-)}
\ee \be \p_-U_{G}=-\frac{1}{16g^2}\left(\oint_\infty\frac{v
dz}{y(z)(z-a_-)}\right)^2+ \frac{S}{2g^2}\oint_\infty\frac{ v
dz}{y(z)(z-a_-)}+\p_-f=-\frac{(\tau_0^-)^2}{64g^2}+\frac{S^2}{g^2d}+\p_-f
\label{Uin} \ee
\be \hat
\Omega_\pm=\sum_{k=0}^\infty\frac{1}{(2k+1)x_\pm^{2k+1}}\frac{\p}{\p\tau_k^{\pm}}
\ee \be \hat \Psi=\oint_{a_+}\frac{\hat
\Omega_+dz}{y(z)(z-a_-)}+\oint_{a_-}\frac{\hat
\Omega_-dz}{y(z)(z-a_-)} =-\frac{4}{\sqrt{d}}
\sum_{k=0}^\infty\left(\alpha_k^-\frac{\p}{\p\tau_k^+}+\beta_k^-\frac{\p}{\p\tau_k^-}\right)
\ee \be \hat \Psi
=-d^{-\frac{3}{2}}\frac{\p}{\p\tau_0^-}+2d^{-\frac{3}{2}}\frac{\p}{\p\tau_0^+}+\dots;\
\ \ \ \ \ \ \ \ \ \ \ \ \ \ \ [\hat \Psi,\tau_0^-]=-d^{-\frac{3}{2}}
\ee \be [\hat U_{K},\tau_0^-]=8g^2\sqrt{d}\hat\Psi \ee \be
f_k^{\pm}=\frac{1}{\sqrt{d}}\left(\oint_{a_{-}}+\oint_{a_{+}}\right)\frac{(\pm
1)^{k+1}y(z)d v}{(z-a_{\pm})^{k+1}}
\ee \be
\tau_\pm^k=\frac{8}{2k+1}\sum_{m=1}^{k+1}\frac{\Gamma(2m-1)(-1)^{m+1}
f_{k-m+1}^\pm}{\Gamma(m)^2 4^m d^{m-1}} \ee \be
\p_-f_k^-=\frac{1}{2d}f_k^--(k+\frac{1}{2})f_{k+1}^- \ee \be
\p_-\tau_k^-=-(k+\frac{3}{2})\tau_{k+1}^-+\frac{2}{2k+1}\frac{\Gamma(2k+3)(-1)^k}{\Gamma(k+2)^2
4^{k+2}d^{k+1}}\tau_0^-=-(k+\frac{3}{2})\tau_{k+1}^-+\beta_k^-\tau_0^-
\ee \be
\p_-\tau_k^+=-\frac{\Gamma(2k+1)(-1)^k}{\Gamma(k+1)^24^kd^{k+1}}\frac{\tau_0^-}{2}=\alpha_k^-\tau_0^-
\ee \be \sum_{k=0}^\infty\frac{\p\tau^+_k}{\p
a_-}\frac{\p}{\p\tau^+_k}+\sum_{k=0}^\infty\frac{\p\tau^-_k}{\p
a_-}\frac{\p}{\p\tau^-_k}=-\sum_{k=0}^\infty\left(k+\frac{3}{2}\right)\tau_{k+1}^-\frac{\p}{\p\tau_k^-}
-\frac{\sqrt{d}}{4}\tau_0^-\hat\Psi \label{ptau} \ee \be
\left[\p_-,\hat\Omega^+(x)\right]=0;\ \ \ \ \ \ \ \ \ \ \ \ \ \
\left[\p_-,\hat\Omega^-(x)\right]=-\frac{\p}{\p x^2}\hat
\Omega^-(x)+\frac{1}{x}\frac{\sqrt{d}}{4}\hat\Psi \ee \be
\p_-\hat U_{K}=8g^2\p_-\left(4\oint_{a_+}\oint_{a_+}\rho^{(0|2)}(a_++x_1^2,a_++x_2^2)x_1x_2dx_1dx_2\hat\Omega_+(x_1)\hat\Omega_+(x_2)\right.\nn\\
\left.-8\oint_{a_+}\oint_{a_-}\rho^{(0|2)}(a_++x_1^2,a_--x_2^2)x_1x_2dx_1dx_2\hat\Omega_+(x_1)
\hat\Omega_-(x_2)+\right.\nn\\
\left.+4\oint_{a_-}\oint_{a_-}\rho^{(0|2)}(a_--x_1^2,a_--x_2^2)x_1x_2dx_1dx_2
\hat\Omega_-(x_1)\hat\Omega_-(x_2)\right) \ee \be
\p_-\oint_{a_+}\oint_{a_+}\rho^{(0|2)}(a_++x_1^2,a_++x_2^2)x_1x_2dx_1dx_2\hat\Omega_+(x_1)
\hat\Omega_+(x_2)=\nn\\
=\oint_{a_+}\oint_{a_+}\left(\p_-\rho^{(0|2)}(a_++x_1^2,a_++x_2^2)\right)x_1x_2dx_1dx_2\hat\Omega_+(x_1)
\hat\Omega_+(x_2)
=-\frac{d}{8}\left(\oint_{a_+}\frac{\hat\Omega^+(x)xdx}{(d+x^2)y(a_++x^2)}\right)^2
\ee \be
\p_-\oint_{a_+}\oint_{a_-}\rho^{(0|2)}(a_++x_1^2,a_--x_2^2)x_1x_2dx_1dx_2\hat\Omega_+(x_1)
\hat\Omega_-(x_2)=\nn\\
=\oint_{a_+}\oint_{a_-}\left(\left(\p_--\frac{\p}{\p
x_2^2}\right)\rho^{(0|2)}(a_++x_1^2,a_--x_2^2)\right)x_1x_2dx_1dx_2\hat\Omega_+(x_1)
\hat\Omega_-(x_2)+\nn\\
+\oint_{a_+}\oint_{a_-}\rho^{(0|2)}(a_++x_1^2,a_--x_2^2)x_1x_2dx_1dx_2\hat\Omega_+(x_1)
\p_-\hat\Omega_-(x_2)=\nn\\
=\frac{d}{8}\left(\oint_{a_+}\frac{\hat\Omega^+(x)xdx}{(d+x^2)y(a_++x^2)}\right)\left(\oint_{a_-}\frac{\hat\Omega^-(x)dx}{xy(a_--x^2)}\right)-\frac{d}{16}\oint_{a_+}\frac{\hat\Omega^+(x)xdx}{(d+x^2)y(a_++x^2)}\hat\Psi
\ee \be
\p_-\oint_{a_-}\oint_{a_-}\rho^{(0|2)}(a_--x_1^2,a_--x_2^2)x_1x_2dx_1dx_2
\hat\Omega_-(x_1)\hat\Omega_-(x_2)=\\=-\frac{d}{8}\left(\oint_{a_-}\frac{\hat\Omega^-(x)dx}{xy(a_--x^2)}
\right)^2+\frac{d}{8}\oint_{a_-}\frac{\hat\Omega^-(x)dx}{xy(a_--x^2)}\hat\Psi
\ee \be \p_-\hat U_{K}=8g^2\left(-\frac{d}{2}\left(
\oint_{a_+}\frac{\hat\Omega^+(x)xdx}{(d+x^2)y(a_++x^2)}+\oint_{a_-}\frac{\hat\Omega^-(x)dx}{xy(a_--x^2)}\right)^2+\right.\nn\\
+\left.\frac{d}{2}\left(
\oint_{a_+}\frac{\hat\Omega^+(x)xdx}{(d+x^2)y(a_++x^2)}+\oint_{a_-}\frac{\hat\Omega^-(x)dx}{xy(a_--x^2)}\right)\hat\Psi\right)=g^2d\hat\Psi^2
\label{Uout} \ee \be [\p_-U_{G},\hat
U_{K}]=-\frac{1}{8d}+\frac{\sqrt{d}}{4}\tau_0^-\hat\Psi \label{com}
\ee \be \frac{1}{2}[[\p_-U_{G},\hat U_{K}],\hat
U_{K}]=-g^2d\hat\Psi^2 \label{com2} \ee Summing up expressions
(\ref{Uin}), (\ref{ptau}), (\ref{Uout}), (\ref{com}) and
(\ref{com2}), one finally gets that the r.h.s. of (\ref{161}) is
$-e^{U_G}e^{\hat U_K}L_{-1}^K(\tau^-)Z_K(\tau^-)Z_K(\tau^+)$ plus an
additional piece that is removed by the proper choice of an
arbitrary multiplicative factor depending only on $S$ and $a_{\pm}$:
\be U_{G}=\dots-\frac{S^2}{g^2}\log{d}-\frac{1}{8}\log{d} \ee

\newpage


\section{Decomposition formula for complex matrix model}
\setcounter{equation}{0}

To
illustrate our generic method in a slightly different situation, in
this section we construct a decomposition formula for the ${\bf
Z}(T)$ projection of the global partition function and consider
current on the sphere {\it with 6 singular points}: two infinities,
two ramification points and two zeros. This decomposition formula
involves the complex matrix model, the Kontsevich one and also
$\tilde Z({\cal T})$: $Z_C(\gotht)\longrightarrow Z_K(\tau)\tilde
Z_K({\cal T})$.

\subsection{Summary}

The global current in this case is \be \hat {\mathbf J}={\hat {\bf
J}^o (z)+\hat {\bf J}^o (-z)\over 2}= \hat{\mathbf J}^o(z|T,0) \ee
\be
\begin{scriptsize}\end{scriptsize}f_{\hat {\mathbf
J}}(z,z'|g^2)=g^2\frac{(y^2y'^2+2S(y^2+y'^2))dzdz'}{(y^2-y'^2)^2yy'}
\ee This global current is equivalent to the complex model current
in the vicinity of infinities and to the Kontsevich current in the
vicinity of ramification points and zeros. The proof is similar to
s.5.3):

\paragraph{1) $\infty_{\pm}$} The equivalence of currents follows from the simple relations
 \be f_{\hat {\mathbf
J}}(z,z'|g^2)\frac{(z^2-z'^2)^2}{dzdz'}\longrightarrow g^2z^2 \ee
\be \frac{\p}{\p z}\left(f_{\hat {\mathbf
J}}(z,z'|g^2)\frac{(z^2-z'^2)^2}{dz dz'}\right)\longrightarrow g^2z
\ee with the times being \be\label{gotht}
\gotht_k=\frac{2}{k}\oint_\infty z^{-2k}\hat{\mathbf J} \ee

\paragraph{2) $\pm 2\sqrt{S}$}

We use the local coordinate $x$ such as \be
z=2\sqrt{S}\left(1+\frac{x^2}{2(4S)^\frac{2}{3}}\right) \ee It is
convenient, because \be y(z)dz=(x^2+O(x^3))dx \ee Similarly to
s.5.3, \be f_{\hat {\mathbf
J}}(z,z'|g^2)\frac{(x^2-x'^2)^2}{dxdx'}\longrightarrow g^2x^2 \ee

\be \frac{\p}{\p x}\left(f_{\hat {\mathbf
J}}(z,z'|g^2)\frac{(x^2-x'^2)^2}{dxdx'}\right)\longrightarrow g^2x
\ee and the times are \be\label{tauC}
\tau_k=\frac{4}{2k+1}\oint_{2\sqrt{S}}x^{-2k-1}\hat {\mathbf J} \ee

\paragraph{3) $0_\pm$}

We use $z$ as a local coordinate \be f_{\hat {\mathbf
J}}(z,z'|g^2)\frac{(z^2-z'^2)^2}{dzdz'}\longrightarrow g^2z^2 \ee
\be \frac{\p}{\p z}\left(f_{\hat {\mathbf
J}}(z,z'|g^2)\frac{(z^2-z'^2)^2}{dzdz'}\right)\longrightarrow g^2z
\ee the times being \be \label{TauC}{\cal
T}_k=\frac{4}{2k+1}\oint_0z^{-2k-1}\hat {\mathbf J} \ee With the
shift \be \Delta \hat {\mathbf J}(z)=-\frac{y(z)dz}{2} \ee there is
an equivalence \be \hat {\mathbf J}(z|g^2)\stackrel{z\rightarrow
\infty}{\sim} \hat
J_C(z|2g^2)\nn\\
\hat {\mathbf J}(z|g^2)\stackrel{z\rightarrow \pm2\sqrt{S}}{\sim} \hat J_K(c|2g^2)\nn\\
\hat {\mathbf J}(z|g^2)\stackrel{z\rightarrow 0_\pm}{\sim}  \hat
J_K(z|2g^2) \ee in the sense of (\ref{jeq}) such that \be
\alpha_\infty =\alpha_a=\alpha_0=1\\
\beta_\infty=\beta_a=\beta_0=2 \ee

According to (\ref{Uinf}), the conjugation operator\footnote{In
order to differ from the case of four singular points, we denote
conjugation operators of this section through $V$ instead of $U$.}
is \be
V_\infty(2g^2)=V_C=\frac{2}{g^4}\oint_\infty\oint_\infty\left(f_{\hat
{\mathbf J}}(z,z'|g^2)-f_C(z,z'|2g^2)\right)\hat
\Omega^C(z|2g^2)\hat \Omega^C(z'|2g^2)+\\+ \oint_\infty
k^\infty(z|2g^2)\hat \Omega^C(z|2g^2) \ee with\be f_{\hat {\mathbf
J}}(z,z'|g^2)-f_C(z,z'|2g^2)=2g^2\rho^{(0|2)}_C(z,z') \ee \be
k^\infty(z|2g^2)=\frac{(z-y(z))dz}{g^2} \ee and the two-point
resolvent $\rho^{(0|2)}_C(z,z')$ here is explicitly given by
(\ref{rho2C}). Since \be \hat
\Omega^C(z|2g^2)=\frac{v_C(z)}{4}+\dots \ee the conjugation operator
at infinity is\be\label{VC}
V_\infty(2g^2)=V_C=\frac{1}{2!2g^2}\oint_\infty\oint_\infty\rho^{(0|2)}_C(z,z')v_C(z)v_C(z')
+\frac{1}{2g^2}\oint_\infty\rho^{(0|1)}_C(z)v_C(z) \ee  with
one-point resolvent $\rho^{(0|1)}_C(z)$ given at (\ref{rho1C}). The
second conjugation operator consist of three parts \be
\label{Vram1}\hat V_{ram}=\hat V_{aa}+\hat V_{a0}+\hat V_{00} \ee
and \be \label{Vram2}\hat
V_{aa}=\frac{2}{g^4}\oint_{2\sqrt{S}}\oint_{2\sqrt{S}}\left(f_{\hat
{\mathbf J}}(z,z'|g^2)-f_K(x,x'|2g^2)\right)\hat
\Omega^K(x|2g^2)\hat \Omega^K(x|2g^2)+
\oint_{2\sqrt{S}}k^a(x|2g^2)\hat \Omega^K(x|2g^2)\\
\hat V_{a0}=\frac{2}{g^4}\oint_{2\sqrt{S}}\oint_{0}f_{\hat {\mathbf
J}}(z,z'|g^2)\hat
\Omega^K(x|2g^2)\hat \Omega^K(z|2g^2)\\
\hat V_{00}=\frac{2}{g^4}\oint_{0}\oint_{0}\left(f_{\hat {\mathbf
J}}(z,z'|g^2)-f_K(z,z'|2g^2)\right)\hat \Omega^K(z|2g^2)\hat
\Omega^K(z'|2g^2)+
\oint_{0}k^0(z|2g^2)\hat \Omega^K(z|2g^2)\\
\ee
where
\be k^a(x|2g^2)=\frac{x^2dx-ydz}{g^2}\\
 k^0(z|2g^2)=\frac{zdz-ydz}{g^2}
\ee Therefore, \be e^{-V_\infty(2g^2)} Z_C(\gotht|2g^2)=e^{\hat
V_{ram}(2g^2)}Z_K(\tau|2g^2)\tilde Z_K({\mathtt t}|2g^2) \ee After
the change \be 2g^2\to g^2 \ee one finally gets \be\label{dfc}
e^{-V_\infty(g^2)} Z_C({\gotht}|g^2)=e^{\hat
V_{ram}(g^2)}Z_K(\tau|g^2)\tilde Z_K(\mathtt t|g^2) \ee where one
should put $ c=2\sqrt{-4S} $ in the definition of $\tilde Z_K$.










\subsection{Comments on moment variables for $Z_C$}

Consider the coordinate transformation \be \tilde z= a\sqrt{ z^2-u}
\ee and look for an identity similar to (\ref{Zt-Ztt}) \be
\label{ZC}Z_C(\gotht_{2k})=e^{U_C}Z_C(\tilde{\gotht}_{k}) \ee This
time we work with the current (\ref{C}) \be \hat
J_C(z)=\frac{1}{2}\sum_{k=0}k{\gotht}_{k}
z^{2k-1}dz+\frac{S}{z}+g^2\sum_{k=1}\frac{1}{z^{2k+1}}\frac{\p}{\p{\gotht}_k}
\ee which invariance (up to conjugation) gives rise to the change of
time variables \be k\tilde {\gotht}_{k}=\oint_\infty \frac{d
v(z)}{\tilde z^{2k}}=\frac{1}{a^{2k}} \sum_{n=k} {\gotht}_{2}
\frac{n!u^{n-k}}{(k-1)!(n-k)!} \ee and the corresponding conjugation
operator is \be e^{U_C},\ \ \ \ U_C=S\sum u^{k}
\gotht_{k}-S^2\log{a} \ee Now one realizes that
$Z_C(\tilde{\gotht})$ depends on $u$ and, therefore, $Z_C$ at the
l.h.s. and the r.h.s. in (\ref{ZC}) can not be the same function.
Indeed, \be \frac{\p Z_C}{\p u}=\frac{\p U_C}{\p
u}+\sum_{k=1}^{\infty}\frac{\p \tilde {\gotht}_{k}}{\p u}\frac{\p
Z_C}{\p \tilde {\gotht}_{k}}= a^2S\tilde
{\gotht}_1+\frac{a^2}{2}\sum_{k=1}^{\infty}(k+1)\tilde
{\gotht}_{k+1}\frac{\p Z_C}{\p \tilde {\gotht}_{k}}\sim L_{-1}Z_C
\ee which is non-zero for the complex model. This means that, for
the complex matrix model, there is only rescaling $z\to az$, which
is generated by \be \exp(\log{a}L_{0}) \ee

\newpage


\app{Explicit expansions of partition functions}

In this Appendix we manifestly list first terms of expansions for
various free energies (partition function) discussed in the paper.
Throughout this and the next Appendices we use {\it only} the time
variables with superscript $\Sigma$, which we omit for the sake of
simplicity. We hope this could not confuse the reader. The most
effective way to get the expansions is to use the corresponding
Virasoro constraints, which lead to a chain of relations that can be
solved recurrently \cite{amm1}.

\sapp{$\bullet$ $Z_G(t)$}

Expansion for the Gaussian free energy can be extracted from
\cite{amm1} for the Gaussian potential ($T_k=\delta_{k,2}/2$) as a
series in $t_k$(we restrict ourselves with values of $k$ ranging
from 0 to 10): \be\label{GF0}F^{(0)}=S^2\log S-{3\over
4}S^2+\left(St_0+S^2t_2+2S^3t_4+5S^4t_6+14S^5t_8
\right)+\\+
{t_1^2\over
2}S+\left(t_2^2+3t_1t_3\right)S^2+\left(6t_3^2+10t_1t_5+8t_2t_4\right)S^3
+\left(18t_4^2+35t_1t_7+30t_2t_6+45t_3t_5\right)S^4+\\
+\left(90t_5^2+126t_1t_9+112t_2t_8+168t_3t_7+144t_4t_6\right)S^5+
O\left(S^6\right)+O\left(t^3\right)\ee
\be\label{GF1}F^{(1)}=-{1\over 12}\log S
+\left(t_4S+10t_6S^2+70t_8S^3
\right)+\\+\left({3\over
2}t_3^2+5t_1t_5+4t_2t_4\right)S+\left(30t_4^2+70t_1t_7+60t_2t_6+60t_3t_5
\right)S^2+\\+\left(300t_5^2+630t_1t_9+560t_2t_8+630t_3t_7+600t_4t_6\right)S^3
+O \left( {S}^{4} \right) +O(t^3) \ee \be\label{GF2}
{F}^{(2)}=-\frac{1}{240S^2}+\left(21St_8+483S^2t_{10}\right)+\nn\\
 \left( 178\,t_{{8}}t_{{2}}+156\,t_{{4}}t_{{6}}+{\frac {165}{2}}\,{t_
{{5}}}^{2}+189\,t_{{1}}t_{{9}}+147\,t_{{3}}t_{{7}} \right) {S}^{1}\nn\\
+ \left(
4760\,t_{{8}}t_{{4}}+2385\,{t_{{6}}}^{2}+4725\,t_{{3}}t_{{9}}+
4830\,t_{{2}}t_{{10}}+4795\,t_{{5}}t_{{7}} \right) {S}^{2}\nn\\
+ \left(
80640\,t_{{10}}t_{{4}}+80640\,t_{{5}}t_{{9}}+80640\,t_{{8}}t_{{6}}+
40670\,{t_{{7}}}^{2} \right) {S}^{3}+O \left( {S}^{4} \right)
+O(t^3) \ee The terms not depending on times in these expressions
come from the normalization volume of the unitary group in
(\ref{Gint}) that can be calculated using the Stirling formula in
(\ref{volume}), see \cite{amm1}.\footnote{(\ref{volume}) is also in
charge of some correcting constant terms that are not consistent
with the genus expansion \cite{amm1}.}

\sapp{$\bullet$ $Z_K(t)$}

Explicit expansion of the function $Z_K(t)$ can be obtained in the
simplest way also from the Virasoro constraints, (\ref{K}) by
iterations \cite{amm1}. Inserting into these constraints $Z_K(t)$ as
a series in $g^2$,\footnote{Note that this expansion differs from
(\ref{loopexK}) by a trivial factor.} \be \log
Z_K(\tau)=\sum_{h=0}^\infty {\cal F}_K^{(p)}(\tau)(8g^2)^{p-1} \ee
One obtains \be
{\cal F}_K^{(0)}(\tau)=1/6\,{\tau_{{0}}}^{3}+\\
+1/4\,{\tau_{{0}}}^{3}\tau_{{1}}+{\frac
{5}{32}}\,{\tau_{{0}}}^{4}\tau_
{{2}}+\\
3/8\,{\tau_{{0}}}^{3}{\tau_{{1}}}^{2}+{\frac
{45}{64}}\,{\tau_{{0}}}^{ 4}\tau_{{1}}\tau_{{2}}+{\frac
{7}{64}}\,{\tau_{{0}}}^{5}\tau_{{3}}+{
\frac {45}{128}}\,{\tau_{{0}}}^{5}{\tau_{{2}}}^{2}+\\
{\frac {9}{16}}\,{\tau_{{0}}}^{3}{\tau_{{1}}}^{3}+{\frac
{135}{64}}\,{ \tau_{{0}}}^{4}{\tau_{{1}}}^{2}\tau_{{2}}+{\frac
{21}{32}}\,{\tau_{{0} }}^{5}\tau_{{1}}\tau_{{3}}+{\frac
{675}{256}}\,{\tau_{{0}}}^{5}\tau_{{ 1}}{\tau_{{2}}}^{2}+{\frac
{21}{256}}\,{\tau_{{0}}}^{6}\tau_{{4}}+{ \frac
{175}{256}}\,{\tau_{{0}}}^{6}\tau_{{2}}\tau_{{3}}+{\frac {1125}{
1024}}\,{\tau_{{0}}}^{6}{\tau_{{2}}}^{3}+\\
{\frac {27}{32}}\,{\tau_{{0}}}^{3}{\tau_{{1}}}^{4}+{\frac
{675}{128}} \,{\tau_{{0}}}^{4}{\tau_{{1}}}^{3}\tau_{{2}}+{\frac
{315}{128}}\,{\tau _{{0}}}^{5}{\tau_{{1}}}^{2}\tau_{{3}}+{\frac
{6075}{512}}\,{\tau_{{0}}
}^{5}{\tau_{{1}}}^{2}{\tau_{{2}}}^{2}+{\frac
{315}{512}}\,{\tau_{{0}}} ^{6}\tau_{{1}}\tau_{{4}}+{\frac
{1575}{256}}\,{\tau_{{0}}}^{6}\tau_{{1
}}\tau_{{2}}\tau_{{3}}+\\+{\frac
{23625}{2048}}\,{\tau_{{0}}}^{6}\tau_{{1 }}{\tau_{{2}}}^{3}+{\frac
{33}{512}}\,{\tau_{{0}}}^{7}\tau_{{5}}+{ \frac
{175}{512}}\,{\tau_{{0}}}^{7}{\tau_{{3}}}^{2}+{\frac {675}{1024}
}\,{\tau_{{0}}}^{7}\tau_{{2}}\tau_{{4}}+{\frac
{7875}{2048}}\,{\tau_{{0 }}}^{7}{\tau_{{2}}}^{2}\tau_{{3}}+{\frac
{16875}{4096}}\,{\tau_{{0}}}^
{7}{\tau_{{2}}}^{4}+\\
{\frac {81}{64}}\,{\tau_{{0}}}^{3}{\tau_{{1}}}^{5}+{\frac
{6075}{512}} \,{\tau_{{0}}}^{4}{\tau_{{1}}}^{4}\tau_{{2}}+{\frac
{945}{128}}\,{\tau _{{0}}}^{5}{\tau_{{1}}}^{3}\tau_{{3}}+{\frac
{42525}{1024}}\,{\tau_{{0
}}}^{5}{\tau_{{1}}}^{3}{\tau_{{2}}}^{2}+{\frac
{2835}{1024}}\,{\tau_{{0 }}}^{6}{\tau_{{1}}}^{2}\tau_{{4}}+{\frac
{33075}{1024}}\,{\tau_{{0}}}^
{6}{\tau_{{1}}}^{2}\tau_{{2}}\tau_{{3}}+\\+{\frac
{70875}{1024}}\,{\tau_{
{0}}}^{6}{\tau_{{1}}}^{2}{\tau_{{2}}}^{3}+{\frac
{297}{512}}\,{\tau_{{0 }}}^{7}\tau_{{1}}\tau_{{5}}+{\frac
{3675}{1024}}\,{\tau_{{0}}}^{7}\tau _{{1}}{\tau_{{3}}}^{2}+{\frac
{14175}{2048}}\,{\tau_{{0}}}^{7}\tau_{{1
}}\tau_{{2}}\tau_{{4}}+{\frac
{23625}{512}}\,{\tau_{{0}}}^{7}\tau_{{1}
}{\tau_{{2}}}^{2}\tau_{{3}}+\\+{\frac
{455625}{8192}}\,{\tau_{{0}}}^{7} \tau_{{1}}{\tau_{{2}}}^{4}+{\frac
{429}{8192}}\,{\tau_{{0}}}^{8}\tau_{ {6}}+{\frac
{11025}{16384}}\,{\tau_{{0}}}^{8}\tau_{{3}}\tau_{{4}}+{ \frac
{10395}{16384}}\,{\tau_{{0}}}^{8}\tau_{{2}}\tau_{{5}}+{\frac {
18375}{4096}}\,{\tau_{{0}}}^{8}\tau_{{2}}{\tau_{{3}}}^{2}+\\+{\frac
{ 70875}{16384}}\,{\tau_{{0}}}^{8}{\tau_{{2}}}^{2}\tau_{{4}}+{\frac
{ 354375}{16384}}\,{\tau_{{0}}}^{8}{\tau_{{2}}}^{3}\tau_{{3}}+{\frac
{ 2278125}{131072}}\,{\tau_{{0}}}^{8}{\tau_{{2}}}^{5}+\dots \ee \be
{\cal F}_K^{(1)}(\tau)=1/16\,\tau_{{1}}+{\frac {5}{32}}\,\tau_{{0}}\tau_{{2}}+\\
{\frac {3}{64}}\,{\tau_{{1}}}^{2}+{\frac
{15}{32}}\,\tau_{{0}}\tau_{{1 }}\tau_{{2}}+{\frac
{35}{128}}\,{\tau_{{0}}}^{2}\tau_{{3}}+{\frac {75}
{128}}\,{\tau_{{0}}}^{2}{\tau_{{2}}}^{2}+\\
{\frac {3}{64}}\,{\tau_{{1}}}^{3}+{\frac
{135}{128}}\,\tau_{{0}}{\tau_ {{1}}}^{2}\tau_{{2}}+{\frac
{315}{256}}\,{\tau_{{0}}}^{2}\tau_{{1}} \tau_{{3}}+{\frac
{225}{64}}\,{\tau_{{0}}}^{2}\tau_{{1}}{\tau_{{2}}}^{ 2}+{\frac
{105}{256}}\,{\tau_{{0}}}^{3}\tau_{{4}}+{\frac {1225}{512}}
\,{\tau_{{0}}}^{3}\tau_{{2}}\tau_{{3}}+{\frac
{375}{128}}\,{\tau_{{0}}
}^{3}{\tau_{{2}}}^{3}+\\
{\frac {27}{512}}\,{\tau_{{1}}}^{4}+{\frac
{135}{64}}\,\tau_{{0}}{\tau _{{1}}}^{3}\tau_{{2}}+{\frac
{945}{256}}\,{\tau_{{0}}}^{2}{\tau_{{1}}} ^{2}\tau_{{3}}+{\frac
{3375}{256}}\,{\tau_{{0}}}^{2}{\tau_{{1}}}^{2}{
\tau_{{2}}}^{2}+{\frac
{315}{128}}\,{\tau_{{0}}}^{3}\tau_{{1}}\tau_{{4 }}+{\frac
{18375}{1024}}\,{\tau_{{0}}}^{3}\tau_{{1}}\tau_{{2}}\tau_{{3
}}+\\+{\frac
{3375}{128}}\,{\tau_{{0}}}^{3}\tau_{{1}}{\tau_{{2}}}^{3}+{ \frac
{1155}{2048}}\,{\tau_{{0}}}^{4}\tau_{{5}}+{\frac {8575}{4096}}\,
{\tau_{{0}}}^{4}{\tau_{{3}}}^{2}+{\frac
{17325}{4096}}\,{\tau_{{0}}}^{ 4}\tau_{{2}}\tau_{{4}}+{\frac
{154875}{8192}}\,{\tau_{{0}}}^{4}{\tau_{ {2}}}^{2}\tau_{{3}}+{\frac
{16875}{1024}}\,{\tau_{{0}}}^{4}{\tau_{{2}}
}^{4}+\\
{\frac {81}{1280}}\,{\tau_{{1}}}^{5}+{\frac
{2025}{512}}\,\tau_{{0}}{ \tau_{{1}}}^{4}\tau_{{2}}+{\frac
{4725}{512}}\,{\tau_{{0}}}^{2}{\tau_{ {1}}}^{3}\tau_{{3}}+{\frac
{10125}{256}}\,{\tau_{{0}}}^{2}{\tau_{{1}}}
^{3}{\tau_{{2}}}^{2}+{\frac
{4725}{512}}\,{\tau_{{0}}}^{3}{\tau_{{1}}} ^{2}\tau_{{4}}+{\frac
{165375}{2048}}\,{\tau_{{0}}}^{3}{\tau_{{1}}}^{2
}\tau_{{2}}\tau_{{3}}+\\
+{\frac {70875}{512}}\,{\tau_{{0}}}^{3}{\tau_{{1}
}}^{2}{\tau_{{2}}}^{3}+{\frac
{17325}{4096}}\,{\tau_{{0}}}^{4}\tau_{{1 }}\tau_{{5}}+{\frac
{77175}{4096}}\,{\tau_{{0}}}^{4}\tau_{{1}}{\tau_{{ 3}}}^{2}+{\frac
{155925}{4096}}\,{\tau_{{0}}}^{4}\tau_{{1}}\tau_{{2}}
\tau_{{4}}+{\frac
{3252375}{16384}}\,{\tau_{{0}}}^{4}\tau_{{1}}{\tau_{
{2}}}^{2}\tau_{{3}}+\\
+{\frac {50625}{256}}\,{\tau_{{0}}}^{4}\tau_{{1}}{
\tau_{{2}}}^{4}+{\frac {3003}{4096}}\,{\tau_{{0}}}^{5}\tau_{{6}}+{
\frac {55125}{8192}}\,{\tau_{{0}}}^{5}\tau_{{3}}\tau_{{4}}+{\frac {
3465}{512}}\,{\tau_{{0}}}^{5}\tau_{{2}}\tau_{{5}}+{\frac
{18375}{512}}
\,{\tau_{{0}}}^{5}\tau_{{2}}{\tau_{{3}}}^{2}+\\
+{\frac {590625}{16384}}\,{\tau_{{0}}}^{5}{\tau_{{2}}}^{2}\tau_{{4}}
+{\frac {4764375}{32768}}\,{
\tau_{{0}}}^{5}{\tau_{{2}}}^{3}\tau_{{3}}+{\frac
{50625}{512}}\,{\tau_ {{0}}}^{5}{\tau_{{2}}}^{5}+\dots \ee \be {\cal
F}_K^{(2)}(\tau)={\frac {105}{2048}}\,\tau_{{4}}+{\frac
{1015}{4096}}\,\tau_{{2}}\tau_{
{3}}+{\frac {525}{2048}}\,{\tau_{{2}}}^{3}+\\
+{\frac {945}{4096}}\,\tau_{{1}}\tau_{{4}}+{\frac
{3045}{2048}}\,\tau_{ {1}}\tau_{{2}}\tau_{{3}}+{\frac
{7875}{4096}}\,\tau_{{1}}{\tau_{{2}}}^ {3}+{\frac
{1155}{4096}}\,\tau_{{0}}\tau_{{5}}+{\frac {7105}{8192}}\,
\tau_{{0}}{\tau_{{3}}}^{2}+{\frac
{3465}{2048}}\,\tau_{{0}}\tau_{{2}} \tau_{{4}}+\\+{\frac
{13125}{2048}}\,\tau_{{0}}{\tau_{{2}}}^{2}\tau_{{3}}
+{\frac {39375}{8192}}\,\tau_{{0}}{\tau_{{2}}}^{4}+\\
+{\frac {2835}{4096}}\,{\tau_{{1}}}^{2}\tau_{{4}}+{\frac
{45675}{8192}} \,{\tau_{{1}}}^{2}\tau_{{2}}\tau_{{3}}+{\frac
{70875}{8192}}\,{\tau_{{ 1}}}^{2}{\tau_{{2}}}^{3}+{\frac
{3465}{2048}}\,\tau_{{0}}\tau_{{1}} \tau_{{5}}+{\frac
{106575}{16384}}\,\tau_{{0}}\tau_{{1}}{\tau_{{3}}}^{
2}+{\frac {51975}{4096}}\,\tau_{{0}}\tau_{{1}}\tau_{{2}}\tau_{{4}}+\\
+{ \frac
{118125}{2048}}\,\tau_{{0}}\tau_{{1}}{\tau_{{2}}}^{2}\tau_{{3}}+
{\frac {826875}{16384}}\,\tau_{{0}}\tau_{{1}}{\tau_{{2}}}^{4}+{\frac
{ 15015}{16384}}\,{\tau_{{0}}}^{2}\tau_{{6}}+{\frac
{112455}{16384}}\,{ \tau_{{0}}}^{2}\tau_{{3}}\tau_{{4}}+{\frac
{3465}{512}}\,{\tau_{{0}}}^
{2}\tau_{{2}}\tau_{{5}}+\\
+{\frac {2002875}{65536}}\,{\tau_{{0}}}^{2}\tau
_{{2}}{\tau_{{3}}}^{2}+{\frac
{496125}{16384}}\,{\tau_{{0}}}^{2}{\tau_ {{2}}}^{2}\tau_{{4}}+{\frac
{433125}{4096}}\,{\tau_{{0}}}^{2}{\tau_{{2 }}}^{3}\tau_{{3}}+{\frac
{4134375}{65536}}\,{\tau_{{0}}}^{2}{\tau_{{2}
}}^{5}+\\
{\frac {4673375}{131072}}\,{\tau_{{0}}}^{3}{\tau_{{3}}}^{3}+{\frac {
337365}{32768}}\,{\tau_{{0}}}^{3}{\tau_{{4}}}^{2}+{\frac
{14175}{8192} }\,{\tau_{{1}}}^{3}\tau_{{4}}+{\frac
{1299375}{1024}}\,{\tau_{{0}}}^{2
}\tau_{{1}}{\tau_{{2}}}^{3}\tau_{{3}}+{\frac
{1012095}{16384}}\,{\tau_
{{0}}}^{2}\tau_{{1}}\tau_{{3}}\tau_{{4}}+\dots \ee \be {\cal
F}_K^{(3)}(\tau)={\frac {25025}{131072}}\,\tau_{{7}}+{\frac
{191205}{262144}}\,{\tau_{{ 4}}}^{2}+{\frac
{193655}{131072}}\,\tau_{{3}}\tau_{{5}}+{\frac {
3570875}{1572864}}\,{\tau_{{3}}}^{3}+{\frac
{385385}{262144}}\,\tau_{{ 2}}\tau_{{6}}+{\frac
{1765575}{131072}}\,\tau_{{2}}\tau_{{3}}\tau_{{4} }+\\+{\frac
{883575}{131072}}\,{\tau_{{2}}}^{2}\tau_{{5}}+{\frac {
18834375}{524288}}\,{\tau_{{2}}}^{2}{\tau_{{3}}}^{2}+{\frac
{6260625}{ 262144}}\,{\tau_{{2}}}^{3}\tau_{{4}}+{\frac
{37996875}{524288}}\,{\tau _{{2}}}^{4}\tau_{{3}}+{\frac
{34453125}{1048576}}\,{\tau_{{2}}}^{6}+\dots \ee

\sapp{$\bullet$ $\tilde Z_K$}
$\tilde Z_K$ is given by definition as the solution to the Virasoro
constraints \be \hat{\tilde L}_{(n)}^K \tilde
Z_K=0~~~~n=0\dots\infty \ee \be \hat{\tilde
L}_{(n)}^K=\sum_{k>0}\left(k+\frac{1}{2}\right)\left({\cal
T}_k-c\delta_{k,0}\right) \frac{\p}{\p {\cal
T}_{k+n}}+g^2\sum_{a+b=n-1}\frac{\p^2}{\p {\cal T}_a\p {\cal
T}_b}+\frac{\delta_{n,0}}{16} \ee $\tilde Z_K$ possesses the very
simple genus expansion, that is, \be
\tilde {\cal F}^{(0)}=0\\
\tilde {\cal F}^{(1)}=-\frac{1}{8}\log(c-{\cal T}_0)\\
\tilde {\cal F}^{(2)}=-\frac{9{\cal T}_1}{32({\cal T}_0-c)^3}\\
\dots \ee

\sapp{$\bullet$ $Z_C(\gotht)$}

In order to get the expansion of the partition function of the
complex matrix model, one can again use the Virasoro constraints,
(\ref{C}). Similarly to the case Gaussian model \cite{amm1}, they
can be recast in the form of loop equations which being expanded in
the $g$-series lead to recurrent relations for the resolvents (the
details can be found in \cite{amm1})
\be\rho^{(k|m)}_C(z_1,...,z_m)\equiv \sum_{n_1,...,n_m} {1\over
z_1^{n_1+1}...z_m^{n_m+1}}{\partial^{n_1+...+n_m}
F_C^{(k)}\over\partial \gotht_{n_1}...\partial \gotht_{n_m}}\\
\log Z_C(\gotht|g^2)=\sum_{h=0}^\infty g^{2h-2}{F}_C^{(h)} \ee
which, for a generic polynomial potential $W(z)$ looks like \be
\frac{W_C'(z)\rho_{W_C}^{(p|m+1)}(z,z_1,\ldots,z_m)}{2} -
f_{W_C}^{(p|m+1)}(z|z_1,\ldots,z_m) = \nn \\ = \sum_q
\sum_{m_1+m_2=m}
\rho_{W_C}^{(q|m_1+1)}(z,z_{i_1},\ldots,z_{i_{m_1}})
\rho_{W_C}^{(p-q|m_2+1)}(z,z_{j_1},\ldots,z_{j_{m_2}}) +\\+
\sum_{i=1}^m \frac{\partial}{\partial z_i}
\frac{z\rho_{W_C}^{(p|m)}(z,z_1,\ldots,\check z_i,\ldots,z_m) -
z_1\rho_W^{(p|m)}(z_1,\ldots,z_m)}{2(z^2-z_i^2)} +
\hat\nabla(z)\rho_{W_C}^{(p-1|m+1)}(z,z_1,\ldots,z_m).
\label{recrel1} \ee \be
\rho^{(0|1)}_C(z)=\frac{\frac{W'(z)}{2}-y_C(z)}{2} \ee \be
y_C(z)^2=\frac{W'^2}{4}-4\check R(z) F^{(0)}_C \ee where $\check
R(z)$ is the check operator introduced in \cite{amm23}, etc. For the
Gaussian complex model with $W=z^2$ \be\label{rho1C}
\rho_C^{(0|1)}(z)=\frac{z-\sqrt{z^2-4S}}{2} \ee \be
y(z_1)\rho_C^{(0|2)}(z_1,z_2) =
\partial_{z_2}\frac{z_1\rho_C^{(0|1)}(z_1)-z_2\rho_C^{(0|1)}(z_2)}
{2(z_1^2-z_2^2)} \ee \be
\rho^{(0|2)}_C(z_1,z_2)=\frac{1}{2(z_1^2-z_2^2)^2}\left(
\frac{z_1^2z_2^2-2S(z_1^2+z_2^2)}{y_1y_2}-z_1z_2\right)
\label{rho2C}\ee (see, for example \cite{Ambjorn:1990ji}) \be
y(z_1)\rho_C^{(1|1)}(z_1) = \rho_C^{(0|2)}(z_1,z_1) \ee \be
\rho^{(1|1)}_C(z)=\frac{S^2}{z^2y(z)^5} \ee \be
y(z_1)\rho_C^{(0|3)}(z_1,z_2,z_3) =
2\rho_C^{(0|2)}(z_1,z_2)\rho_C^{(0|2)}(z_1,z_3) + \nn \\ +
\partial_{z_2}\frac{z_1\rho_C^{(0|2)}(z_1,z_3)-z_2\rho_C^{(0|2)}(z_2,z_3)}
{2(z_1^2-z_2^2)} +
\partial_{z_3}\frac{z_1\rho_C^{(0|2)}(z_1,z_2)-z_3\rho_C^{(0|2)}(z_2,z_3)}
{2(z_1^2-z_3^2)} \ee \be \rho^{(0|3)}_C(z_1,z_2,z_3)=2\,{\frac
{{S}^{2}}{ \left( {{ z_1}}^{2}-4\,S \right) ^{3/2}
 \left( {{ z_3}}^{2}-4\,S \right) ^{3/2} \left( {{ z_2}}^{2}-4
\,S \right) ^{3/2}}} \ee \be y(z_1)\rho_C^{(1|2)}(z_1,z_2) =
2\rho_C^{(0|2)}(z_1,z_2)\rho_C^{(1|1)}(z_1) +
\rho_C^{(0|3)}(z_1,z_1,z_2) +
\partial_{z_2}\frac{z_1\rho_C^{(1|1)}(z_1)-z_2\rho_C^{(1|1)}(z_2)}
{2(z_1^2-z_2^2)} \ee \be
\rho_C^{(1|2)}(z_1,z_2)=\frac{S^2}{z_1^2z_2^2y_1^7y_2^7}\left(3z_2^2z_1^6+3z_2^6z_1^2
+2z_2^4z_1^4\right.\nn\\
\left.-2(z_1^6+17z_1^4z_2^2+17z_1^2z_2^4+z_2^6)S
+8(3z_1^4+19z_1^2z_2^2+3z_2^4)S^2-96(z_1^2+z_2^2)S^3+128S^4 \right)
\ee \be y(z_1)\rho_C^{(2|1)}(z_1) =
\left(\rho_C^{(1|1)}(z_1)\right)^2 + \rho_C^{(1|2)}(z_1,z_1) \ee \be
\rho_C^{(2|1)}(z)=\frac{(9S^2-8z^2S+8z^4)S^2}{z^4y(z)^{11}} \ee \be
{F}_C^{(0)}=\gotht_0S+{S}^{2}\gotht_{{1}}+2\,{S}^{3}\gotht_{{2}}+5\,{S}^{4}\gotht_{{3}}+14\,{S}^{5}\gotht_{{4}}+
42\,{S}^{6}\gotht_{{5}}+132\,{S}^{7}\gotht_{{6}}+\dots+\nn\\
\frac{1}{2}\,{\gotht_{{1}}}^{2}{S}^{2}+4\,\gotht_{{1}}\gotht_{{2}}{S}^{3}+
\left( 15\,\gotht_{{3} }\gotht_{{1}}+9\,{\gotht_{{2}}}^{2} \right)
{S}^{4}+ \left( 56\,\gotht_{{4}}\gotht_{{1}}+
72\,\gotht_{{3}}\gotht_{{2}} \right) {S}^{5}+ \left(
150\,{\gotht_{{3}}}^{2}+210\,\gotht_{
{5}}\gotht_{{1}}+280\,\gotht_{{4}}\gotht_{{2}} \right) {S}^{6}+\nn\\
+ \left( 1080\,\gotht_{{5}}\gotht
_{{2}}+1200\,\gotht_{{4}}\gotht_{{3}}+792\,\gotht_{{6}}\gotht_{{1}}
\right) {S}^{7}+
 \left( 4725\,\gotht_{{3}}\gotht_{{5}}+4158\,\gotht_{{2}}\gotht_{{6}}+2450\,{\gotht_{{4}}}^{2}
 \right) {S}^{8}+\nn\\
+ \left(
18480\,\gotht_{{3}}\gotht_{{6}}+19600\,\gotht_{{4}}\gotht_{{5}}
 \right) {S}^{9}+ \left( 39690\,{\gotht_{{5}}}^{2}+77616\,\gotht_{{4}}\gotht_{{6}}
 \right) {S}^{10}+317520\,\gotht_{{5}}\gotht_{{6}}{S}^{11}+640332\,{\gotht_{{6}}}^{2}
{S}^{12}+\dots+\nn\\
+\frac{1}{3}\,{\gotht_{{1}}}^{3}{S}^{2}+6\,\gotht_{{2}}{\gotht_{{1}}}^{2}{S}^{3}+
\left( 30\,
\gotht_{{3}}{\gotht_{{1}}}^{2}+36\,{\gotht_{{2}}}^{2}\gotht_{{1}}
\right) {S}^{4}+ \left(
140\,\gotht_{{4}}{\gotht_{{1}}}^{2}+360\,\gotht_{{3}}\gotht_{{2}}\gotht_{{1}}+72\,{\gotht_{{2}}}^{3}
 \right) {S}^{5}+\nn\\
+ \left(
900\,\gotht_{{1}}{\gotht_{{3}}}^{2}+630\,\gotht_{{5}}{\gotht_{{1}
}}^{2}+1080\,\gotht_{{3}}{\gotht_{{2}}}^{2}+1680\,\gotht_{{4}}\gotht_{{2}}\gotht_{{1}}
\right)
{S}^{6}+\nn\\
+ \left(
8400\,\gotht_{{4}}\gotht_{{3}}\gotht_{{1}}+5400\,\gotht_{{2}}{\gotht_{{3}}}^{2}
+7560\,\gotht_{{5}}\gotht_{{2}}\gotht_{{1}}+5040\,\gotht_{{4}}{\gotht_{{2}}}^{2}
\right) {S}^{7
}+\nn\\
+ \left(
22680\,\gotht_{{5}}{\gotht_{{2}}}^{2}+50400\,\gotht_{{4}}\gotht_{{2}}\gotht_{{3}}+
19600\,{\gotht_{{4}}}^{2}\gotht_{{1}}+9000\,{\gotht_{{3}}}^{3}+37800\,\gotht_{{5}}\gotht_{{3}}\gotht
_{{1}} \right) {S}^{8}+\nn\\
+ \left( 176400\,\gotht_{{4}}\gotht_{{5}}\gotht_{{1}}+126000\,\gotht
_{{4}}{\gotht_{{3}}}^{2}+226800\,\gotht_{{5}}\gotht_{{3}}\gotht_{{2}}+117600\,\gotht_{{2}}{\gotht_{{
4}}}^{2} \right) {S}^{9}+\nn\\
+ \left( 1058400\,\gotht_{{4}}\gotht_{{2}}\gotht_{{5}}+588000
\,{\gotht_{{4}}}^{2}\gotht_{{3}}+396900\,{\gotht_{{5}}}^{2}\gotht_{{1}}+567000\,\gotht_{{5}}{\gotht_
{{3}}}^{2} \right) {S}^{10}+\nn\\
+ \left( {\frac {2744000}{3}}\,{\gotht_{{4}}}^{3
}+2381400\,{\gotht_{{5}}}^{2}\gotht_{{2}}+5292000\,\gotht_{{4}}\gotht_{{5}}\gotht_{{3}}
 \right) {S}^{11}+ \left( 11907000\,\gotht_{{3}}{\gotht_{{5}}}^{2}+12348000\,{\gotht_
{{4}}}^{2}\gotht_{{5}} \right) {S}^{12}+\nn\\
+55566000\,\gotht_{{4}}{\gotht_{{5}}}^{2}{S}^{
13}+83349000\,{\gotht_{{5}}}^{3}{S}^{14}+\dots \ee \be
{F}_C^{(1)}=\gotht_{{3}}{S} ^{2}+10\,\gotht_{{4}}{S}^{3}+
70\,{S}^{4}\gotht_{{5}}+420\,{S}^{5}\gotht_{{6}}+2310\,{S}^{6}\gotht_{{7}}+12012\,{S}^{7}\gotht_{{8}}+
60060\,{S}^{8}\gotht_{{9}}+\dots+\nn\\
+\left( 3\,\gotht_{{1}}\gotht_{{3}}+{\gotht_{{2}}}^{2} \right)
{S}^{2}+ \left( 36\,\gotht
_{{2}}\gotht_{{3}}+40\,\gotht_{{1}}\gotht_{{4}} \right) {S}^{3}+
\left( 174\,{\gotht_{{3}}
}^{2}+350\,\gotht_{{1}}\gotht_{{5}}+360\,\gotht_{{2}}\gotht_{{4}} \right) {S}^{4}+\nn\\
+ \left(
2800\,\gotht_{{2}}\gotht_{{5}}+2520\,\gotht_{{1}}\gotht_{{6}}+2760\,\gotht_{{4}}\gotht_{{3}}
 \right) {S}^{5}+ \left( 18900\,\gotht_{{6}}\gotht_{{2}}+19110\,\gotht_{{5}}\gotht_{{3}}+
9500\,{\gotht_{{4}}}^{2}+16170\,\gotht_{{1}}\gotht_{{7}} \right) {S}^{6}+\nn\\
+ \left(
96096\,\gotht_{{8}}\gotht_{{1}}+120400\,\gotht_{{5}}\gotht_{{4}}+116424\,\gotht_{{7}}\gotht_{{2}}+
120456\,\gotht_{{6}}\gotht_{{3}} \right) {S}^{7}+\nn\\
+ \left( 357700\,{\gotht_{{5}}}^{2}+
709632\,\gotht_{{7}}\gotht_{{3}}+715680\,\gotht_{{6}}\gotht_{{4}}+672672\,\gotht_{{8}}\gotht_{{2}}
 \right) {S}^{8}+\nn\\
+ \left(
4057200\,\gotht_{{6}}\gotht_{{5}}+4047120\,\gotht_{{7}}\gotht_{{4
}}+3974256\,\gotht_{{8}}\gotht_{{3}} \right) {S}^{9}+ \left(
21999120\,\gotht_{{8}}\gotht_
{{4}}+11086740\,{\gotht_{{6}}}^{2}+22152900\,\gotht_{{7}}\gotht_{{5}}
\right) {S}^{10
}+\nn\\
+ \left(
117237120\,\gotht_{{8}}\gotht_{{5}}+117588240\,\gotht_{{7}}\gotht_{{6}}
\right) {S}^{11}+ \left(
304371144\,{\gotht_{{7}}}^{2}+607999392\,\gotht_{{8}}\gotht_{{6}}
 \right) {S}^{12}+\nn\\
+3085546464\,\gotht_{{8}}\gotht_{{7}}{S}^{13}+7688496816\,{\gotht_{{
8}}}^{2}{S}^{14}+\dots \ee \be
{F}_C^{(2)}=8\,\gotht_{{5}}{S}^{2}+168\,\gotht_{{6}}{S}^{3}+2121\,{S}^{4}\gotht_{{7}}+
20790\,{S}^{5}\gotht_{{8}}+
174174\,{S}^{6}\gotht_{{9}}+\nn\\
+1309308\,\gotht_{{10}}{S}^{7}+9087078\,{
S}^{8}\gotht_{{11}}+59306676\,\gotht_{{12}}{S}^{9}+\dots \ee



\sapp{$\bullet$ $Z_G(t^\pm)$}

In order to obtain the Gaussian free energy in terms of the ACKM
moments\footnote{In \cite{ACKM} the notations used are $t^+_k=J_k$
and $t^-_k=M_k$.}, one suffices to use manifest expressions
$F^{(p)}$ in terms of the standard times (see the beginning of this
Appendix) and insert there manifest expressions for the times
through moment variables. E.g., one can use explicit formula
(\ref{GF2}) for the genus two free energy (which we actually need
below to check explicitly the corresponding decomposition formula,
which we check starting from the genus 2 free energy only) to obtain
the well-known answer
 \be {F}^{(2)}={\frac
{1}{256}}\, \left( -{\frac {1448}{15}}\,{\frac {1}{{t^-_{{1}}}^
{2}d}}-80\,{\frac {1}{t^+_{{1}}dt^-_{{1}}}}-{\frac
{1448}{15}}\,{\frac {1} {{t^+_{{1}}}^{2}d}} \right){d}^{-3} -{ \frac
{35}{384}}\,{\frac {t^-_{{4}}}{d{t^-_{{1}}}^{3}}}+{\frac {43}
{192}}\,{\frac {t^-_{{3}}}{{d}^{2}{t^-_{{1}}}^{3}}}\nn\\
+{\frac {1}{256} }\, \left( -{\frac
{1448}{15}}\,{t^-_{{1}}}^{-3}-12\,{\frac {1}{t^+_{{1}}{
t^-_{{1}}}^{2}}} \right) t^-_{{2}}{d}^{-3}+{\frac {35}{384}}\,{
\frac {t^+_{{4}}}{d{t^+_{{1}}}^{3}}}+{\frac {43}{192}}\,{\frac {{g}
^{2}t^+_{{3}}}{{d}^{2}{t^+_{{1}}}^{3}}}\nn\\
+{\frac {1}{256}}\, \left( {\frac {
1448}{15}}\,{t^+_{{1}}}^{-3}+12\,{\frac
{1}{{t^+_{{1}}}^{2}t^-_{{1}}}}
 \right) t^+_{{2}}{d}^{-3}+{\frac {29}{128}}\,{\frac {t^-_{{2}}t^-_{{3}}}{d{t^-_{{1}}}^{4}}}-{\frac {11}{40}}\,{\frac {{t^-_{{2}}
}^{2}}{{d}^{2}{t^-_{{1}}}^{4}}}+{\frac {1}{64}}\,{\frac
{t^+_{{2}}t^-
_{{2}}}{{d}^{2}{t^+_{{1}}}^{2}{t^-_{{1}}}^{2}}}\nn\\
-{\frac {29}{128}}\,{\frac
{t^+_{{2}}t^+_{{3}}}{d{t^+_{{1}}}^{4}}}-{\frac {11}{40}}\,{\frac
{{t^+_ {{2}}}^{2}}{{d}^{2}{t^+_{{1}}}^{4}}}-{\frac
{21}{160}}\,{\frac {{
g}^{2}{t^-_{{2}}}^{3}}{d{t^-_{{1}}}^{5}}}+{\frac {21}{160}}\,{\frac
{{t^+_{{ 2}}}^{3}}{d{t^+_{{1}}}^{5}}} \label{ansmom} \ee  Indeed,
one can easily check this is correct by solving (\ref{usl}) and
(\ref{A+-}) perturbatively in $t$ and $S$ and then substituting them
into (\ref{GF2}): \be
A_+=2\sqrt{S}+t_{{1}}+2\,t_{{2}}t_{{1}}+2\,t_{{2}}\sqrt {S}+ \left(
3\,{t_{{2}}}^{2} +6\,t_{{1}}t_{{3}} \right) \sqrt {S}+6\,t_{{3}}S+
\left( 24\,t_{{1}}t_
{{4}}+24\,t_{{2}}t_{{3}} \right) S\nn\\
+12\,t_{{4}}{S}^{3/2}+ \left( 60\,t_
{{1}}t_{{5}}+60\,t_{{2}}t_{{4}}+36\,{t_{{3}}}^{2} \right) {S}^{3/2}+30
\,t_{{5}}{S}^{2}
+ \left( 216\,t_{{4}}t_{{3}}+180\,t_{{5}}t_{{2}}+180\,
t_{{1}}t_{{6}} \right) {S}^{2}\nn\\
+60\,t_{{6}}{S}^{5/2}+ \left( 420\,t_{{2
}}t_{{6}}+540\,t_{{3}}t_{{5}}+252\,{t_{{4}}}^{2}+420\,t_{{1}}t_{{7}}
 \right) {S}^{5/2}+140\,t_{{7}}{S}^{3}\nn\\
+ \left( 1440\,t_{{3}}t_{{6}}+
1120\,t_{{7}}t_{{2}}+1440\,t_{{5}}t_{{4}}+1120\,t_{{1}}t_{{8}}
 \right) {S}^{3}\nn\\
+280\,t_{{8}}{S}^{7/2}+ \left( 3360\,t_{{3}}t_{{7}}+
2520\,t_{{8}}t_{{2}}+3240\,t_{{4}}t_{{6}}+1800\,{t_{{5}}}^{2}+2520\,t_
{{1}}t_{{9}} \right) {S}^{7/2}+630\,t_{{9}}{S}^{4}\nn\\
+ \left( 8400\,t_{{7
}}t_{{4}}+6300\,t_{{1}}t_{{10}}+8400\,t_{{3}}t_{{8}}+6300\,t_{{9}}t_{{
2}}+9000\,t_{{5}}t_{{6}} \right) {S}^{4}+1260\,t_{{10}}{S}^{9/2}\nn\\
+ \left( 18900\,t_{{3}}t_{{9}}+13860\,t_{{2}}t_{{10}}+9900\,{t_{{6}}}^{
2}+21000\,t_{{5}}t_{{7}}+18480\,t_{{8}}t_{{4}} \right) {S}^{9/2}\nn\\
+ \left( 45360\,t_{{3}}t_{{10}}+50400\,t_{{5}}t_{{8}}+50400\,t_{{7}}t_{
{6}}+45360\,t_{{9}}t_{{4}} \right) {S}^{5}+O(S^{11/2})
\ee
\be
A_-=-2\sqrt{S}+t_{{1}}+2\,t_{{2}}t_{{1}}-2\,t_{{2}}\sqrt {S}+ \left( -3\,{t_{{2}}}^{2
}-6\,t_{{1}}t_{{3}} \right) \sqrt {S}+6\,t_{{3}}S
+ \left( 24\,t_{{1}}t
_{{4}}+24\,t_{{2}}t_{{3}} \right) S\nn\\
-12\,t_{{4}}{S}^{3/2}+ \left( -60\,
t_{{2}}t_{{4}}-60\,t_{{1}}t_{{5}}-36\,{t_{{3}}}^{2} \right) {S}^{3/2}+
30\,t_{{5}}{S}^{2}+ \left( 216\,t_{{4}}t_{{3}}+180\,t_{{5}}t_{{2}}+180
\,t_{{1}}t_{{6}} \right) {S}^{2}\nn\\
-60\,t_{{6}}{S}^{5/2}+ \left( -420\,t_
{{1}}t_{{7}}-252\,{t_{{4}}}^{2}-540\,t_{{3}}t_{{5}}-420\,t_{{2}}t_{{6}
} \right) {S}^{5/2}+140\,t_{{7}}{S}^{3}\nn\\
+ \left( 1440\,t_{{3}}t_{{6}}+
1120\,t_{{7}}t_{{2}}+1440\,t_{{5}}t_{{4}}+1120\,t_{{1}}t_{{8}}
 \right) {S}^{3}\nn\\
-280\,t_{{8}}{S}^{7/2}+ \left( -1800\,{t_{{5}}}^{2}-
3360\,t_{{3}}t_{{7}}-2520\,t_{{8}}t_{{2}}-2520\,t_{{1}}t_{{9}}-3240\,t
_{{4}}t_{{6}} \right) {S}^{7/2}+630\,t_{{9}}{S}^{4}\nn\\
+ \left( 8400\,t_{{
7}}t_{{4}}+6300\,t_{{1}}t_{{10}}+8400\,t_{{3}}t_{{8}}+6300\,t_{{9}}t_{
{2}}+9000\,t_{{5}}t_{{6}} \right) {S}^{4}-1260\,t_{{10}}{S}^{9/2}\nn\\
+
 \left( -13860\,t_{{2}}t_{{10}}-21000\,t_{{5}}t_{{7}}-18480\,t_{{8}}t_
{{4}}-18900\,t_{{3}}t_{{9}}-9900\,{t_{{6}}}^{2} \right) {S}^{9/2}\nn\\
+
 \left( 45360\,t_{{3}}t_{{10}}+50400\,t_{{5}}t_{{8}}+50400\,t_{{7}}t_{
{6}}+45360\,t_{{9}}t_{{4}} \right) {S}^{5} \ee and \be
t^-_1=-1+6\,t_{{1}}t_{{3}}+2\,t_{{2}}+ \left(
6\,t_{{2}}t_{{3}}+24\,t_{{1}}t
_{{4}}+6\,t_{{3}} \right) \sqrt {S}\nn\\
+ \left( 36\,{t_{{3}}}^{2}+48\,t_{{
2}}t_{{4}}+120\,t_{{1}}t_{{5}}+24\,t_{{4}} \right) S+ \left( 360\,t_{{
1}}t_{{6}}+60\,t_{{5}}+180\,t_{{4}}t_{{3}}+180\,t_{{5}}t_{{2}}
 \right) {S}^{3/2}\nn\\
+ \left( 720\,t_{{2}}t_{{6}}+288\,{t_{{4}}}^{2}+180
\,t_{{6}}+1260\,t_{{1}}t_{{7}}+900\,t_{{3}}t_{{5}} \right) {S}^{2}\nn\\
+
 \left( 1800\,t_{{5}}t_{{4}}+420\,t_{{7}}+2100\,t_{{7}}t_{{2}}+2340\,t
_{{3}}t_{{6}}+3360\,t_{{1}}t_{{8}} \right) {S}^{5/2}\nn\\
+ \left( 6720\,t_{
{8}}t_{{2}}+1120\,t_{{8}}+5760\,t_{{4}}t_{{6}}+8400\,t_{{3}}t_{{7}}+
10080\,t_{{1}}t_{{9}}+3600\,{t_{{5}}}^{2} \right) {S}^{3}\nn\\
+ \left(
15960\,t_{{7}}t_{{4}}+17640\,t_{{9}}t_{{2}}+2520\,t_{{9}}+25200\,t_{{1
}}t_{{10}}+16200\,t_{{5}}t_{{6}}+21000\,t_{{3}}t_{{8}} \right) {S}^{7/
2}\nn\\
+ \left( 64260\,t_{{3}}t_{{9}}+6300\,t_{{10}}+21600\,{t_{{6}}}^{2}+
47040\,t_{{8}}t_{{4}}+54600\,t_{{5}}t_{{7}}+50400\,t_{{2}}t_{{10}}
 \right) {S}^{4}\nn\\
+ \left( 113400\,t_{{7}}t_{{6}}+126000\,t_{{5}}t_{{8}}
+154980\,t_{{10}}t_{{3}}+120960\,t_{{9}}t_{{4}} \right) {S}^{9/2}\nn\\
+
 \left( 332640\,t_{{4}}t_{{10}}+302400\,t_{{8}}t_{{6}}+176400\,{t_{{7}
}}^{2}+378000\,t_{{9}}t_{{5}} \right) {S}^{5} \ee \be
t^-_2=\left.3\,t_{{3}}+12\,t_{{1}}t_{{4}}+ \left(
16\,t_{{4}}+80\,t_{{1}}t_{{5}}
+16\,t_{{2}}t_{{4}} \right) \sqrt {S}\right.\nn\\
+ \left( 140\,t_{{5}}t_{{2}}+420
\,t_{{1}}t_{{6}}+72\,t_{{4}}t_{{3}}+70\,t_{{5}} \right) S+ \left( 96
\,{t_{{4}}}^{2}+720\,t_{{2}}t_{{6}}+240\,t_{{6}}+1680\,t_{{1}}t_{{7}}-
480\,t_{{3}}t_{{5}} \right) {S}^{3/2}\nn\\
+ \left( 3080\,t_{{7}}t_{{2}}+
1200\,t_{{5}}t_{{4}}+2520\,t_{{3}}t_{{6}}+770\,t_{{7}}+6160\,t_{{1}}t_
{{8}} \right) {S}^{2}\nn\\
+ \left( 4800\,t_{{4}}t_{{6}}+20160\,t_{{1}}t_{{
9}}+2400\,{t_{{5}}}^{2}+2240\,t_{{8}}+11200\,t_{{8}}t_{{2}}+10080\,t_{
{3}}t_{{7}} \right) {S}^{5/2}\nn\\
+ \left( 63000\,t_{{1}}t_{{10}}+16800\,t
_{{5}}t_{{6}}+36960\,t_{{3}}t_{{8}}+20160\,t_{{7}}t_{{4}}+6300\,t_{{9}
}+37800\,t_{{9}}t_{{2}} \right) {S}^{3}\nn\\
+ \left( 61600\,t_{{5}}t_{{7}}
+21600\,{t_{{6}}}^{2}+69440\,t_{{8}}t_{{4}}+117600\,t_{{2}}t_{{10}}+
16800\,t_{{10}}+120960\,t_{{3}}t_{{9}} \right) {S}^{7/2}\nn\\
+ \left(
204400\,t_{{5}}t_{{8}}+151200\,t_{{7}}t_{{6}}+234360\,t_{{9}}t_{{4}}+
378000\,t_{{10}}t_{{3}} \right) {S}^{4}\nn\\
+ \left( 436800\,t_{{6}}t_{{8}
}+235200\,{t_{{7}}}^{2}+655200\,t_{{9}}t_{{5}}+715680\,t_{{4}}t_{{10}}
 \right) {S}^{9/2}+O(S^5)
\ee \be t^-_3= 20\,t_{{1}}t_{{5}}+4\,t_{{4}}+ \left(
30\,t_{{5}}+180\,t_{{1}}t_{{6}}+
30\,t_{{5}}t_{{2}} \right) \sqrt {S}\nn\\
+ \left( 1092\,t_{{1}}t_{{7}}+120
\,t_{{3}}t_{{5}}+312\,t_{{2}}t_{{6}}+156\,t_{{6}} \right) S+ \left(
5152\,t_{{1}}t_{{8}}+644\,t_{{7}}+180\,t_{{5}}t_{{4}}+1080\,t_{{3}}t_{
{6}}+1932\,t_{{7}}t_{{2}} \right) {S}^{3/2}\nn\\
+ \left( 2352\,t_{{8}}+
21168\,t_{{1}}t_{{9}}+600\,{t_{{5}}}^{2}+6552\,t_{{3}}t_{{7}}+1872\,t_
{{4}}t_{{6}}+9408\,t_{{8}}t_{{2}} \right) {S}^{2}\nn\\
+ \left( 39060\,t_{{9
}}t_{{2}}+6300\,t_{{5}}t_{{6}}+11592\,t_{{7}}t_{{4}}+30912\,t_{{3}}t_{
{8}}+7812\,t_{{9}}+78120\,t_{{1}}t_{{10}} \right) {S}^{5/2}\nn\\
+ \left(
24360\,t_{{10}}+127008\,t_{{3}}t_{{9}}+146160\,t_{{2}}t_{{10}}+9360\,{
t_{{6}}}^{2}+56448\,t_{{8}}t_{{4}}+35560\,t_{{5}}t_{{7}} \right) {S}^{
3}\nn\\
+ \left( 234360\,t_{{9}}t_{{4}}+468720\,t_{{10}}t_{{3}}+83160\,t_{{7
}}t_{{6}}+158760\,t_{{8}}t_{{5}} \right) {S}^{7/2}\nn\\
+ \left( 647640\,t_{
{9}}t_{{5}}+876960\,t_{{10}}t_{{4}}+152880\,{t_{{7}}}^{2}+325920\,t_{{
8}}t_{{6}} \right) {S}^{4}+O(S^{9/2}) \ee \be
t^-_4=30\,t_{{1}}t_{{6}}+5\,t_{{5}}+ \left(
336\,t_{{1}}t_{{7}}+48\,t_{{2}}t
_{{6}}+48\,t_{{6}} \right) \sqrt {S}\nn\\
+ \left( 180\,t_{{3}}t_{{6}}+2352
\,t_{{1}}t_{{8}}+588\,t_{{7}}t_{{2}}+294\,t_{{7}} \right) S+ \left(
288\,t_{{4}}t_{{6}}+4224\,t_{{8}}t_{{2}}+2016\,t_{{3}}t_{{7}}+12672\,t
_{{1}}t_{{9}}+1408\,t_{{8}} \right) {S}^{3/2}\nn\\
+ \left( 900\,t_{{5}}t_{{
6}}+14112\,t_{{3}}t_{{8}}+5814\,t_{{9}}+23256\,t_{{9}}t_{{2}}+3528\,t_
{{7}}t_{{4}}+58140\,t_{{1}}t_{{10}} \right) {S}^{2}\nn\\
+ \left( 10080\,t_{
{5}}t_{{7}}+76032\,t_{{3}}t_{{9}}+25344\,t_{{8}}t_{{4}}+1440\,{t_{{6}}
}^{2}+21600\,t_{{10}}+108000\,t_{{2}}t_{{10}} \right) {S}^{5/2}\nn\\
+\left( 21840\,t_{{7}}t_{{6}}+70560\,t_{{8}}t_{{5}}+139536\,t_{{9}}t_{
{4}}+348840\,t_{{10}}t_{{3}} \right) {S}^{3}\nn\\
+ \left( 380160\,t_{{9}}t_
{{5}}+133440\,t_{{8}}t_{{6}}+648000\,t_{{10}}t_{{4}}+47040\,{t_{{7}}}^
{2} \right) {S}^{7/2}+O(S^4) \ee Similarly, \be
t^+_1=-1+6\,t_{{1}}t_{{3}}+2\,t_{{2}}- \left(
6\,t_{{2}}t_{{3}}+24\,t_{{1}}t
_{{4}}+6\,t_{{3}} \right) \sqrt {S}\nn\\
+ \left( 36\,{t_{{3}}}^{2}+48\,t_{{
2}}t_{{4}}+120\,t_{{1}}t_{{5}}+24\,t_{{4}} \right) S- \left(
360\,t_{{
1}}t_{{6}}+60\,t_{{5}}+180\,t_{{4}}t_{{3}}+180\,t_{{5}}t_{{2}}
 \right) {S}^{3/2}\nn\\
+ \left( 720\,t_{{2}}t_{{6}}+288\,{t_{{4}}}^{2}+180
\,t_{{6}}+1260\,t_{{1}}t_{{7}}+900\,t_{{3}}t_{{5}} \right) {S}^{2}\nn\\
-
 \left( 1800\,t_{{5}}t_{{4}}+420\,t_{{7}}+2100\,t_{{7}}t_{{2}}+2340\,t
_{{3}}t_{{6}}+3360\,t_{{1}}t_{{8}} \right) {S}^{5/2}\nn\\
+ \left( 6720\,t_{
{8}}t_{{2}}+1120\,t_{{8}}+5760\,t_{{4}}t_{{6}}+8400\,t_{{3}}t_{{7}}+
10080\,t_{{1}}t_{{9}}+3600\,{t_{{5}}}^{2} \right) {S}^{3}\nn\\
- \left(
15960\,t_{{7}}t_{{4}}+17640\,t_{{9}}t_{{2}}+2520\,t_{{9}}+25200\,t_{{1
}}t_{{10}}+16200\,t_{{5}}t_{{6}}+21000\,t_{{3}}t_{{8}} \right)
{S}^{7/
2}\nn\\
+ \left( 64260\,t_{{3}}t_{{9}}+6300\,t_{{10}}+21600\,{t_{{6}}}^{2}+
47040\,t_{{8}}t_{{4}}+54600\,t_{{5}}t_{{7}}+50400\,t_{{2}}t_{{10}}
 \right) {S}^{4}\nn\\
- \left( 113400\,t_{{7}}t_{{6}}+126000\,t_{{5}}t_{{8}}
+154980\,t_{{10}}t_{{3}}+120960\,t_{{9}}t_{{4}} \right) {S}^{9/2}\nn\\
+
 \left( 332640\,t_{{4}}t_{{10}}+302400\,t_{{8}}t_{{6}}+176400\,{t_{{7}
}}^{2}+378000\,t_{{9}}t_{{5}} \right) {S}^{5} \ee \be
t^+_2=\left.3\,t_{{3}}+12\,t_{{1}}t_{{4}}- \left(
16\,t_{{4}}+80\,t_{{1}}t_{{5}}
+16\,t_{{2}}t_{{4}} \right) \sqrt {S}\right.\nn\\
+ \left( 140\,t_{{5}}t_{{2}}+420
\,t_{{1}}t_{{6}}+72\,t_{{4}}t_{{3}}+70\,t_{{5}} \right) S- \left( 96
\,{t_{{4}}}^{2}+720\,t_{{2}}t_{{6}}+240\,t_{{6}}+1680\,t_{{1}}t_{{7}}-
480\,t_{{3}}t_{{5}} \right) {S}^{3/2}\nn\\
+ \left( 3080\,t_{{7}}t_{{2}}+
1200\,t_{{5}}t_{{4}}+2520\,t_{{3}}t_{{6}}+770\,t_{{7}}+6160\,t_{{1}}t_
{{8}} \right) {S}^{2}\nn\\
- \left( 4800\,t_{{4}}t_{{6}}+20160\,t_{{1}}t_{{
9}}+2400\,{t_{{5}}}^{2}+2240\,t_{{8}}+11200\,t_{{8}}t_{{2}}+10080\,t_{
{3}}t_{{7}} \right) {S}^{5/2}\nn\\
+ \left( 63000\,t_{{1}}t_{{10}}+16800\,t
_{{5}}t_{{6}}+36960\,t_{{3}}t_{{8}}+20160\,t_{{7}}t_{{4}}+6300\,t_{{9}
}+37800\,t_{{9}}t_{{2}} \right) {S}^{3}\nn\\
- \left( 61600\,t_{{5}}t_{{7}}
+21600\,{t_{{6}}}^{2}+69440\,t_{{8}}t_{{4}}+117600\,t_{{2}}t_{{10}}+
16800\,t_{{10}}+120960\,t_{{3}}t_{{9}} \right) {S}^{7/2}\nn\\
+ \left(
204400\,t_{{5}}t_{{8}}+151200\,t_{{7}}t_{{6}}+234360\,t_{{9}}t_{{4}}+
378000\,t_{{10}}t_{{3}} \right) {S}^{4}\nn\\
-\left( 436800\,t_{{6}}t_{{8}
}+235200\,{t_{{7}}}^{2}+655200\,t_{{9}}t_{{5}}+715680\,t_{{4}}t_{{10}}
 \right) {S}^{9/2}+O(S^5)
\ee \be t^+_3= 20\,t_{{1}}t_{{5}}+4\,t_{{4}}- \left(
30\,t_{{5}}+180\,t_{{1}}t_{{6}}+
30\,t_{{5}}t_{{2}} \right) \sqrt {S}\nn\\
+ \left( 1092\,t_{{1}}t_{{7}}+120
\,t_{{3}}t_{{5}}+312\,t_{{2}}t_{{6}}+156\,t_{{6}} \right) S- \left(
5152\,t_{{1}}t_{{8}}+644\,t_{{7}}+180\,t_{{5}}t_{{4}}+1080\,t_{{3}}t_{
{6}}+1932\,t_{{7}}t_{{2}} \right) {S}^{3/2}\nn\\
+ \left( 2352\,t_{{8}}+
21168\,t_{{1}}t_{{9}}+600\,{t_{{5}}}^{2}+6552\,t_{{3}}t_{{7}}+1872\,t_
{{4}}t_{{6}}+9408\,t_{{8}}t_{{2}} \right) {S}^{2}\nn\\
- \left( 39060\,t_{{9
}}t_{{2}}+6300\,t_{{5}}t_{{6}}+11592\,t_{{7}}t_{{4}}+30912\,t_{{3}}t_{
{8}}+7812\,t_{{9}}+78120\,t_{{1}}t_{{10}} \right) {S}^{5/2}\nn\\
+ \left(
24360\,t_{{10}}+127008\,t_{{3}}t_{{9}}+146160\,t_{{2}}t_{{10}}+9360\,{
t_{{6}}}^{2}+56448\,t_{{8}}t_{{4}}+35560\,t_{{5}}t_{{7}} \right)
{S}^{
3}\nn\\
- \left( 234360\,t_{{9}}t_{{4}}+468720\,t_{{10}}t_{{3}}+83160\,t_{{7
}}t_{{6}}+158760\,t_{{8}}t_{{5}} \right) {S}^{7/2}\nn\\
+ \left( 647640\,t_{
{9}}t_{{5}}+876960\,t_{{10}}t_{{4}}+152880\,{t_{{7}}}^{2}+325920\,t_{{
8}}t_{{6}} \right) {S}^{4}+O(S^{9/2}) \ee \be
t^+_4=30\,t_{{1}}t_{{6}}+5\,t_{{5}}- \left(
336\,t_{{1}}t_{{7}}+48\,t_{{2}}t
_{{6}}+48\,t_{{6}} \right) \sqrt {S}\nn\\
+ \left( 180\,t_{{3}}t_{{6}}+2352
\,t_{{1}}t_{{8}}+588\,t_{{7}}t_{{2}}+294\,t_{{7}} \right) S- \left(
288\,t_{{4}}t_{{6}}+4224\,t_{{8}}t_{{2}}+2016\,t_{{3}}t_{{7}}+12672\,t
_{{1}}t_{{9}}+1408\,t_{{8}} \right) {S}^{3/2}\nn\\
+ \left( 900\,t_{{5}}t_{{
6}}+14112\,t_{{3}}t_{{8}}+5814\,t_{{9}}+23256\,t_{{9}}t_{{2}}+3528\,t_
{{7}}t_{{4}}+58140\,t_{{1}}t_{{10}} \right) {S}^{2}\nn\\
- \left( 10080\,t_{
{5}}t_{{7}}+76032\,t_{{3}}t_{{9}}+25344\,t_{{8}}t_{{4}}+1440\,{t_{{6}}
}^{2}+21600\,t_{{10}}+108000\,t_{{2}}t_{{10}} \right) {S}^{5/2}\nn\\
+\left(
21840\,t_{{7}}t_{{6}}+70560\,t_{{8}}t_{{5}}+139536\,t_{{9}}t_{
{4}}+348840\,t_{{10}}t_{{3}} \right) {S}^{3}\nn\\
- \left( 380160\,t_{{9}}t_
{{5}}+133440\,t_{{8}}t_{{6}}+648000\,t_{{10}}t_{{4}}+47040\,{t_{{7}}}^
{2} \right) {S}^{7/2}+O(S^4) \ee These expressions, indeed, convert
(\ref{ansmom}) into (\ref{GF2}).

\newpage

\app{Explicit checks of relations between various
partition functions}

\sapp{Experimental check of the decomposition formula}

We check here the decomposition formula (\ref{df}) of s.6
explicitly for the first several terms. To this end, we use the
non-standard genus expansion of s.2.5.3.

For variables ${t'}$ the decomposition formula acquires the form \be
Z_G(t'|S)e^{- U_{G}}=e^{-\hat
U_{K}}Z_K\left(-\frac{g\tau^+}{S}\left|\frac{g^2}{4S^2}\right)Z_K
\left(\frac{g\tau^-}{S}\right|\frac{g^2}{4S^2}\right) \ee with \be
\hat
U_{G}=\frac{1}{g}\oint\rho^{(0|1)}(z){{v'}}(z)+\frac{1}{2!}\oint\rho^{(0|2)}(z_1,z_2){{v'}}(z_1){{v'}}(z_2)
\ee and $U_K$ defined in (\ref{UK}).
We check this formula in the leading order in $g$. For doing this,
we need expressions for integrals of the simplest correlation
functions \be
\oint\rho^{(1|1)}(z)v'(z)dz=\frac{1}{16S}(\tau^-_1-\tau^+_1)-\frac{1}{24S}(\tau^-_0-\tau^+_0)
\ee and \be \frac{1}{3!}\oint\rho^{(0|3)}(z_1,z_2,z_3){{v'}}(z_1)
{{v'}}(z_2){{v'}}(z_3)=\frac{1}{2 S
3!}\left((\tau^-_0)^3-(\tau^+_0)^3\right) \ee The last equality
follows from \be
\rho^{(0|3)}(z_1,z_2,z_3)=\frac{2S(z_1z_2+z_1z_3+z_2z_3+4S)}{y(z_1)^3y(z_2)^3y(z_3)^3}=\\=
\frac{\sqrt{S}}{2y(z_1)y(z_2)y(z_3)}\left(\frac{1}{(z_1-2\sqrt{S})(z_2-2\sqrt{S})(z_3-2\sqrt{S})}
-\frac{1}{(z_1+2\sqrt{S})(z_2+2\sqrt{S})(z_3+2\sqrt{S})}\right) \ee
and \be
\oint\frac{v'(z)dz}{y(z)(z\pm2\sqrt{S})}=\frac{1}{\sqrt{S}}\tau^{\pm}_0
\ee Substituting the explicit expansion of the Kontsevich
$\tau$-function (see Appendix I) \be
\log{Z_K(\tau|\hbar)}=\frac{1}{8\hbar}\left(\frac{\tau_0^3}{3!}+\frac{\tau_0^3\tau_1}{2!2}+\dots\right)+\frac{1}{16}\tau_1
+\frac{5}{32}\tau_0\tau_2+\dots \ee into the decomposition formula
one gets \be e^{-\hat
U_{K}}Z_K(-\frac{g\tau^+}{S}|\frac{g^2}{4S^2})Z_K(\frac{g\tau^-}{S}|\frac{g^2}{4S^2})=
\left(1-\frac{1}{24}\left(\left(\frac{\p}{\p\tau^+_0}\right)^2+\left(\frac{\p}{\p\tau^-_0}\right)^2+\dots\right)\right)\times\\
\times\left(1-\frac{1}{3!}\frac{g(\tau^+_0)^3}{2S}-\frac{1}{16}\frac{g\tau^+_1}{S}+\frac{5}{32}
\frac{g^2\tau^+_0(\tau^+_2+\frac{16S}{5!g})}{S^2}+\dots\right)\times\\\times\left(1+\frac{1}{3!}\frac{g(\tau^-_0)^3}{2S}+
\frac{1}{16}\frac{g\tau^-_1}{S}+\frac{5}{32}
\frac{g^2\tau^-_0(\tau^-_2-\frac{16S}{5!g})}{S^2}+\dots\right)=\nn\\=
1+g\left(\frac{1}{2 S
3!}\left((\tau^-_0)^3-(\tau^+_0)^3\right)+\frac{1}{16S}(\tau^-_1-\tau^+_1)-\frac{1}{24S}(\tau^-_0-\tau^+_0)\right)+O(g^2)
\ee

\sapp{Non-trivial dependence on $t_0$ and the decomposition
formula}

Here we check the decomposition formula in the case of non-trivial
$t_0$-dependence, (\ref{dft0}) (see s.6.2) for the non-standard
genus expansion of s.2.5.3. That is, we check that the
$t_0$-dependence factorizes out from the formula.

Following \cite{amm1}, one can construct a generic branch of the
partition function. In this paper, we do not study dependence on
coefficients of the potential $W(z)$ and, hence, all the derivatives
of $F^{(k)}$ w.r.t. these coefficients should be considered as
constants determining a concrete solution.

\paragraph{1)} For the genus 1 free energy, one considers \be
{{F}}_{(1)}=\oint_\infty\rho^{(0|1)}(z){{v'}}(z)dz=\oint_\infty\rho_G^{(0|1)}(z){{v'}}(z)dz
\ee where the one-point resolvent $\rho^{(0|1)}(z)$ depends on the
curve only, but not on a particular branch of the partition function
chosen, and this term is canceled by $U_{G}$.

\paragraph{2)} The genus 2 free energy is determined by the genus zero
two-point function \be
{{F}}_{(2)}({{t'}}|S)=\frac{1}{2!}\oint\rho^{(0|2)}(z_1,z_2){{v'}}(z_1){{v'}}(z_2)dz_1dz_2=
\frac{1}{2}\oint\left(\rho^{(0|2)}_G(z_1,z_2)-\frac{\alpha}{2y(z_1)y(z_2)}\right){{v'}}(z_1){{v'}}(z_2)
dz_1dz_2\ee where $\alpha$ is an arbitrary constant. In terms of
check-operators of \cite{amm23} it can be written as \be \check R
y(z)=\frac{\alpha}{y(z)} \ee

From the generic construction one concludes that, for any solution
$Z$, the only dependence on $t_0$ comes in the combination $T_{-1}$,
(\ref{T-1}). The operator \be e^{(t_0-T_{-1})\frac{\p}{\p t_0}} \ee
just transform $T_{-1}$ into $t_0$, \be e^{(t_0-T_{-1})\frac{\p}{\p
t_0}}e^{-\hat
U_{\infty}}\exp{\left(g{\mathcal{F}}_{(1)}+g^2{\mathcal{F}}_{(2)}\right)}=e^{(t_0-T_{-1})\frac{\p}{\p
t_0}}e^{\frac{-\alpha T_{-1}^2}{4}}=e^{-\frac{\alpha t_0^2}{4}} \ee

\paragraph{3)} The genus 3 free energy is \be
{{F}}_{(3)}({{t'}}|S)=\frac{1}{3!}\oint\rho^{(0|3)}(z_1,z_2,z_3){{v'}}(z_1)
{{v'}}(z_2){{v'}}(z_3)dz_1dz_2dz_3+\oint\rho^{(1|1)}(z){{v'}}(z)dz
\ee

\be \oint\rho^{(0|3)}(z_1,z_2,z_3){{v'}}(z_1)
{{v'}}(z_2){{v'}}(z_3)dz_1dz_2dz_3=\oint\left( \frac{2S (z_1z_2 +
z_2z_3 + z_3z_1 + 4S)
}{y_G^3(z_1)y_G^3(z_2)y_G^3(z_3)}\right.-\nn\\
-\frac{\alpha}{2y(z_1)y(z_2)y(z_3)}\left(\frac{z_2z_3+4S}{y(z_2)^2y(z_3)^2}+
\frac{z_1z_3+4S}{y(z_1)^2y(z_3)^2}+\frac{z_2z_1+4S}{y(z_2)^2y(z_1)^2}\right)+\nn\\
\left.+\frac{\alpha^2}{2y(z_1)y(z_2)y(z_3)}\left(\frac{1}{y(z_1)^2}+\frac{1}{y(z_3)^2}+\frac{1}{y(z_3)^2}\right)
-\frac{\check R \alpha}{2y(z_1)y(z_2)y(z_3)} \right){{v'}}(z_1)
{{v'}}(z_2){{v'}}(z_3)dz_1dz_2dz_3=\nn\\=
2S(3\widetilde{S}_0^2\widetilde{T}_0+4S\widetilde{T}_0^3)-
\frac{3\alpha
T_{-1}}{2}(\widetilde{S}_0^2+4S\widetilde{T}_0^2)+\frac{3\alpha^2T_{-1}^2\widetilde{T}_0}{2}
-\frac{\check R \alpha T_{-1}^3}{2} \ee and \be
e^{(t_0-T_{-1})\frac{\p}{\p
t_0}}\oint\rho^{(0|3)}(z_1,z_2,z_3){{v'}}(z_1)
{{v'}}(z_2){{v'}}(z_3)dz_1dz_2dz_3=\nn\\=
2S(3\widetilde{S}_0^2\widetilde{T}_0+4S\widetilde{T}_0^3)-
\frac{3\alpha
t_0}{2}(\widetilde{S}_0^2+4S\widetilde{T}_0^2)+\frac{3\alpha^2t_0^2\widetilde{T}_0}{2}
-\frac{\check R \alpha t_0^3}{2} \ee \be e^{(t_0-T_{-1})\frac{\p}{\p
t_0}}\oint\rho^{(1|1)}(z){{v'}}(z)dz= S\widetilde{T}_1-\frac{\alpha
\widetilde{T}_0}{2}+(\check R F^{(1)})t_0 \ee where the times
$\widetilde{T}_k$, $\widetilde{S}_k$ are defined in (\ref{Tt}), e.g.
\be \widetilde{T}_0=\oint\frac{v(z)dz}{y(z)^3}=\frac{T_0}{8S} \ee
and $\check R F^{(1)}$, $\check R\alpha$ are considered as
independent constants parameterizing a solution $Z_{eG}$ to the
Virasoro constraints. After commuting the operators \be
\exp{\left(-4\frac{\p}{\p t_0}\frac{\p}{\p
T_0}\right)}\exp{\left(-\frac{\alpha t_0^2}{4}\right)}=
\exp{\left(-\frac{\alpha t_0^2}{4}+2\alpha t_0\frac{\p}{\p
T_0}-4\alpha\frac{\p^2}{\p T_0^2} \right)}\exp{\left(-4\frac{\p}{\p
t_0}\frac{\p }{\p T_0}\right)} \ee one obtains \be
\exp\left(-\frac{\alpha t_0^2}{4}+2\alpha t_0\frac{\p}{\p
T_0}-4\alpha\frac{\p^2}{\p T_0^2}\right)\exp\left(-4\frac{\p}{\p
t_0}\frac{\p }{\p T_0}\right)\times\\\times
\exp\left(\frac{1}{3!}\oint\rho^{(0|3)}(z_1,z_2,z_3){{v'}}(z_1)
{{v'}}(z_2){{v'}}(z_3)dz_1dz_2dz_3+\oint\rho^{(1|1)}(z){{v'}}\right)=\nn\\
=\frac{1}{3!}\exp{\left(-\frac{\alpha t_0^2}{4}+2\alpha
t_0\frac{\p}{\p T_0}-4\alpha\frac{\p^2}{\p T_0^2} \right)}
\left(1+2S(3\widetilde{S}_0^2\widetilde{T}_0+4S\widetilde{T}_0^3)-
\frac{3\alpha
t_0}{2}(\widetilde{S}_0^2+4S\widetilde{T}_0^2)+\right.\\+\left.\frac{3\alpha^2t_0^2\widetilde{T}_0}{2}
-\frac{\check R \alpha t_0^3}{2}+\frac{3\alpha
T_0}{8S}-\frac{3\alpha^2t_0}{2S} +3!(S\widetilde{T}_1+(\check R
F^{(1)})t_0) +\dots\right)=
\nn\\
=\frac{1}{3!}\exp{\left(-\frac{\alpha
t_0^2}{4}-4\alpha\frac{\p^2}{\p T_0^2} \right)}
\left(1+2S(3\widetilde{S}_0^2\widetilde{T}_0+4S\widetilde{T}_0^3)+\frac{\alpha^3t_0^3}{8S}
-\frac{\check R \alpha t_0^3}{2}+\right.\\+\left.\frac{3\alpha
T_0}{8S}-\frac{3\alpha^2t_0}{4S} +3!(S\widetilde{T}_1+(\check R
F^{(1)})t_0)
+\dots\right)=\nn\\
=\frac{1}{3!}\exp{\left(-\frac{\alpha t_0^2}{4}\right)}
\left(1+2S(3\widetilde{S}_0^2\widetilde{T}_0+4S\widetilde{T}_0^3)+\frac{\alpha^3t_0^3}{8S}
-\frac{\check R \alpha t_0^3}{2}-\frac{3\alpha^2t_0}{4S}
+3!(S\widetilde{T}_1+(\check R F^{(1)})t_0)
+\dots\right)=\nn\\
=\frac{1}{3!}\exp{\left(-\frac{\alpha
t_0^2}{4}\right)}(1+\frac{3\widetilde{S}_0^2T_0}{4}+\frac{T_0^3}{64S}+3!S\widetilde{T}_1\dots)(1+
\frac{\alpha^3t_0^3}{8S}-\frac{\check R \alpha
t_0^3}{2}-\frac{3\alpha^2t_0}{4S} +3!(\check R F^{(1)})t_0\dots) \ee
Therefore, one, indeed, obtains that the dependence on $t_0$
factories out from the decomposition formula (\ref{dft0}), the
function $Z(t_0)$ being $\exp{\left(-\frac{\alpha t_0^2}{4}\right)}
(1+\frac{\alpha^3t_0^3}{8S}-\frac{\check R \alpha
t_0^3}{2}-\frac{3\alpha^2t_0}{4S} +3!(\check R F^{(1)})t_0\dots)$.

\sapp{Check of the decomposition formula in terms of ACKM moments}

Here we check the decomposition formula (\ref{dfk}) of s.7. For the
sake of simplicity, we check it starting only from the genus 2 free
energy, since the genus 0 and genus 1 free energies are not
polynomials even in terms of moments, see s.3, and therefore, their
direct check is highly involved. The Gaussian free energy $F^{(2)}$
as a function of moments $t^\pm_k$ can be found in the previous
Appendix, formula (\ref{ansmom}). One can also calculate the same
quantity using the decomposition formula. Using the coordinate
system: \be
z-A_-=-x_-^2\\
z-A_+=x_+^2 \ee and proceeding with calculations described in s.7,
one obtains \be F^{(2)}={\frac
{1}{256d^{3}}}\,\left(-1225\,{d}^{2}{G_+}^{-4}{\tau^+_
{{2}}}^{2}+100\,{d}^{2}{G_-}^{-2}{G_+}^{-2}\tau^-_{{2}} \tau^+_{{2}}
-100\,d{G_-}^{-2}{G_+}^{-1}\tau^-_{{2}}\right.\nn\\
+420\,{d}^{3}{G_+}^{-3}\tau^+_{{4}}+2030\,{d}^{3}{G_+}^{-4}
\tau^+_{{2}}\tau^+_{{3}}+276\,{G_-}^{-1}{G_+}^{-1}+
2100\,{d}^{3}{G_+}^{-5}{\tau^+_{{2}}}^{3}-560\,{d}^{2}{G_-}^{-3}\tau^-_{{3}}\nn\\
-100\,d{G_-}^{-1}{G_+}^{-2}\tau^+
_{{2}}+2030\,{d}^{3}{G_-}^{-4}\tau^-_{{2}}\tau^-_{{3}}-175
\,{G_+}^{-2}-175\,{G_-}^{-2}+525\,d{G_+}^{-3}\tau^+
_{{2}}\nn\\
-1225\,{d}^{2}{G_-}^{-4}{\tau^-_{{2}}}^{2}+525\,d{G_-}^{-3}\tau^-_{{2}}+420\,{d}^{3}{G_-}^{-3}\tau^-_{{4}
}+2100\,{d}^{3}{G_-}^{-5}{\tau^-_{{2}}}^{3}-\left.560\,{d}^{2}{{
G_+}}^{-3}\tau^+_{{3}}\right) g^2\ee where \be d=A_+-A_- \ee

To compare it with the answer in terms of ACKM moments (\ref{GF2}),
one should substitute $\tau^\pm$ with their expressions in terms of
$t^{\pm}_k$ \be\label{explJM}
t^+_{{1}}=-3/4\,{\frac {\tau^-_{{1}}}{\sqrt {d}}}\\
t^-_{{1}}=3/4\,{\frac {\tau^+_{{1}}}{\sqrt {d}}}\\
t^+_{{2}}=1/2\, \left( 5/2\,\tau^-_{{2}}-3/4\,{\frac {\tau^-_{{
1}}}{d}} \right) {\frac {1}{\sqrt {d}}}\\
t^-_{{2}}=1/2\, \left( 5/2\,\tau^+_{{2}}-3/4\,{\frac {\tau^+_{{
1}}}{d}} \right) {\frac {1}{\sqrt {d}}}\\
t^+_{{3}}=-1/2\, \left( 7/2\,\tau^-_{{3}}-5/4\,{\frac {\tau^-_{{2
}}}{d}}+{\frac {9}{16}}\,{\frac {\tau^-_{{1}}}{{d}^{2}}} \right) {
\frac {1}{\sqrt {d}}}\\
t^-_{{3}}=1/2\, \left( 7/2\,\tau^+_{{3}}-5/4\,{\frac {\tau^+_{{
2}}}{d}}+{\frac {9}{16}}\,{\frac {\tau^+_{{1}}}{{d}^{2}}} \right)
{\frac {1}{\sqrt {d}}}\\
t^+_{{4}}=1/2\, \left( 9/2\,\tau^-_{{4}}-7/4\,{\frac {\tau^-_{{
3}}}{d}}+{\frac {15}{16}}\,{\frac {\tau^-_{{2}}}{{d}^{2}}}-{\frac
{15}{32}}\,{\frac {\tau^-_{{1}}}{{d}^{3}}} \right) {\frac {1}{
\sqrt {d}}}\\
t^-_{{4}}=1/2\, \left( 9/2\,\tau^+_{{4}}-7/4\,{\frac {\tau^+_{{
3}}}{d}}+{\frac {15}{16}}\,{\frac {\tau^+_{{2}}}{{d}^{2}}}-{\frac
{15}{32}}\,{\frac {\tau^+_{{1}}}{{d}^{3}}} \right) {\frac {1}{ \sqrt
{d}}} \ee Then, one immediately comes to formula (\ref{GF2}).

\sapp{Check of the decomposition formula for $Z_C\longrightarrow
Z_K\tilde Z_K$} We check here the decomposition formula (\ref{dfc})
for $Z_C(\gotht)\longrightarrow Z_K(\tau)\tilde Z_K({\cal T})$ up to
$g^2$. Let us denote the contribution of degree $k$ in times into
the genus $l$ free energy of the complex matrix model as
$r^{(l|k)}$. Then, \be\label{1}
r^{(0|3)}=\frac{1}{3!}\oint_\infty\oint_\infty\oint_\infty\rho^{(0|3)}_C(z_1,z_2,z_3) v^\Sigma_C(z_1)v^\Sigma_C(z_2)v^\Sigma_C(z_3)=\nn\\
=\frac{2S^2}{3!}\left(\oint_\infty\frac{v^\Sigma_C(z)dz}{y(z)^3}\right)^3=
\frac{2S^2}{3!}\left(\oint_a\frac{v^\Sigma_C(z)dz}{y(z)^3}\right)^3=
\frac{2S^2}{3!}\left(\frac{\tau_0}{(4S)^\frac{2}{3}}\right)^3=\frac{\tau_0^3}{3!8}
\ee \be\label{2}
r^{(1|1)}=\oint_\infty\rho^{(1|1)}_C(z)v^\Sigma_C(z)=\left(\oint_a+\oint_0\right)\rho^{(1|1)}_C(z)v^\Sigma_C(z)=
\frac{\tau_1}{16}-\frac{13\cdot2^\frac{2}{3}\tau_0}{512S^\frac{2}{3}}+\frac{{\cal
T}_0}{16\sqrt{-4S}} \ee
\be\label{3}
r^{(1|2)}=\frac{1}{2!}\oint_\infty\oint_\infty\rho^{(1|2)}_C(z_1,z_2)v^\Sigma_C(z_1)v^\Sigma_C(z_2)=
\frac{1}{2!}\left(\oint_a\oint_a+2\oint_a\oint_0+\oint_0\oint_0\right)\rho^{(1|2)}_C(z_1,z_2)v^\Sigma_C(z_1)v^\Sigma_C(z_2)=\nn\\
={\frac {123}{16384}}\,{\frac
{2^{\frac{1}{3}}{\tau_{{0}}}^{2}}{{S}^{4/3}}}-{ \frac
{21}{512}}\,{\frac
{{2}^{2/3}\tau_{{0}}\tau_{{1}}}{{S}^{2/3}}}+{\frac
{5}{32}}\,\tau_{{0}}\tau_{{2}}+{\frac {3}{64}}\,{\tau_{{1}}}^{2}-
\frac{{\cal T}_0^2}{64\cdot4S} \ee
Note now that in the l.h.s. of the decomposition formula (\ref{dfc})
there are no terms $r^{(0|1)}$ and $r^{(0|2)}$, since they are
canceled by the corresponding contributions from the conjugation
operator $e^{U_C}$. Therefore, one should compare sum of the three
terms above, (\ref{1}), (\ref{2}) and (\ref{3}) with the r.h.s. of
the decomposition formula (\ref{dfc}): \be e^{\hat
V_{ram}}Z_K(\tau)\tilde{Z}_K({\cal T})=
\left(1-\frac{2^\frac{2}{3}}{16\cdot5gS^\frac{2}{3}}
\frac{\p}{\p\tau_2}+ \frac{2^\frac{1}{3}}{512\cdot7gS^\frac{4}{3}}
\frac{\p}{\p\tau_3}- \frac{3\cdot
2^\frac{2}{3}}{16S^\frac{2}{3}}\frac{\p^2}{\p
\tau_0^2}+\right.\\\left.+\frac{1}{2}\left(\frac{3\cdot
2^\frac{2}{3}}{16S^\frac{2}{3}}\frac{\p^2}{\p
\tau_0^2}\right)^2+\frac{5\cdot
2^{\frac{1}{3}}}{256S^{\frac{4}{3}}}\frac{\p^2}{\p\tau_0\p\tau_1}-
\frac{2^\frac{1}{3}}{S^\frac{1}{3}\sqrt{-S}}\frac{\p^2}{\p\tau_0\p{\cal
T}_0}\dots \right)\times\nn\\\times
\left(1+g\left(\frac{\tau_0^3}{3!8}+\frac{\tau_1}{16}+g\frac{5\tau_0\tau_2}{32}\right)+\frac{g^2}{2}\left(\frac{\tau_0^3}{3!8}+\frac{\tau_1}{16}+g\frac{5\tau_0\tau_2}{32}\right)^2\right.+\nn\\
+\left.g^2\left(\frac{1}{4}\,{\tau_{{0}}}^{3}\tau_{{1}}+g{\frac
{5}{32}}\,{\tau_{{0}}}^{4}\tau_ {{2}}+{\frac
{3}{64}}\,{\tau_{{1}}}^{2}+g{\frac {15}{32}}\,\tau_{{0}}\tau_{{1
}}\tau_{{2}}+g{\frac
{35}{128}}\,{\tau_{{0}}}^{2}\tau_{{3}}+g^2{\frac {75}
{128}}\,{\tau_{{0}}}^{2}{\tau_{{2}}}^{2}\right)+O(g^3)\right)\times
\nn\\\times
\left(1+g\frac{{\cal T}_0}{16\sqrt{-4S}}-g^2(\frac{{\cal T}_0^2}{64\cdot4S}+\frac{{\cal T}_0^2}{16\cdot16\cdot2\cdot4S})+O(g^3)\right)=\nn\\
=\hbox{const}+\left( -{\frac {13}{512}}\,{\frac {{
2}^{2/3}\tau_{{0}}}{{S}^{2/3}}}+\frac{1}{16}\,\tau_{{1}}+\frac{1}{48}\,{\tau_{{0}}}^{3}+\frac{{\cal
T}_0}{16\sqrt{-4S}}
 \right) g+ \left( {\frac {2137}{262144}}\,{\frac {2^{\frac{1}{3}}{\tau_{{0}
}}^{2}}{{S}^{4/3}}}+{\frac {25}{512}}\,{\tau_{{ 1}}}^{2}
-\right.\\-{\frac {349}{8192}}\,{\frac
{{2}^{2/3}\tau_{{1}}\tau_{{0}}}{{S}^{ 2/3}}}+{\frac
{5}{32}}\,\tau_{{2}}\tau_{{0}}-{\frac {37}{24576}}\,{ \frac
{{2}^{2/3}{\tau_{{0}}}^{4}}{{S}^{2/3}}}+{\frac
{25}{768}}\,{\tau_{{0}}
}^{3}\tau_{{1}}+\frac{t_0^6}{4608} +\nn\\
+\left.\left( -{\frac {13}{512}}\,{\frac {{
2}^{2/3}\tau_{{0}}}{{S}^{2/3}}}+\frac{1}{16}\,\tau_{{1}}+\frac{1}{48}\,{\tau_{{0}}}^{3}\right)\frac{{\cal T}_0}{16\sqrt{-4S}}-\left(\frac{{\cal T}_0^2}{64\cdot4S}+\frac{{\cal T}_0^2}{16\cdot16\cdot2\cdot4S}\right)\right)g^2+O(g^3)=\nn\\
=\hbox{const}+g(r^{(1|1)}+r^{(0|3)})+g^2\left(\frac{(r^{(1|1)})^2
}{2!}+r^{(1|2)}+O(t^4)\right)+O(g^3) \ee

\section*{Acknowledgements}

We are grateful to I.Kostov who has provided us with an improved version of
\cite{Kostov2}.

This work was partially supported by the Federal Program of the
Russian Ministry of Industry, Science and Technology No
40.052.1.1.1112, by the grants RFBR 03-02-17373 (Alexandrov), RFBR
04-02-16538a (Mironov), RFBR 04-02-16880 (Morozov), by the Grant of
Support for the Scientific Schools 8004.2006.2, NWO project
047.011.2004.026, INTAS grant 05-1000008-7865
and ANR-05-BLAN-0029-01 project "Geometry and Integrability in
Mathematical Physics" (A.M.'s).


\newpage

\begin{thebibliography}{12}

\bibitem{amm1}  A.~Alexandrov, A.~Mironov and A.~Morozov,
Partition functions of matrix models as the first special functions
of string theory. I: Finite size Hermitean 1-matrix model,
Int.J.Mod.Phys. {\bf A19} (2004) 4127; Teor.\ Mat.\ Fiz.\  {\bf 142}
(2005) 419 (Theor.Math.Phys. {\bf 142} (2005) 349), hep-th/0310113.

\bibitem{M} A.~Morozov,
Challenges of matrix models,
  hep-th/0502010\\
A.~Mironov, Matrix Models vs. Matrix Integrals, Theor.Math.Phys.
{\bf 146} (2005) 63, hep-th/0506158.

\bibitem{HM} M.L.~Mehta, {\sl Random matrices},
2nd ed., Academic Press, New York, 1991\\
E.~Br\'ezin, C.~Itzykson, G.~Parisi and J.-B.~Zuber, Planar
diagrams, Commun.Math.Phys. {\bf 59} (1978) 35\\
D.~Bessis, A new method in the combinatorics of the topological
expansion, Commun.Math.Phys. {\bf 69} (1979) 147\\
D.~Bessis, C.~Itzykson and J.-B.~Zuber, Quantum field theory
techniques in graphical enumeration, Adv.Appl.Math. {\bf 1} (1980) 109\\
C.~Itzykson and J.-B.~Zuber, The planar approximation. II,
J.Math.Phys. {\bf 21} (1980) 411.

\bibitem{GMMMO} A.~Gerasimov, A.~Marshakov, A.~Mironov, A.~Morozov and A.~Orlov,
Matrix Models Of 2-D Gravity And Toda Theory,
  Nucl.Phys. {\bf B357} (1991) 565.

\bibitem{UFN3} A.~Morozov,
Integrability and matrix models,
Phys.Usp.(UFN) {\bf 37} (1994) 1, hep-th/9303139; hep-th/9502091\\
A.~Mironov, $2d$ gravity and matrix models. I. $2d$ gravity,
Int.J.Mod.Phys. {\bf A9} (1994) 4355, hep-th/9312212; Matrix models
of two-dimensional gravity, Phys.Part.Nucl. {\bf 33} (2002) 537.

\bibitem{DV} R.~Dijkgraaf and C.~Vafa,
Matrix Models, Topological Strings, and Supersymmetric Gauge
Theories, Nucl.Phys. {\bf B644} (2002) 3-20, hep-th/0206255; On
Geometry and Matrix Models, Nucl.Phys. {\bf B644} (2002) 21-39,
hep-th/0207106; Perturbative Derivation of Mirror Symmetry,
hep-th/0208048.

\bibitem{ChMi} L.~Chekhov and A.~Mironov, Matrix models vs.
Seiberg-Witten/Whitham theories, Phys.Lett. {\bf B552}
(2003) 293, hep-th/0209085.

\bibitem{Ito4} H.~Itoyama and A.~Morozov, The Dijkgraaf-Vafa prepotential in the context of general Seiberg-Witten
theory, Nucl.Phys. {\bf B657} (2003) 53,  hep-th/0211245.

\bibitem{IM} H.~Itoyama and A.~Morozov, Experiments with the WDVV equations for the gluino-condensate prepotential: the cubic (two-cut)
case, Phys.Lett. {\bf B555} (2003) 287, hep-th/0211259; Calculating
Gluino-Condensate Prepotential, Prog.Theor.Phys. {\bf 109} (2003)
433, hep-th/0212032; Gluino-Condensate (CIV-DV) Prepotential from
its Whitham-Time Derivatives, Int.J.Mod.Phys. {\bf A18} (2003) 5889,
hep-th/0301136.

\bibitem{ChMMV} L.~Chekhov, A.~Marshakov, A.~Mironov and D.~Vasiliev, DV and
WDVV, Phys.Lett. {\bf B562} (2003) 323, hep-th/0301071.

\bibitem{ChMMV2}  L.~Chekhov, A.~Marshakov, A.~Mironov and D.~Vasiliev, Complex Geometry of Matrix
Models, Proc. Steklov Inst.Math. {\bf 251} (2005) 254,
hep-th/0506075.

\bibitem{Ey} B.~Eynard, All genus correlation functions for the hermitian 1-matrix model, JHEP {\bf 0411} (2004) 031, hep-th/0407261.

\bibitem{amm23} A.~Alexandrov, A.~Mironov and A.~Morozov,
Solving Virasoro constraints in matrix models,
  Fortsch.Phys. {\bf 53} (2005) 512, hep-th/0412205; Unified description of correlators in non-Gaussian phases of Hermitean
matrix model, Int.J.Mod.Phys. {\bf A21} (2006) 2481, hep-th/0412099.

\bibitem{EyCh}  L.~Chekhov and B.~Eynard,
Hermitean matrix model free energy: Feynman graph technique for all
genera, hep-th/0504116.

\bibitem{amm4} A.~Alexandrov, A.~Mironov and A.~Morozov,
M-Theory of Matrix Models, hep-th/0605171.

\bibitem{Th} G.~Bonnet, F.~David and B.~Eynard, Breakdown of universality in multi-cut matrix models, J.Phys. {\bf A33}
(2000) 6739, cond-mat/0003324\\
A.~Klemm, M.~Mari\~no and S.~Theisen, JHEP {\bf 0303} (2003) 051,
Gravitational corrections in supersymmetric gauge theory and matrix
models, hep-th/0211216.

\bibitem{mmmm}  Yu.~Makeenko, A.~Marshakov, A.~Mironov and A.~Morozov,
Continuum versus discrete Virasoro in one-matrix models, Nucl.Phys.
{\bf B356} (1991) 574.

\bibitem{Ch}  L.~Chekhov,
Matrix Models and Geometry of Moduli Spaces, hep-th/9509001.

\bibitem{Kostov2} I.K.~Kostov,
Conformal field theory techniques in random matrix models,
hep-th/9907060; an improved version (unpublished).

\bibitem{Ko} M.L.~Kontsevich,
Theory of intersections on moduli space of curves,
Funk.Anal.Prilozh. {\bf 25} (1991) v.2, p.50 (in Russian).

\bibitem{GKM} S.~Kharchev, A.~Marshakov, A.~Mironov, A.~Morozov and A.~Zabrodin,
Towards unified theory of $2d$ gravity, Nucl.Phys. {\bf B380} (1992)
181-240, hep-th/9201013; Unification of All String Models with
$c<1$, Phys.Lett. {\bf B275} (1992) 311-314, hep-th/9111037.

\bibitem{GKMKM} S.~Kharchev, A.~Marshakov, A.~Mironov and A.~Morozov, Generalized Kontsevich Model Versus Toda Hierarchy and Discrete Matrix
Models, Nucl. Phys. {\bf B397} (1993) 339, hep-th/9203043;
Landau-Ginzburg Topological Theories in the Framework of GKM and
Equivalent Hierarchies, Mod.Phys.Lett. {\bf A8} (1993) 1047-1062,
hep-th/9208046; Generalized Kazakov-Migdal-Kontsevich Model: group
theory aspects, Int.J.Mod.Phys. {\bf A10} (1995) 2015,
hep-th/9312210.

\bibitem{Kom} P.~Di Francesco, C.~Itzykson and J.-B.~Zuber,
Polynomial averages in the Kontsevich model, Commun.Math.Phys. {\bf
151} (1993) 193-219, hep-th/9206090.

\bibitem{BPST} A.~Belavin, A.~Polyakov, A.~Schwarz and Yu.~Tyupkin,  Pseudoparticle Solutions Of The Yang-Mills
Equations, Phys.Lett. {\bf B59} (1975) 85-87.

\bibitem{CGD} C.~Callan, R.~Dashen and D.~Gross,  Toward A Theory Of The Strong Interactions, Phys.Rev. {\bf D17}
(1978) 2717.

\bibitem{mmt} A.~Mironov, A.~Morozov and T.~Tomaras, On the Need for Phenomenological Theory of P-Vortices or Does
Spaghetti Confinement Pattern Admit Condensed-Matter Analogies? J.Exp.Theor.Phys.
{\bf 101} (2005) 331-340.

\bibitem{discvir} A.~Gerasimov, A.~Marshakov, A.~Mironov, A.~Morozov and
A.~Orlov,  Matrix models of 2-D gravity and Toda theory, Nucl.Phys. {\bf B357} (1991) 565\\
F.~David,  Loop Equations And Nonperturbative Effects In Two-Dimensional Quantum Gravity, Mod.Phys.Lett. {\bf A5} (1990) 1019\\
A.~Mironov and A.~Morozov,  On the origin of Virasoro constraints in matrix models: Lagrangian approach, Phys.Lett. {\bf B252} (1990) 47-52\\
J.~Ambj{\o}rn and Yu.~Makeenko,  Properties Of Loop Equations For The Hermitean Matrix Model And For Two-Dimensional Quantum Gravity, Mod.Phys.Lett. {\bf A5} (1990) 1753\\
H.~Itoyama and Y.~Matsuo,  Noncritical Virasoro algebra of $d < 1$
matrix model and quantized string field, Phys.Lett. {\bf 255B}
(1991) 202.

\bibitem{tH} G.~'t~Hooft,
A planar diagram theory for strong interactions, Nucl.Phys. {\bf
B72} (1974) 461.

\bibitem{KN}  I.M.~Krichever and S.P.~Novikov,
 Algebras of Virasoro type, Riemann surfaces and the structure of soliton
 theory, Funct.Anal.Appl.  {\bf 21} (1987) 126-142;
Virasoro-Gelfand-Fuks Type Algebras, Riemann Surfaces, Operator's
Theory Of Closed Strings, J.Geom.Phys. {\bf 5} (1988) 631-661
Virasoro-type algebras, Riemann surfaces and strings in
Minkowsky space, Funct.Anal.Appl. 21 No.4 (1987) 294-307; Algebras
of Virasoro type, the energy-momentum tensor, and operator
expansions on Riemann surfaces, Funct.Anal.Appl. {\bf 23} (1989)
19-33.

\bibitem{KNdop} R.~Dick,
Global Expansions Of Holomorphic Differentials On Punctured Riemann
Surfaces, 
Lett.Math.Phys. {\bf 18} (1989) 255\\
 M.~Schlichenmaier, Krichever-Novikov Algebras For More Than Two Points, Lett.Math.Phys.
{\bf 19} (1990) 151; Krichever-Novikov Algebras For More Than Two
Points: Explicit Generators, Lett.Math.Phys. {\bf 19}
  (1990) 327.

\bibitem{Grass} A.~Morozov,  String Theory And The Structure Of Universal Module Space,
Phys.Lett. {\bf B196} (1987) 325.

\bibitem{SFT}
E.~Witten,  Noncommutative Geometry And String Field Theory,
Nucl.Phys. {\bf B268} (1985) 253;  Interacting Field Theory Of Open
Superstrings, {\it ibid.} {\bf B276} (1986) 291;
Quantum Background Independence In String Theory, hep-th/9306122\\
B.~Zwiebach, Quantum open string theory with manifest closed string
factorization, Phys.Lett. {\bf B256} (1991) 22-29;  Interpolating
string field theories, Mod.Phys.Lett. {\bf A7} (1992) 1079-1090;
 Oriented open - closed string theory revisited, Annals Phys. {\bf 267} (1998) 193-248;
 Closed string field theory: An Introduction, hep-th/9305026\\
A.~Sen and B.~Zwiebach,  A Proof of local background independence of
classical closed string field theory, Nucl.Phys. {\bf B414} (1994)
649-714;  Quantum background independence of closed string field
theory, {\it ibid.} {\bf B423} (1994) 580-630;  Background
independent algebraic structures in closed string field theory,
Commun.Math.Phys. {\bf 177} (1996) 305-326.

\bibitem{MMMWDVV} A.~Marshakov, A.~Mironov and A.~Morozov, WDVV-like equations in N=2 SUSY Yang-Mills
Theory, Phys. Lett. {\bf B389} (1996) 43, hep-th/9607109; WDVV
Equations from Algebra of Forms, Mod.Phys.Lett. {\bf A12} (1997)
773-787, hep-th/9701014; More Evidence for the WDVV Equations in N=2
SUSY Yang-Mills Theories, Int.J.Mod.Phys. {\bf A15} (2000)
1157-1206, hep-th/9701123.

\bibitem{FFRS} V.~Knizhnik,  Analytic Fields On Riemannian Surfaces, Phys.Lett. {\bf B180} (1986) 247;
 Analytic Fields On Riemann Surfaces. 2, Comm.Math.Phys. {\bf 112} (1987) 567;  Multiloop amplitudes in the
theory of quantum strings and complex geometry, Sov.Phys.Uspekhi,
{\bf 32} (1989) {\it \#3},
945, in Russian Edition: vol.159, p.451\\
D.~Lebedev and A.~Morozov,  Statistical Sums Of Strings On
Hyperelliptic Surfaces,
Nucl.Phys. {\bf B302} (1988) 163\\
A.~Morozov,  Two Loop Statsum Of Superstring, Nucl.Phys. {\bf B303}
(1988) 342.

\bibitem{SW} N.~Seiberg and E.~Witten, Monopole Condensation, And Confinement
In N=2 Supersymmetric Yang-Mills Theory, Nucl.Phys. {\bf B426}
(1994) 19; Erratum-{\it ibid.} {\bf B430} (1994) 485-486.

\bibitem{GKMMM} A.~Gorsky, I.~Krichever, A.~Marshakov, A.~Mironov and A.~Morozov,
Integrability and exact Seiberg-Witten solution, Phys.Lett. {\bf
B355} (1995) 466-477, hep-th/9505035.

\bibitem{DW}  R.~Donagi and E.~Witten,
Supersymmetric Yang-Mills Theory And Integrable Systems, Nucl.Phys.
{\bf B460} (1996) 299, hep-th/9510101.

\bibitem{Ito13} H.~Itoyama and A.~Morozov,
Integrability and Seiberg-Witten theory; curves and periods,
Nucl.Phys., {\bf B477} (1996) 855-877, hep-th/9511126; Prepotential
and the Seiberg- Witten theory, Nucl.Phys., {\bf B491} (1997)
529-573, hep-th/9512161; Integrability and Seiberg-Witten theory,
hep-th/9601168.

\bibitem{GM} A.~Gorsky and A.~Mironov, Integrable Many-Body Systems and Gauge Theories,
hep-th/0011197.

\bibitem{MMM} A.~Marshakov, A.~Mironov and A.~Morozov, On Equivalence of Topological and Quantum 2d Gravity,
Phys. Lett. {\bf B274} (1992) 280, hep-th/9201011.

\bibitem{MomK} C.~Itzykson and J.B.~Zuber,
Combinatorics of the modular group. 2. The Kontsevich integrals,
 Int.J.Mod.Phys. {\bf A7} (1992) 5661, hep-th/9201001.

\bibitem{ACKM} J.~Ambj\"orn, L.~Chekhov and Yu.~Makeenko, Higher Genus Correlators from the Hermitian One-Matrix
Model, Phys.Lett. {\bf B282} (1992) 341-348, hep-th/9203009\\
J.~Ambj\"orn, L.~Chekhov, C.F.~Kristjansen and Yu.~Makeenko, Matrix
Model Calculations beyond the Spherical Limit, Nucl.Phys. {\bf B404}
(1993) 127-172; Erratum-{\it ibid.} {\bf B449} (1995) 681,
hep-th/9302014.

\bibitem{Giv} A.~Givental, Semisimple Frobenius structures at higher genus,
math.AG/0008067.

\bibitem{Hori}
T.~Eguchi, Y.~Yamada and S.K.~Yang, On The Genus Expansion In The
Topological String Theory,
 Rev.Math.Phys.  {\bf 7} (1995) 279, hep-th/9405106.

\bibitem{Eguchi}
 T.~Eguchi and S.K.~Yang,
The Topological $CP^1$ model and the large N matrix integral,
 Mod.Phys.Lett. {\bf A9} (1994) 2893, hep-th/9407134\\
T.~Eguchi, K.~Hori and S.K.~Yang, Topological sigma models and large
N matrix integral,
 Int.J.Mod.Phys. {\bf A10} (1995) 4203, hep-th/9503017\\
K.~Hori, Constraints for topological strings in $D\ge 1$,
 Nucl.Phys. {\bf B439} (1995) 395, hep-th/9411135.

\bibitem{TSTCP1}
J.S.~Song and Y.S.~Song, Notes from the underground: A propos of
Givental's conjecture, J.Math.Phys. {\bf 45} (2004) 4539, hep-th/0103254\\
A.~Alexandrov, Givental formula in terms of Virasoro operators,
 J.Math.Phys. {\bf 44} (2003) 5268, hep-th/0205261.

\bibitem{cmamo} T.R.~Morris, $2d$ Quantum Gravity, Multicritical
Matter and Complex Matrices, FERMILAB-PUB-90-136-T\\
S.~Dalley, C.V.~Johnson and T.~Morris,  Multicritical complex matrix
models and nonperturbative 2-D quantum gravity, Nucl.Phys. {\bf
B368} (1992) 625\\
S.~Dalley, C.V.~Johnson, T.~Morris and A.~Watterstam,  Unitary
matrix models and 2-D quantum gravity, Mod.Phys.Lett.
{\bf A7} (1992) 2753, hep-th/9206060\\
R.~Lafrance and R.C.~Meyers, Flows for rectangular matrix models,
Mod.Phys.Lett. {\bf A9} (1994) 101, hep-th/9308113\\
P.~Di Francesco, Rectangular Matrix Models and Combinatorics of
Colored Graphs, Nucl.Phys. {\bf B648} (2003) 461, cond-mat/0207682.

\bibitem{Ambjorn:1990ji}
  J.~Ambjorn, J.~Jurkiewicz and Y.M.~Makeenko,
Multiloop Correlators For Two-Dimensional Quantum Gravity,
Phys.Lett. {\bf B251} (1990) 517.

\end{thebibliography}
\end{document}